\documentclass[aps,prd,final,twocolumn,letterpaper,superscriptaddress,nofootinbib]{revtex4-1}

\usepackage{graphicx}                       
\usepackage[outdir=./]{epstopdf}
\usepackage{amsmath}
\usepackage{amssymb}
\usepackage{amsthm}
\usepackage{bbold}

\usepackage[usenames,dvipsnames,table]{xcolor}
\usepackage{array}  
\usepackage{overpic}
\usepackage{placeins} 
\usepackage{fancyhdr}
\usepackage{colortbl}
\usepackage{hyperref} 
\hypersetup{colorlinks = true, linkcolor = blue, breaklinks = true}

\newcommand{\eq}[1]{(\ref{#1})}
\newcommand{\Eq}[1]{Eq.~\eq{#1}}
\newcommand{\Eqs}[1]{Eqs.~\eq{#1}}
\newcommand{\Fig}[1]{Fig.~\ref{#1}}
\newcommand{\Sec}[1]{Sec.~\ref{#1}}
\newcommand{\Ref}[1]{Ref.~\cite{#1}}
\newcommand{\Refs}[1]{Refs.~\cite{#1}}
\newcommand{\App}[1]{Appendix~\ref{#1}}

\newcommand{\eg}{{e.g., }}
\newcommand{\ie}{{i.e., }}

\newcommand{\mc}[1]{\mathcal{#1}}

\newcommand{\msf}[1]{\mathsf{#1}}

\newcommand{\mbb}[1]{\mathbb{#1}}

\newcommand{\oper}[1]{\smash{\hat{#1}}}
\newcommand{\pd}{\partial}

\newcommand{\dd}{\mathrm{d}}
\newcommand{\fourier}[1]{\smash{\widetilde{#1}}}
\newcommand{\bra}[1]{\langle#1 |}
\newcommand{\ket}[1]{|#1 \rangle}
\newcommand{\braket}[2]{\langle#1 |  #2 \rangle}

\newcommand{\R}{\mbb{R}}
\newcommand{\dubDot}{\text{\large :}}
\newcommand{\Sp}{\text{Sp}(2N,\R)}
\newcommand{\Tr}{\text{tr}}
\newcommand{\nablaNEW}{\widetilde{\nabla}}
\newcommand{\NIMT}[1]{\mc{N}^{}_{#1}}
\newcommand{\PMT}[1]{\mc{P}^{}_{#1}}
\newcommand{\IdentOp}{\oper{\mbb{1}}}

\newcommand{\Vect}[1]{{\boldsymbol{\rm #1}}}
\newcommand{\VectOp}[1]{\oper{\Vect{#1}}}

\newcommand{\Mat}[1]{\msf{#1}}
\newcommand{\IMat}[1]{\Mat{I}_{#1}}
\newcommand{\OMat}[1]{\Mat{0}_{#1}}
\newcommand{\TpMat}{\Mat{T}_+}
\newcommand{\TmMat}{\Mat{T}_-}

\newcommand{\nullFrac}{\vphantom{\frac{}{}}}

\graphicspath{{../NIMT_Figs/}}

\begin{document}
\setlength{\parskip}{0pt}


\title{Pseudo-differential representation of the metaplectic transform and its application to fast algorithms}
\author{N. A. Lopez}
\affiliation{Department of Astrophysical Sciences, Princeton University, Princeton, New Jersey 08544, USA}
\author{I. Y. Dodin}
\affiliation{Department of Astrophysical Sciences, Princeton University, Princeton, New Jersey 08544, USA}
\affiliation{Princeton Plasma Physics Laboratory, Princeton, NJ 08543, USA}

\begin{abstract}

The metaplectic transform (MT), also known as the linear canonical transform, is a unitary integral mapping which is widely used in signal processing and can be viewed as a generalization of the Fourier transform. For a given function $\psi$ on an $N$-dimensional continuous space $\Vect{q}$, the MT of $\psi$ is parameterized by a rotation (or more generally, a linear symplectic transformation) of the $2N$-dimensional phase space $(\Vect{q},\Vect{p})$, where $\Vect{p}$ is the wavevector space dual to $\Vect{q}$. Here, we derive a pseudo-differential form of the MT. For small-angle rotations, or near-identity transformations of the phase space, it readily yields asymptotic \textit{differential} representations of the MT, which are easy to compute numerically. Rotations by larger angles are implemented as successive applications of $K \gg 1$ small-angle MTs. The algorithm complexity scales as $O(K N^3 N_p)$, where $N_p$ is the number of grid points. We present a numerical implementation of this algorithm and discuss how to mitigate the associated numerical instabilities.

\end{abstract}

\maketitle

\pagestyle{fancy}
\lhead{Lopez \& Dodin}
\rhead{Pseudo-differential metaplectic transform}
\thispagestyle{empty}


\section{Introduction}

Suppose a signal described by a square-integrable function $\psi$ of some continuous coordinate $\Vect{q}$. Like in quantum mechanics, one can introduce a `state vector' $\ket{\psi}$ such that $\psi$ be the projection of $\ket{\psi}$ onto the coordinate axis. Correspondingly, $\psi$'s Fourier image $\fourier{\psi}$ can be viewed as the projection of $\ket{\psi}$ onto the wavevector axis $\Vect{p}$, or equivalently, onto the coordinate axis obtained via rotation of the original phase space $(\Vect{q},\Vect{p})$ by $\pi/2$. But one can also introduce rotations by different angles or, most generally, linear symplectic transformations of the original phase space. Suppose a phase space $(\Vect{Q},\Vect{P})$ obtained via such transformation of $(\Vect{q},\Vect{p})$. One can then obtain $\Psi$, the projection of $\ket{\psi}$ onto the new coordinate space $\Vect{Q}$, and relate it to the original projection $\psi$ by a linear unitary mapping. This mapping is called the \textit{metaplectic transform} (MT)~\cite{Littlejohn86a,deGosson06}\footnote{It is also sometimes called the \textit{linear canonical transform}.}. It subsumes the Fourier transform as a special case and represents one of the pillars of modern phase space analysis used in many applications~\cite{Tracy93,Tracy07,Gopinathan08,Camara11,Bazarov12,Child14}.

To accommodate these applications, a number of numerical algorithms have been proposed which efficiently compute the MT on both 1-dimensional (1-D) and 2-D configuration spaces~\cite{Ozaktas96,Hennelly05b, Healy10, Koc10a, Ding12, Pei16,Sun18a}. Many of them are reviewed in \Ref{Healy18}. Despite this multitude, however, there also exist applications for which suitable MT algorithms have yet to be designed. In particular, consider the modeling of electromagnetic waves in media with slowly-varying parameters. Such waves are usually described by the equations of geometrical optics~\cite{Tracy14}, but this approach fails near reflection points, where the local wavenumber goes to zero and its derivative becomes singular. The MT provides a means to reinstate geometrical optics near reflection points, because a simple rotation of the phase space can make the wavenumber nonzero again~\cite{Littlejohn85} (see \Sec{SecRot}). It is convenient to perform such rotations consecutively along the ray trajectory at small angles; the corresponding MTs will be near-identity. Since the existing algorithms treat the MT as an integral transform, they are not optimal for computing the MT in this limit. A differential representation would be advantageous but remains to be developed.

Here, we propose an algorithm which closes this gap, as it is specifically tailored to computing near-identity MTs. We start by deriving a general pseudo-differential form of the MT. For small-angle phase-space rotations, or more generally, for any near-identity symplectic transformations of the phase space, this readily yields asymptotic \textit{differential} representations of the MT, which are easy to compute numerically. Rotations by larger angles can be implemented as successive applications of $K \gg 1$ small-angle MTs. We show that the algorithm complexity scales as $O(K N^3 N_p)$, where $N$ is the dimension of the configuration space and $N_p$ is the number of grid points. This means that our algorithm allows computing the MT in linear time, which is a faster scaling than other published MT algorithms~\cite{Healy18}, albeit with a potentially-large prefactor. We then assess the stability of our algorithm, discuss ways to optimize its performance, and present a numerical implementation.

The paper is organized as follows. In \Sec{SecMET}, we introduce the MT in a familiar setting of elementary quantum mechanics. In \Sec{SecPMT}, we derive the pseudo-differential representation of the MT from its integral representation, and we also discuss its possible truncations. In \Sec{SecIterNIMT}, we describe how the near-identity MT can be used in an iterative algorithm to perform cumulative MTs which are not near-identity. We also discuss the computational complexity and stability of such an algorithm, and demonstrate how it can be used to simulate quadratic Hamiltonian systems. In \Sec{SecRot}, we outline briefly how our new algorithm can feature in a ray-tracing code to resolve caustics, using Airy's equation as an example. In \Sec{SecCONCL}, we present our main conclusions. Auxiliary calculations are presented in appendices.


\section{Metaplectic transforms and their integral representations}
\label{SecMET}


\subsection{Special case: a quantum harmonic oscillator and its propagator as an MT}
\label{SecMETQHO}

To better understand what the MT is, let us first consider an elementary problem from quantum mechanics, namely, the quantum harmonic oscillator (QHO). The QHO is described by the Schr\"odinger equation%
\footnote{In the following, we adopt the operator notation that is standard in quantum-mechanical literature and also in optics~\Ref{Stoler81}. \textbf{Bold} font denotes vectors, \textsf{sans serif} font denotes matrices, and $\doteq$ denotes definitions.}
\begin{equation}
    i\pd_t\ket{\psi_t} = \oper{H}\ket{\psi_t} \, ,
    \quad
    \oper{H} \doteq (\oper{p}^2 + \oper{q}^2 )/2 \, .
    \label{QHO}
\end{equation}

\noindent Equation \eq{QHO} has the solution $\ket{\psi_t} = \oper{M}_t \ket{\psi_0}$, where $\ket{\psi_0}$ is an initial wavefunction and the propagator $\oper{M}_t$ is a unitary operator given by
\begin{equation}
    \oper{M}_t = \exp\left(-i\oper{H}t\right) \, .
    \label{metOPERATOR}
\end{equation}
 
An interesting property of $\oper{M}_t$ is revealed by switching from the Schr\"odinger representation to the Heisenberg representation, in which the wavefunction is fixed but $\oper{q}$ and $\oper{p}$ evolve in time as governed by~\cite{Shankar94}
\begin{subequations}
    \label{HeisEQ}
    \begin{align}
        \pd_t (\oper{M}_t^\dagger \oper{q} \oper{M}_t ) &= i \oper{M}_t^\dagger \left[\oper{H},\oper{q} \right] \oper{M}_t = \oper{M}_t^\dagger \oper{p} \oper{M}_t \, ,\\
        \pd_t (\oper{M}_t^\dagger \oper{p} \oper{M}_t ) &= i \oper{M}_t^\dagger \left[\oper{H},\oper{p} \right] \oper{M}_t = -\oper{M}_t^\dagger \oper{q} \oper{M}_t \, .
    \end{align}
\end{subequations}
 
\noindent The coordinate and momentum operators of the QHO are seen to satisfy the same Hamilton's equations that describe a classical harmonic oscillator~\cite{Goldstein02}. The solution to \Eqs{HeisEQ} is therefore
\begin{equation}
    \oper{Q} = \cos(t) \oper{q} + \sin(t) \oper{p} \, ,
    \quad
    \oper{P} = -\sin(t) \oper{q} + \cos(t) \oper{p} \, ,
    \label{HeisSOL}
\end{equation}

\noindent where we introduced
\begin{equation}
    \oper{Q} \doteq \oper{M}_t^\dagger \oper{q} \oper{M}_t \, ,
    \quad
    \oper{P} \doteq \oper{M}_t^\dagger \oper{p} \oper{M}_t \, .
\end{equation}

\noindent Equations \eq{HeisSOL} can be considered as a mapping $(\oper{q},\oper{p}) \mapsto (\oper{Q},\oper{P})$ which is a phase-space rotation by angle $t$. The unitary propagator $\oper{M}_t$ that effects this rotation is called a metaplectic operator%
\footnote{Here, $\oper{M}_t$ also acts as the \textit{fractional Fourier transform} operator, up to a phase.}.

The metaplectic operator also induces a mapping between the \textit{projections} of $\ket{\psi_0}$ onto the original coordinate axis $q$ and onto the new axis $Q$. The former is defined as $\psi(x) \doteq \braket{q(x)}{\psi_0}$, where $\ket{q(x)}$ is the eigenvector of $\oper{q}$ corresponding to the eigenvalue $x$. Likewise, the projection onto $Q$ is $\Psi(y) \doteq \braket{Q(y)}{\psi_0}$, where $\ket{Q(y)}$ is the eigenvector of $\oper{Q}$ corresponding to the eigenvalue $y$. We assume the usual normalization, $\braket{q(x)}{q(y)} = \braket{Q(x)}{Q(y)} = \delta(x-y)$, so
\begin{equation}
    \int \dd x \, \ket{q(x)}\bra{q(x)} = \int \dd x \, \ket{Q(x)}\bra{Q(x)} = \IdentOp
\end{equation}

\noindent and $\ket{Q(x)} = \oper{M}_t^\dagger \ket{q(x)}$. (Here $\IdentOp$ is a unit operator.) Then,
\begin{align}
    \Psi(y) &= \int \dd x \, \braket{Q(y)}{q(x)} \braket{q(x)}{\psi_0}\nonumber\\
    &= \int \dd x \, \bra{q(y)}\oper{M}_t \ket{q(x)} \psi_0(x) \, .
    \label{MT0}
\end{align}

\noindent Note that the right-hand side of \eq{MT0} is the same as $\psi_t(y) \doteq \bra{q(y)}\oper{M}_t \ket{\psi_0}$, because in our example $\oper{M}_t$ is the propagator. Hence, for the QHO considered here, the MT can be equivalently understood as the evolution of the wavefunction in the Schr\"odinger representation, $\ket{\psi_0} \mapsto \ket{\psi_t}$, or as the evolution of the projection basis in the Heisenberg representation, $\bra{q(y)} \mapsto \bra{Q(y)}$.

Finally, let us notice the following. As is well-known, the eigenvalues of the QHO Hamiltonian are~\cite{Shankar94}
\begin{equation}
	\oper{H}\ket{n} = (n + 1/2)\ket{n} \, ,
	\label{QHOmet}
\end{equation}

\noindent with $n$ an integer and $\ket{n}$ the $n$-th eigenstate of $\oper{H}$; hence, the specific MT considered in \Eq{metOPERATOR} can also be represented as
\begin{equation}
    \oper{M}_t = \exp\left(-it/2 \right) \sum_{n=0}^\infty \exp\left(-int \right)\ket{n}\bra{n} \, .
    \label{metDIAG}
\end{equation}

\begin{figure}
	\includegraphics[width=0.9\linewidth,trim={2mm 5mm 13mm 2mm},clip]{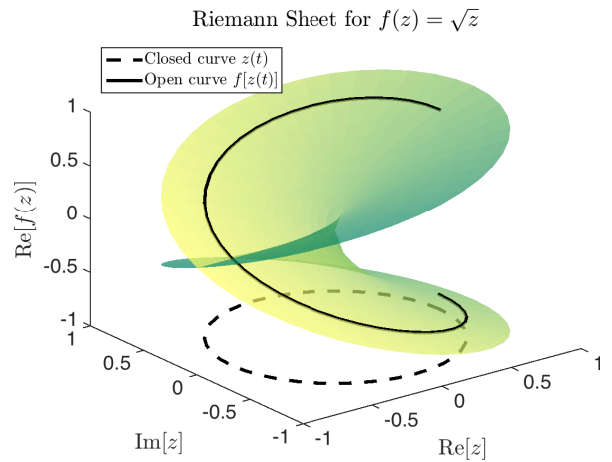}
	\caption{The Riemann sheet (colored) of the function $f(z) \doteq \sqrt{z}$ illustrates the relationship between the family $\{\oper{M}_t\}$ (more generally, the metaplectic group) and the family of all phase-space rotations (more generally, the symplectic group) for all~$t$. As depicted in the figure, $f$ maps a closed curve on the $z$ plane (dashed) to a closed curve on the Riemann sheet only if the winding number is even. Likewise, it takes two rotation periods for $\oper{M}_t$ to return to its original value $\oper{M}_0 = \IdentOp$. In a more general formulation, the metaplectic group forms a double-cover of the symplectic group. For details, see \Ref{Littlejohn86a}.}
	\label{riemann}
\end{figure}

\noindent A notable aspect of this formula is that it takes not one but \textit{two} rotation periods ($t=4\pi$) for $\oper{M}_t$ to return to its original value $\oper{M}_0 = \IdentOp$. More generally, $\oper{M}_{2\pi n} = \IdentOp$ for even $n$ yet $\oper{M}_{2\pi n} = -\IdentOp$ for odd $n$. Hence, the same identity transformation on phase space [governed by \Eq{HeisSOL}] can be effected by two distinct metaplectic operators, $\pm \IdentOp$. This double-valuedness also holds for arbitrary rotation angles, and is in fact a general property of the MT. This is illustrated by analogy with the behavior of the complex function $f(z)\doteq \sqrt{z}$ in \Fig{riemann}.


\subsection{General definition of the MT}

A more general definition of the MT is as follows. Let $\VectOp{q}$ and $\VectOp{p}$ be respectively the $N$-dimensional coordinate and momentum operators. Consider
\begin{equation}
    \VectOp{Q} \doteq \oper{M}^\dagger \VectOp{q} \oper{M}\, ,
    \quad
    \VectOp{P} \doteq \oper{M}^\dagger \VectOp{p} \oper{M}\, ,
\end{equation}

\noindent where $\oper{M}$ is a unitary operator such that
\begin{equation}
    \begin{pmatrix}
        \VectOp{Q}\\[0.5mm]
        \VectOp{P}
    \end{pmatrix} = \Mat{S}
    \begin{pmatrix}
        \VectOp{q}\\
        \VectOp{p}
    \end{pmatrix} \, ,
    \quad
    \Mat{S} =
    \begin{pmatrix}
        \Mat{A} & \Mat{B}\\
        \Mat{C} & \Mat{D}
    \end{pmatrix} \, ,
    \label{CanonTRANS}
\end{equation}

\noindent and $\Mat{S}$ is real and \textit{symplectic}. The latter means that
\begin{equation}
    \Mat{S}
    \begin{pmatrix}
        \OMat{N} & \IMat{N}\\
        -\IMat{N} & \OMat{N}
    \end{pmatrix} \Mat{S}^\intercal = 
    \begin{pmatrix}
        \OMat{N} & \IMat{N}\\
        -\IMat{N} & \OMat{N}
    \end{pmatrix} \, ,
    \label{SymplecS}
\end{equation}

\noindent which implies (cf. \App{AppABCDderiv}) Luneburg's relations~\cite{Luneburg64}
\begin{subequations}
    \label{SymplecABCD}
    \begin{align}
        \label{symplec1}
        \Mat{A} \Mat{D}^\intercal - \Mat{B} \Mat{C}^\intercal &= \IMat{N} \, ,\\
        \label{symplec2}
        \Mat{A}^\intercal \Mat{D} - \Mat{C}^\intercal \Mat{B} &= \IMat{N} \, ,\\
        \label{symplec3}
        \Mat{A} \Mat{B}^\intercal - \Mat{B} \Mat{A}^\intercal &= \OMat{N} \, ,\\
        \label{symplec4}
        \Mat{B}^\intercal \Mat{D} - \Mat{D}^\intercal \Mat{B} &= \OMat{N} \, ,\\
        \label{symplec5}
        \Mat{C}^\intercal \Mat{A} - \Mat{A}^\intercal \Mat{C} &= \OMat{N} \, ,\\
        \label{symplec6}
        \Mat{D} \Mat{C}^\intercal - \Mat{C} \Mat{D}^\intercal &= \OMat{N} \, ,
    \end{align}
\end{subequations}

\noindent where $\OMat{N}$ and $\IMat{N}$ denote respectively the $N\times N$ null and identity matrices\footnote{Note that at $N=1$, \Eqs{symplec3}-\eq{symplec6} are satisfied automatically, and \Eqs{symplec1} and \eq{symplec2} are equivalent to $\det \Mat{S} = 1$; hence, a $2 \times 2$ matrix is symplectic if and only if it has unit determinant.}. Then, $\oper{M}$ is called the metaplectic operator corresponding to the chosen $\Mat{S}$.

Like in the previous section, we now define the MT as the mapping between a given function $\psi$ on the coordinate space associated with $\VectOp{q}$ and the projection of the corresponding state vector $\ket{\psi}$ on the coordinate space associated with $\VectOp{Q}$. Again, this leads to\footnote{Analogous to the Schr\"odinger and Heisenberg representations of time evolution, there exists in the general case a distinction between whether $\oper{M}$ transforms the wavefunction (`active' representation) or transforms the projection basis (`passive' representation). In our discussion, we assume the passive representation.}
\begin{subequations}
    \begin{align}
        \Psi(\Vect{y}) &= \int \dd \Vect{x} \, U(\Vect{y},\Vect{x}) \psi(\Vect{x}) \, ,\\
        U(\Vect{y},\Vect{x}) &= \braket{\Vect{Q}(\Vect{y})}{\Vect{q}(\Vect{x})} = \bra{\Vect{q}(\Vect{y})}\oper{M} \ket{\Vect{q}(\Vect{x})} \, .
    \end{align}
\end{subequations}

\noindent To calculate $U$, let us consider the top row of \Eq{CanonTRANS}, $\VectOp{Q} = \Mat{A}\VectOp{q} + \Mat{B}\VectOp{p}$, and apply $\bra{\Vect{Q}(\Vect{y})}$ from the left and $\ket{\Vect{q}(\Vect{x})}$ from the right. Using the eigenvalue relations along with
\begin{equation}
    \bra{\Vect{Q}(\Vect{y})} \VectOp{p} \ket{\Vect{q}(\Vect{x})} = \left[\bra{\Vect{q}(\Vect{x})} \VectOp{p} \ket{\Vect{Q}(\Vect{y})}\right]^* = i\partial_\Vect{x} U(\Vect{y},\Vect{x})
\end{equation}

\noindent leads to a differential equation~\cite{Littlejohn86a,Moshinsky71}
\begin{equation}
    \Vect{y} U(\Vect{y},\Vect{x}) = \left(\Mat{A} \Vect{x} + i \Mat{B} \pd_{\Vect{x}} \right) U(\Vect{y},\Vect{x})  \, ,
\end{equation}

\noindent which can be solved to yield
\begin{equation}
    U(\Vect{y},\Vect{x}) = f(\Vect{y}) \, e^{\frac{i}{2} \Vect{x}^\intercal \Mat{B}^{-1}\Mat{A}\Vect{x} - i \Vect{x}^\intercal \Mat{B}^{-1} \Vect{y}} \, .
    \label{xSOL}
\end{equation}

\noindent Doing the same with the bottom row of \Eq{CanonTRANS} leads to
\begin{equation}
    \pd_{\Vect{y}} U(\Vect{y},\Vect{x}) = \left(i\Mat{C} \Vect{x} - \Mat{D} \pd{\Vect{x}}\right) U(\Vect{y},\Vect{x}) \, .
    \label{pde2}
\end{equation}

\noindent Using \Eqs{SymplecABCD}, \eq{xSOL}, and \eq{pde2} determines $f(\Vect{y})$ up to a multiplicative constant:
\begin{equation}
    f(\Vect{y}) = \alpha \, e^{\frac{i}{2} \Vect{y}^\intercal \Mat{D}\Mat{B}^{-1}\Vect{y}} \, .
\end{equation}

Normalization determines the constant $\alpha$ up to a phase. The phase requires more involved analysis to determine, and the result is not unique: there exist two possible phases which differ by $\pi$. This ambiguity is required to ensure that the metaplectic operators form a group, but results in a one-to-two correspondence between the symplectic and the metaplectic groups~\cite{Littlejohn86a}. In other words, changing the overall sign of a metaplectic operator does not change the resulting phase-space transformation, which \Eqs{HeisSOL} and \eq{metDIAG} demonstrate for the QHO example. As discussed in \Sec{SecMETQHO} (and also related to the general Bohr-Sommerfeld rule~\cite{Shankar94}), the sign ambiguity becomes important when one considers a family of transformations parameterized by some path variable $t$. A closed trajectory in the space of symplectic matrices, $\Mat{S}_t$, results in a closed trajectory in the space of metaplectic operators only for even winding numbers. In contrast, for odd winding numbers $\oper{M}_t$ changes sign, just like the function $f(z) \doteq \sqrt{z}$ changes sign each time $z$ encircles the origin in the complex plane~\cite{Littlejohn86a} (see \Fig{riemann}).

Including the phase and sign ambiguity, the final result for the transformation is~\cite{Collins70,Moshinsky71,Littlejohn86a}
\begin{align}
    \Psi(\Vect{Q}) &= \pm\frac{ e^{\frac{i}{2}\Vect{Q}^\intercal \Mat{D} \Mat{B}^{-1} \Vect{Q}}}{(2\pi i)^{\frac{N}{2}}\sqrt{\det{\Mat{B}}}}\nonumber\\
    &\hspace{4mm} \times \int \dd\Vect{q}~ e^{\frac{i}{2}\Vect{q}^\intercal \Mat{B}^{-1}\Mat{A} \Vect{q} - i\Vect{q}^\intercal \Mat{B}^{-1} \Vect{Q}} \, \psi(\Vect{q}) \, ,
    \label{metTRANS}
\end{align}

\noindent where $\Mat{B}^{-1}\Mat{A}$ and $\Mat{D} \Mat{B}^{-1}$ are symmetric due to \Eqs{symplec3} and \eq{symplec4}. Equation \eq{metTRANS} defines $\Psi(\Vect{Q})$ as the MT image of $\psi(\Vect{q})$. In writing \Eq{metTRANS}, we have dropped the $\Vect{x}$ and $\Vect{y}$ notation in favor of $\Vect{q}$ and $\Vect{Q}$, as there is no longer any risk of ambiguity, and our branch cut convention restricts all complex phases to the interval $(-\pi,\pi]$.


\section{Pseudo-differential representation of the Metaplectic Transform}
\label{SecPMT}

Here, we develop a pseudo-differential representation of \Eq{metTRANS}. This representation is particularly useful when $\Mat{A}^{-1}\Mat{B}$ is small, because then the MT can be approximated by a finite-order differential transform, which is easier to evaluate than the integral transform of \Eq{metTRANS}. Specifically, we proceed as follows. Using the substitution $\Vect{u} \doteq \Vect{q} - \Mat{A}^{-1} \Vect{Q}$, \Eq{metTRANS} can be re-written as
\begin{align}
	\Psi(\Vect{Q}) &= \pm \frac{e^{i\Vect{Q}^\intercal \Mat{G} \Vect{Q}}}{(2\pi i)^{\frac{N}{2}}\sqrt{\det{\Mat{B}}}}\nonumber\\
	&\hspace{4mm} \times \int \dd\Vect{u}~ e^{i \Vect{u}^\intercal \Lambda^{-1} \Vect{u}} \, \psi(\Mat{A}^{-1} \Vect{Q} + \Vect{u}) \, ,
	\label{NIMTtrans}
\end{align}

\noindent where we have defined the matrices
\begin{equation}
    \Mat{G} \doteq \Mat{C} \Mat{A}^{-1}/2 \, , \quad \Lambda \doteq 2\Mat{A}^{-1} \Mat{B} \, .
\end{equation}

\noindent Notably, both $\Mat{G}$ and $\Lambda$ are symmetric per \Eqs{symplec3} and \eq{symplec5}. In the following, we shall assume that $\Lambda$ is small. This assumption is not strictly necessary, since the final result is convergent for all values of $\|\Lambda \|$ and thereby possesses a natural analytic continuation; however, it aids intuition in the forthcoming derivation.


\subsection{1-D case}
\label{SecPMT1D}

Let us first consider the $1$-D case ($N = 1$) for simplicity. Since $\Lambda^{-1}$ is assumed large, only small values of $u$ will contribute to the integral of \Eq{NIMTtrans}. Therefore, we can expand the function $\psi\left(Q/A+u\right)$ around the point $u = 0$ as
\begin{equation}
	\psi\left(\frac{Q}{A} + u\right) = \sum_{n=0}^{\infty} \frac{u^n}{n!} \, \psi^{(n)}\left(\frac{Q}{A}\right) \, ,
	\label{Taylor1D}
\end{equation}

\noindent where $\psi^{(n)}(Q/A)$ is the $n$-th derivative of $\psi(q)$ evaluated at $q = Q/A$. (Here, we assume that $\psi$ is smooth, but we shall revisit this assumption below.) Hence,
\begin{align}
	&\int_{-\infty}^{\infty} \dd u~ e^{i \Lambda^{-1} u^2} \, \psi\left(\frac{Q}{A} + u\right)\nonumber\\ 
	&\sim \sum_{n=0}^\infty \frac{1}{n!} \, \psi^{(n)}\left(\frac{Q}{A}\right) \int_{-\infty}^{\infty} \dd u~u^n \, e^{i \Lambda^{-1} u^2} \, .
	\label{asymINT}
\end{align}

\noindent By parity, all integrals with odd powers of $u$ are identically zero, so the sum can be written solely in terms of even powers as
\begin{align}
	&\sum_{n=0}^\infty \frac{1}{n!} \, \psi^{(n)}\left(\frac{Q}{A}\right) \int_{-\infty}^{\infty} \dd u~u^n \, e^{i \Lambda^{-1} u^2}\nonumber\\
	&= \sum_{n=0}^\infty \frac{1}{(2n)!} \, \psi^{(2n)}\left(\frac{Q}{A}\right) \int_{-\infty}^{\infty} \dd u~u^{2n} \, e^{i \Lambda^{-1} u^2} \, .
	\label{asymEVEN}
\end{align}

Let us introduce a dummy multiplicative variable $s$, which will eventually be taken to unity. Then, since
\begin{equation}
	\pd^n_s \, e^{i s \Lambda^{-1} u^2} = \left(-i \Lambda \right)^{-n} u^{2n} \, e^{i s \Lambda^{-1} u^2} \, ,
\end{equation}

\noindent we obtain
\begin{align}
	\int_{-\infty}^{\infty} \dd u~u^{2n} \, e^{i \Lambda^{-1} u^2} &= \frac{\Lambda^n}{i^n} \left.\pd^n_s \int_{-\infty}^{\infty} \dd u~ e^{i s \Lambda^{-1} u^2}\right|_{s=1}\nonumber\\
	&= \frac{\Lambda^n}{i^n} \left.\pd^n_s \left(\sqrt{\pi \Lambda i}~s^{-\frac{1}{2}}\right)\right|_{s=1}\nonumber\\
	&= \frac{\Lambda^n}{i^n} \sqrt{\pi \Lambda i} \, \frac{\Gamma\left(\frac{1}{2}\right)}{\Gamma\left(\frac{1}{2}-n\right)} \, ,
	\label{1DFeynman}
\end{align}

\noindent where the first line invokes Leibniz's rule, the final equality follows from the binomial theorem~\cite{Olver10}, and $\Gamma(z)$ is the gamma function~\cite{Olver10}. By combining \Eqs{asymINT}, \eq{asymEVEN}, and \eq{1DFeynman}, we obtain the asymptotic representation
\begin{align}
	&\int_{-\infty}^{\infty} \dd u~ e^{i \Lambda u^2}\psi\left(\frac{Q}{A} + u\right)  \nonumber\\
	&\sim \sum_{n=0}^\infty  \frac{\Lambda^n}{i^n} \frac{\Gamma\left(\frac{1}{2}\right) \sqrt{\pi \Lambda i}}{\Gamma\left(2n+1\right)\Gamma\left(\frac{1}{2}-n\right)} \, \psi^{(2n)}\left(\frac{Q}{A}\right) \, .
	\label{NIMTunsimple}
\end{align}

\noindent Finally, using well-known properties of the gamma function yields the pseudo-differential representation of the MT in $1$-D:
\begin{subequations}
    \label{NIMT1D}
	\begin{equation}
	    \label{NIMTseries1D}
		\Psi(Q) = \pm \frac{e^{iG Q^2}}{\sqrt{A}} \, \sum_{n=0}^\infty \frac{\left(i \Lambda/4 \right)^n}{n!} \, \psi^{(2n)}\left(\frac{Q}{A}\right) \, ,
	\end{equation}
	
	\noindent or symbolically,
	\begin{equation}
	    \label{NIMTsymb1D}
		\Psi(Q) = \pm\left. \frac{e^{iG Q^2}}{\sqrt{A}} \, \exp \left(\frac{i\Lambda}{4} \, \pd^2_q \right) \psi \left(q \right)\right|_{q=Q/A} \, .
	\end{equation}
\end{subequations}

\noindent We can also express \Eq{NIMTsymb1D} in an equivalent vector form $\ket{\Psi} = \oper{M} \ket{\psi}$, where $\oper{M}$ is the manifestly-unitary MT operator given as
\begin{equation}
    \oper{M} \doteq \pm (A)^{-1/2} \oper{D}_{A} \, e^{\frac{i}{2} AC \oper{q}^2} \, e^{-i \frac{\Lambda}{4} \oper{p}^2} \, ,
    \label{NIMToper1D}
\end{equation}

\noindent and $\oper{D}_A$ is the inverse dilation operator defined via its effect in the spatial representation, $\bra{q}\oper{D}_A \ket{\psi} \doteq \psi(q/A)$.

We call \Eqs{NIMT1D} the 1-D pseudo-differential metaplectic transform (PMT). Although the above derivation assumes smooth $\psi$, the final result can be understood more generally, which is why the asymptotic relation has been replaced with an exact equality. As shown in Appendix B, the operator \eq{NIMToper1D} has exactly the same kernel as the original integral MT \eq{metTRANS} and exists on the space of all functions which have a well-defined Fourier transform; i.e., smoothness of $\psi$ is not required. In this sense, \Eq{NIMTsymb1D} should be understood not as a symbolic representation of the series \eq{NIMTseries1D} (whose convergence depends on details of $\psi$) but rather as a symbolic representation of the integral MT \eq{metTRANS}. This new representation is advantageous in that it is compact, and facilitates asymptotic expansions of the MT to any order of $\Lambda$.

Let us also discuss the case when $\Lambda$ is small and $\psi$ is smooth enough so \Eq{NIMTseries1D}  can be approximated with a truncated series. We define the $m$-th order near-identity metaplectic transform (NIMT) as the truncation of \Eq{NIMTseries1D} that neglects all terms with $n > m$. This nomenclature is chosen because up to a phase, the limit $B \to 0$ reduces \Eqs{NIMT1D} to a scaled-identity operation. Also, to be connected with the identity, we explicitly choose the overall $+$ sign when performing NIMT truncations. Decreasing $m$ will increase the locality of the truncated transformation, because the necessary stencil width to compute the $m$-th order NIMT will decrease. This enables the $m$-th order NIMT to be performed pointwise, as the transformed function evaluated at some point $Q = Q_0$ depends only on the original function and its first $2m$ derivatives evaluated at the corresponding point $q = q_0(Q_0)$.

When the order is not specified, the `NIMT' refers solely to the first-order NIMT,
\begin{equation}
    \Psi(Q) \approx \frac{e^{i\frac{C}{2A}Q^2}}{\sqrt{A}}\left[\psi\left(\frac{Q}{A} \right) + \frac{iB}{2A} \, \psi''\left(\frac{Q}{A} \right)\right] \, ,
    \label{NIMT1DTrunc}
\end{equation}

\noindent as it is the lowest-order truncation that remains practical. (The truncation at $m=0$ is too simplified to yield an accurate representation of the MT, regardless the smoothness of $\psi$.) We shall make use of \Eq{NIMT1DTrunc} in \Sec{SecIterNIMT}.


\subsection{N-D case}

The generalization from 1-D to the arbitrary $N$-D case is straightforward. We consider again the integral of \Eq{NIMTtrans}. Since $\Lambda$ is a symmetric matrix, by the spectral theorem it can be diagonalized. Let us enumerate with subscripts $j \in \{1,\ldots,N \}$ vector components with respect to the diagonalizing basis of $\Lambda$. Then,
\begin{align}
	&\int \dd\Vect{u} \, e^{i \Vect{u}^\intercal \Lambda^{-1} \Vect{u}} \, \psi(\Mat{A}^{-1}\Vect{Q} + \Vect{u}) \nonumber\\
	&= \int \dd u_1 \, e^{i \lambda_1^{-1} u_1^2} \ldots \dd u_N \, e^{i \lambda_N^{-1} u_N^2} \,\psi(\Mat{A}^{-1}\Vect{Q} + \Vect{u}) \, ,
\end{align}

\noindent where $\lambda_j$ is the $j$-th eigenvalue of $\Lambda$. As before, $\psi(\Mat{A}^{-1}\Vect{Q} + \Vect{u})$ is expanded around $\Vect{u} = 0$. In multiple dimensions, this expansion is written as
\begin{align}
	&\psi(\Mat{A}^{-1}\Vect{Q} + \Vect{u})\nonumber\\
	&= \sum_{n_1 = 0}^\infty \ldots \sum_{n_N = 0}^\infty \frac{u_1^{n_1}}{n_1!}\ldots \frac{u_N^{n_N}}{n_N!} \, \psi^{(n_1, \ldots , n_N)} (\Mat{A}^{-1}\Vect{Q} ) \, ,
\end{align}

\noindent with the shorthand notation 
\begin{equation}
	\psi^{(n_1,\ldots , n_N)} (\Mat{A}^{-1}\Vect{Q}) \doteq \left. \frac{\partial^{n_1+ \ldots +n_N}}{\partial q_1^{n_1} \ldots \partial q_N^{n_N}} \psi(\Vect{q}) \right|_{\Vect{q} = \Mat{A}^{-1}\Vect{Q}}
\end{equation}

\noindent denoting the derivatives of $\psi$ along the eigenvectors of $\Mat{A}^{-1}\Mat{B}$. The integral therefore becomes
\begin{align}
	&\int \dd\Vect{u}~ e^{i \Vect{u}^\intercal \Lambda^{-1} \Vect{u}} \, \psi(\Mat{A}^{-1}\Vect{Q} + \Vect{u})\nonumber\\
	&\sim \sum_{n_1 = 0}^\infty \ldots \sum_{n_N = 0}^\infty \frac{1}{\left(2n_1\right)! \ldots \left(2n_N\right)!} \, \psi^{(2n_1, \ldots , 2n_N)} (\Mat{A}^{-1}\Vect{Q})\nonumber\\
	&\times \int \dd u_1~u_1^{2n_1} e^{\frac{i}{2}\lambda_1^{-1} u_1^2} \ldots \int \dd u_N~ u_N^{2n_N}e^{\frac{i}{2}\lambda_N^{-1} u_N^2} \, ,
\end{align}

\noindent where once again, the summation has been restricted to even integers by parity considerations.

The remaining integrals are of the same form as those in \Eq{1DFeynman}. Hence, the $N$-D PMT is ultimately obtained as
\begin{subequations}
    \begin{align}
        \label{NIMTseriesND}
	    \Psi(\Vect{Q}) &= \pm\frac{e^{i \Vect{Q}^\intercal \Mat{G} \Vect{Q}}}{\sqrt{\det{\Mat{A}}}} \, \sum_{n_1=0}^\infty \ldots \sum_{n_N=0}^\infty \frac{\left(i \lambda_1/4 \right)^{n_1} \ldots \left(i \lambda_N/4 \right)^{n_N}}{n_1! \ldots n_N!}\nonumber\\
	    &\hspace{4mm} \times \psi^{(2n_1, \ldots , 2n_N)} \left(\Mat{A}^{-1}\Vect{Q}\right) \, ,
    \end{align}

    \noindent or symbolically,
    \begin{equation}
        \Psi(\Vect{Q}) = \pm\left. \frac{e^{i\Vect{Q}^\intercal \Mat{G} \Vect{Q}}}{\sqrt{\det{\Mat{A}}}} \, \exp\left(\frac{i}{4} \, \Lambda \dubDot \nabla \nabla \right) \psi \right|_{\Vect{q} = \Mat{A}^{-1}\Vect{Q}} \, ,
        \label{NIMTsymbND}
    \end{equation}
\end{subequations}

\noindent where the notation $\Lambda \dubDot \nabla \nabla$ denotes the double dot product between $\Lambda$ and the Hessian operator $\nabla \nabla$; i.e., $\Lambda:\nabla\nabla = \Lambda_{ab}(\pd_{x_a})(\pd_{x_b})$ summed over common indices. In this case, the equivalent operator representation for \Eq{NIMTsymbND} uses
\begin{equation}
    \oper{M} \doteq \pm \sqrt{\det \Mat{A}^{-1}} \, \oper{D}_{\Mat{A}} \, e^{\frac{i}{2} \VectOp{q}^\intercal \Mat{A}^\intercal \Mat{C} \VectOp{q}} \, e^{-i \VectOp{p}^\intercal\frac{\Lambda}{4} \VectOp{p}} \, ,
    \label{NIMToperND}
\end{equation}

\noindent and $\bra{\Vect{q}}\oper{D}_{\Mat{A}} \ket{\psi} \doteq \psi(\Mat{A}^{-1}\Vect{q})$. Retaining only the terms corresponding to $\sum_{j=1}^N n_j = 0$ and $\sum_{j=1}^N n_j = 1$ in \eq{NIMTseriesND}, the $N$-D NIMT is
\begin{align}
	\Psi(\Vect{Q}) \approx \frac{e^{\frac{i}{2}\Vect{Q}^\intercal \Mat{C}\Mat{A}^{-1} \Vect{Q}}}{\sqrt{\det{\Mat{A}}}}\left[ \psi + \frac{i}{2} \Tr \left( \Mat{A}^{-1}\Mat{B} \, \nabla\nabla\psi \right) \right] \, ,
	\label{NIMTNDTrunc}
\end{align}

\noindent where the term in brackets is evaluated at $\Vect{q} = \Mat{A}^{-1}\Vect{Q}$, and the overall $+$ sign is assumed, as in \Sec{SecPMT1D}. Since matrix operations can be computationally expensive when $N$ is large, \App{AppMat} provides some low-order approximations for $\det{\Mat{A}}$, $\Mat{A}^{-1}$, $\Mat{A}^{-1}\Mat{B}$, and $\Mat{C} \Mat{A}^{-1}$ for use when $\Mat{S}$ is near-identity. We also provide auxiliary calculations when $\psi(\Vect{q})$ is eikonal in \App{AppEik}.


\section{Finite transformations via iterated Near-Identity transformations}
\label{SecIterNIMT}

The pseudo-differential representation of the MT naturally gives rise to an iterative algorithm: successive applications of the NIMT can compute a finite transformation from a sequence of near-identity transformations. To see this, consider the MT of a function $\psi$ that results from the symplectic matrix $\Mat{S}$, which may be the result of a single optical operation or a cascade of operations. As the symplectic group is topologically connected, it is always possible to find a smooth trajectory of symplectic matrices $\Mat{S}_t$ with parameterization $t$ such that $\Mat{S}_0 = \IMat{2N}$ and $\Mat{S}_T = \Mat{S}$ at some final $T$.

Let us discretize $\Mat{S}_t$ with a uniform step size $\Delta t \doteq T/K$ such that $\forall~k\in \{1, 2, \ldots, K\}$, the matrix $\Mat{S}_{(k-1)\Delta t}^{-1} \Mat{S}_{k\Delta t}$ is near-identity. Then, since
\begin{equation}
	\Mat{S} = \Mat{S}_T = \Mat{S}_{\Delta t}^{} \, \Mat{S}_{\Delta t}^{-1} \, \Mat{S}_{2\Delta t}^{} \, \Mat{S}_{2\Delta t}^{-1} \, \ldots \, \Mat{S}_{(K-1)\Delta t}^{-1} \, \Mat{S}_{K\Delta t}^{} \, ,
\end{equation}

\noindent one can compute the MT associated with $\Mat{S}$ by iteratively applying the NIMT: first the NIMT associated with $\Mat{S}_{\Delta t}^{}$ (which is near-identity by definition), next the NIMT associated with $\Mat{S}_{\Delta t}^{-1} \, \Mat{S}_{2\Delta t}^{}$, and so forth until finally, the NIMT associated with $\Mat{S}_{T-\Delta t}^{-1} \, \Mat{S}_T$. Hence,
\begin{equation}
	\Psi \approx \NIMT{\Mat{S}_{T-\Delta t}^{-1} \Mat{S}_T^{}}\left\{\ldots \NIMT{\Mat{S}_{\Delta t}^{-1} \Mat{S}_{2\Delta t}^{} }\left[\NIMT{\Mat{S}_{\Delta t}^{}}\left(\psi \right)  \right] \right\} \, ,
	\label{PathDiscr}
\end{equation}

\noindent where $\NIMT{\Mat{S}}$ is the NIMT associated with symplectic matrix~$\Mat{S}$.

Note that the discretization of $\Mat{S}_t$ by itself does not incur any errors, so the accuracy of \Eq{PathDiscr} depends solely on the truncation order of the NIMT. Another advantage of this approach is that the algorithm is independent of the dimensionality. One only needs to adjust the size of $\Mat{S}$ when changing from, say, a $1$-D application to a $3$-D application. This is not true for other numerical MT algorithms in the literature, which can only handle up to $2$-D and are explicitly different depending on whether $\Mat{S}$ is `separable' or `nonseparable'~\cite{Koc10a,Ding12,Pei16}. Such restrictions do not arise with the iterated NIMT.


\subsection{Computational efficiency}

Let us estimate the computational efficiency of the iterated NIMT. We should first emphasize that although the NIMT appears to require interpolation, this is not strictly necessary. Suppose that $\psi(\Vect{q})$ is only known on a discrete set of points $\{\Vect{q}_k\}$. The discretization of $\psi(\Vect{q})$ can be used to inform the discretization of $\Psi(\Vect{Q})$ by evaluating the NIMT only at the corresponding points $\{\Vect{Q}_k \doteq \Mat{A}\Vect{q}_k \}$. No interpolation is required, unless, one needs to evaluate $\Psi(\Vect{Q})$ off-grid. In that case, either the discrete set $\{\psi(\Vect{q}_k)\}$ must be interpolated and transformed, or the discrete set $\{\Psi(\Vect{Q}_k) \}$ must be interpolated. For this reason, and because interpolation efficiency is highly implementation-specific, we do not account for interpolation in our runtime estimate.

From \Eq{NIMT1DTrunc}, evaluating $\Psi(Q)$ at $N_p$ discrete points using the $1$-D NIMT requires only $O(N_p)$ floating-point operations (FLOPs). For the $N$-D case, this estimate becomes $O(N^3 N_p)$, since each evaluation includes a matrix multiplication~\cite{Trefethen97}. Thus, the NIMT always scales linearly with the number of sample points, independent of dimensionality. The iterated NIMT remains `fast' with respect to the number of sample points, since the FLOP count scales as $O(K N^3 N_p)$, with $K$ the number of iterations. The linear scaling is faster than many of the other MT algorithms found in the literature~\cite{Healy18}, which scale as $O(N_p \log N_p)$.


\subsection{Computational stability}

\begin{figure}
	\includegraphics[width=0.9\linewidth,trim={4mm 4mm 3mm 3mm},clip]{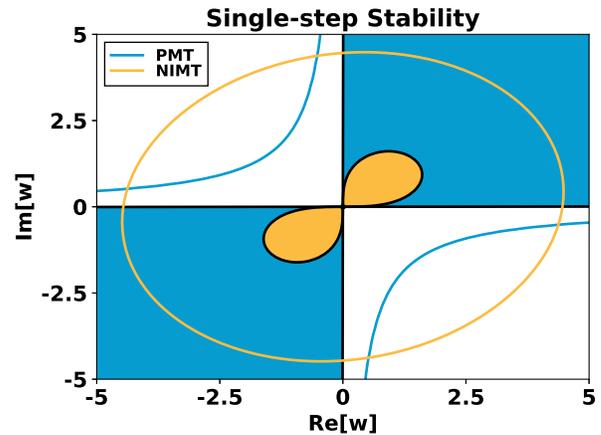}
	\caption{Stability diagrams for the PMT [blue shaded, from \Eq{PMTstabCond}] and the NIMT [orange shaded, from \Eq{NIMTstabCond}]. The solid lines of each color mark the contour at which the magnification factor of the respective algorithm equals 10.}
	\label{1StepStab}
\end{figure}

Although the iterated NIMT scales faster than other published MT algorithms, it may not be as stable. Intuitively, one would expect that refining the discretization of $\Mat{S}_t$ would increase the accuracy of the iterated NIMT, since the magnitude of $\|\Mat{A}^{-1}_j\Mat{B}_j\|$ for each successive $j$-th application of the NIMT would decrease. As the magnitude of $\|\Mat{A}^{-1}_j \Mat{B}_j\|$ decreases, however, the number of iterations required to generate a fixed final transformation increases. Careful analysis is needed to determine if the truncation errors of the iterated NIMT accumulate coherently, which we accomplish by estimating the parameter regimes in which the iterated NIMT is non-unitary. For simplicity, the forthcoming analysis is restricted to $1$-D.

\begin{figure*}[t!]
    \centering
	\begin{overpic}[width=0.44\linewidth,trim={4mm 4mm 3mm 3mm},clip]{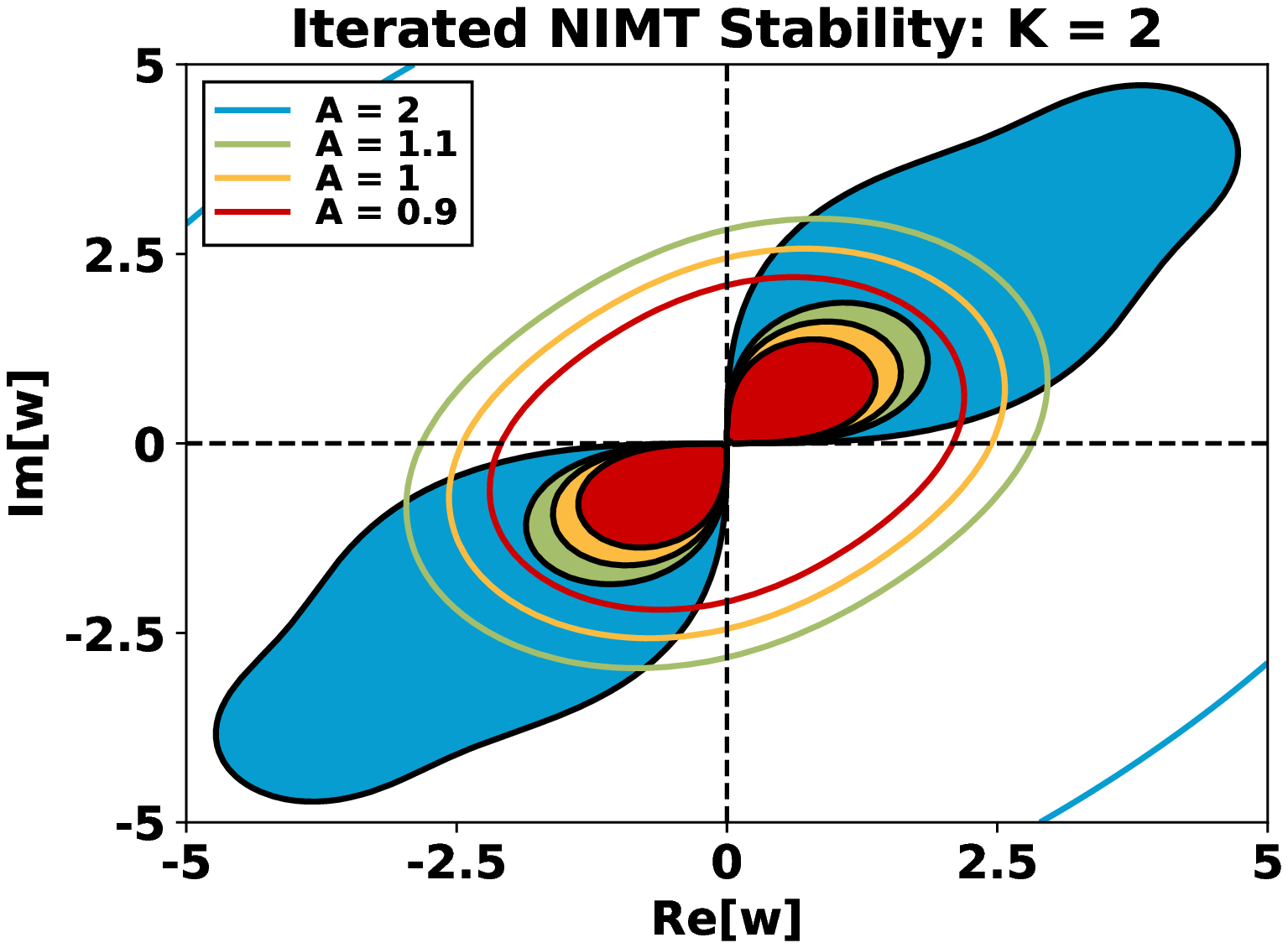}
		\put(90,12){\textbf{\large(a)}}
	\end{overpic}
	\hspace{3mm}
	\begin{overpic}[width=0.44\linewidth,trim={4mm 4mm 3mm 3mm},clip]{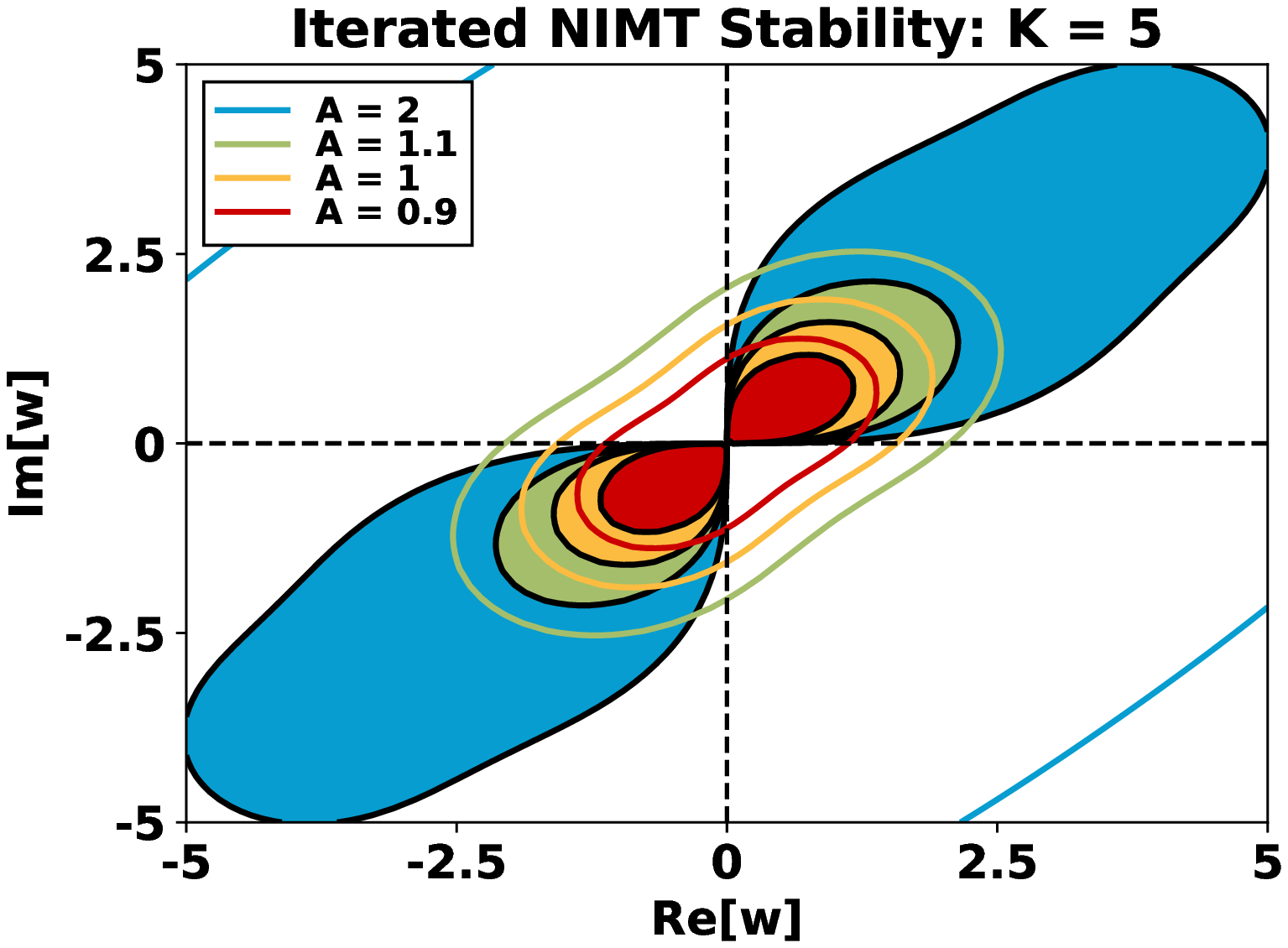}
		\put(90,12){\textbf{\large(b)}}
	\end{overpic}

    \vspace{3mm}
	\begin{overpic}[width=0.44\linewidth,trim={4mm 4mm 3mm 3mm},clip]{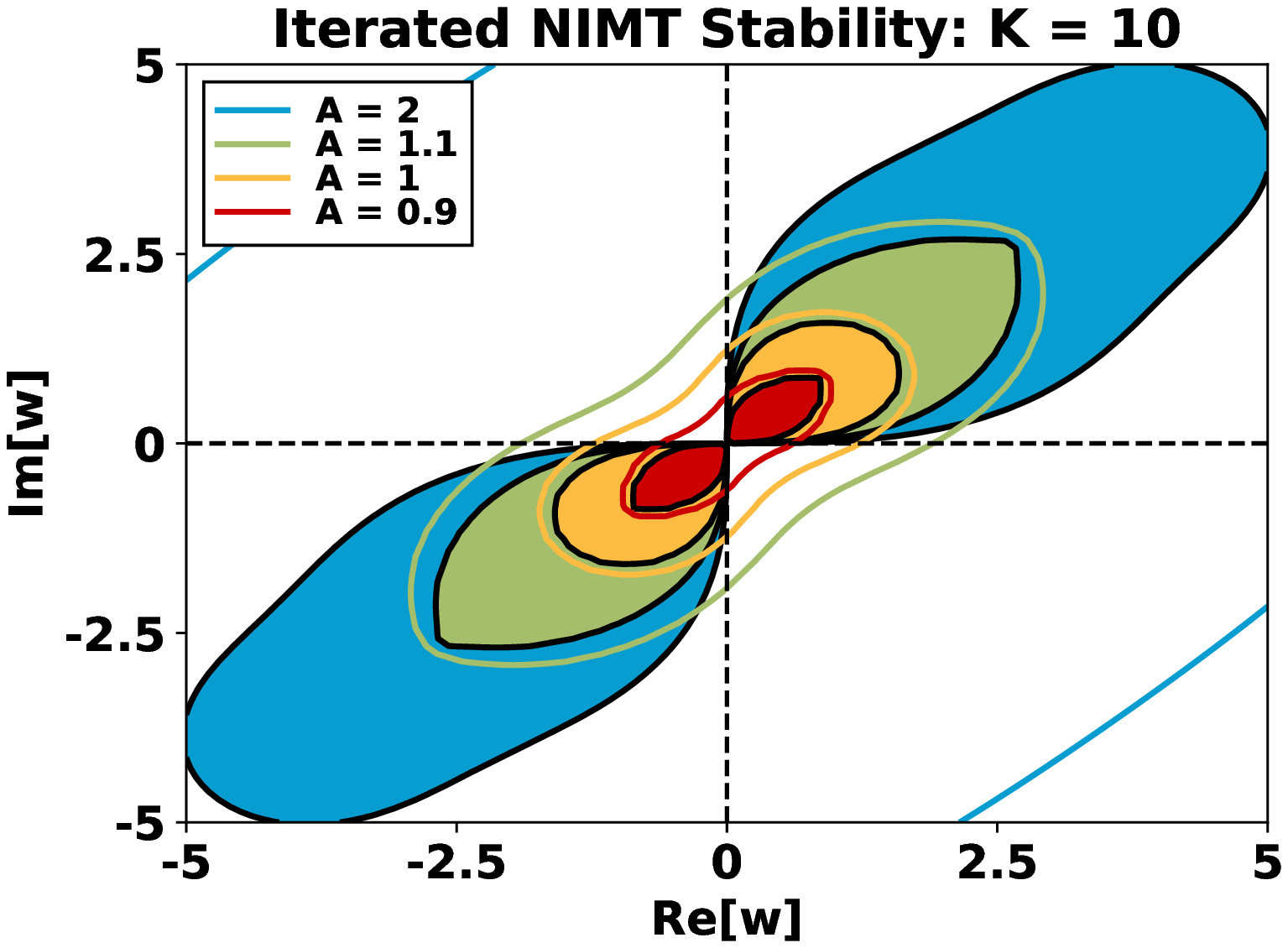}
		\put(90,12){\textbf{\large(c)}}
	\end{overpic}
	\hspace{3mm}
	\begin{overpic}[width=0.44\linewidth,trim={4mm 4mm 3mm 3mm},clip]{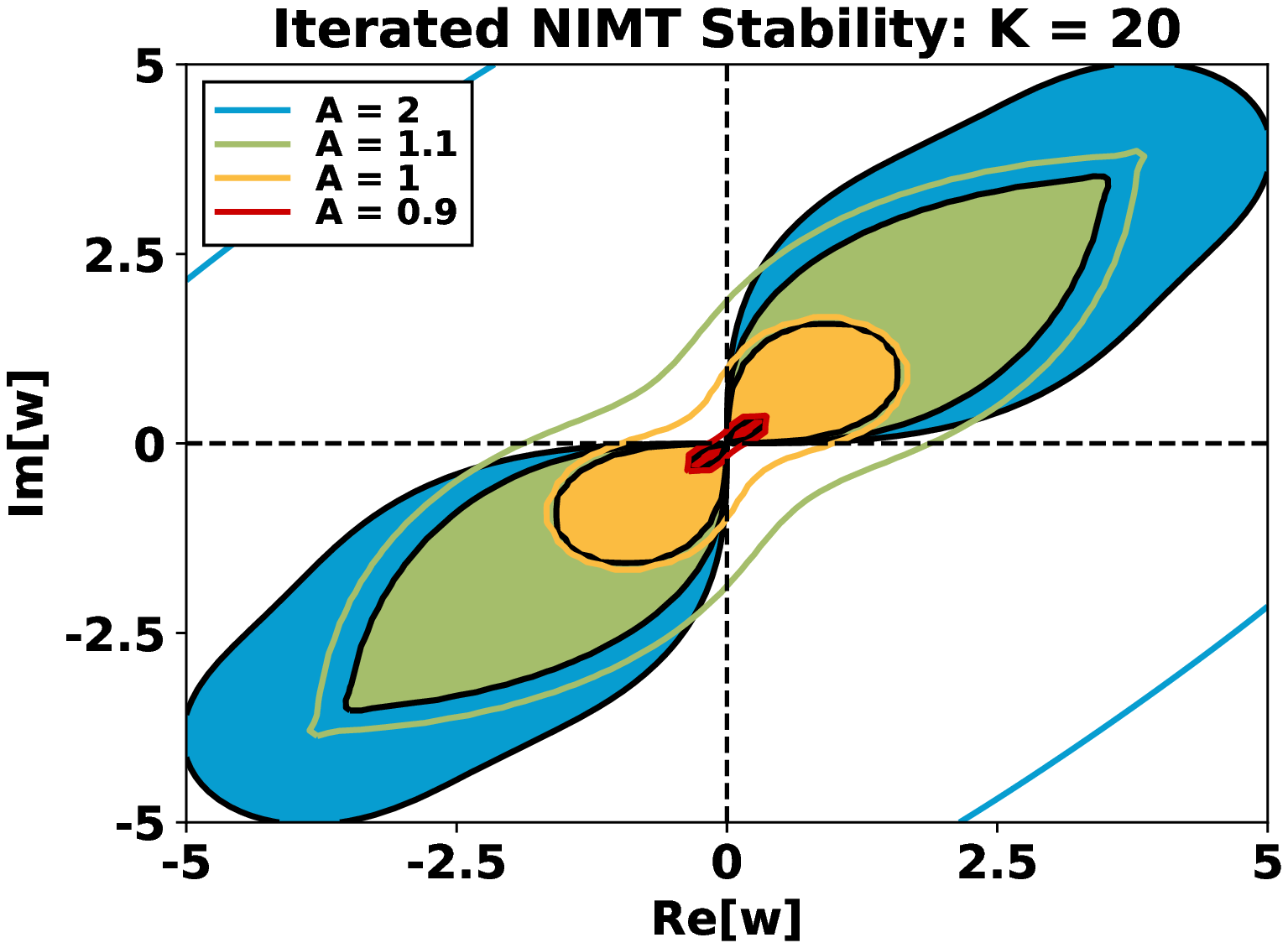}
		\put(90,12){\textbf{\large(d)}}
	\end{overpic}
	\caption{Stability diagrams for the iterated NIMT at various values of the iteration number $K$ and $A$, as computed via \Eq{IterStab}. For each color, the shaded region is the region of stability for the respective value of $A$, while the solid line labels the contour at which the magnification factor equals 10. Note that the near-identity limit corresponds to $A \approx 1$.}
	\label{IterNIMTstab}
\end{figure*}

Let us consider how the PMT and the iterated NIMT transform the single-parameter family of exponential functions $\psi_\kappa(q) \doteq e^{\kappa q}$, with $\kappa$ being complex. Generally speaking, we define an MT algorithm as \textit{stable}, or non-magnifying, if the norms of the transformed function $\Psi_\kappa(Q)$ and the original function $\psi_\kappa(q)$ satisfy $\|\Psi_\kappa(Q)\| \le \| \psi_\kappa(q)\|$; conversely, we define an MT algorithm as \textit{unstable}, or magnifying, if $\|\Psi_\kappa(Q)\| > \| \psi_\kappa(q)\|$. Unitarity corresponds to a strict equality. The ratio $\|\Psi_\kappa(Q)\|/\| \psi_\kappa(q)\|$ is referred to as the \textit{magnification factor}. Additionally, we define an MT algorithm as either \textit{$L^2$-stable} or, respectively, \textit{$L^2$-unitary} if the algorithm is stable or unitary along the entire imaginary $\kappa$ axis. This is because any $L^2$ function can be expanded into Fourier modes; thus, an $L^2$-unitary MT algorithm will be exactly unitary for any $L^2$ function. In our analysis, we shall only consider the class of function norms where $\|e^{ig(Q)}f(Q/A) \| = \sqrt{A} \, \|f(q) \|$ for $g(Q)$ real, an example of which being the $L^2$ norm.

Since $\psi_\kappa'(q) = \kappa \psi_\kappa (q)$, the PMT of $\psi_\kappa(q)$ is
\begin{equation}
	\PMT{\Mat{S}} \left[\psi_\kappa(q) \right] = \frac{e^{i\frac{C}{2A}Q^2}}{\sqrt{A}} \, e^{i\frac{B}{2A}\kappa^2} \, \psi_\kappa\left(\frac{Q}{A} \right) \, ,
\end{equation}

\noindent where $\PMT{\Mat{S}}$ is the PMT for symplectic matrix $\Mat{S}$. Let us define the rescaled variable $w \doteq \kappa \sqrt{B/A}$. Then, the PMT is stable when
\begin{equation}
	\left| e^{\frac{i}{2}w^2} \right| \le 1 \, .
	\label{PMTstabCond}
\end{equation}

\noindent This region of the complex $w$ plane is shown in \Fig{1StepStab}. 
The PMT is stable within the first and third quadrants of the complex plane, and is unitary along the real and imaginary $w$ axes. Hence, the PMT is $L^2$-unitary. Interestingly, the PMT is not unitary on its entire domain. This is because the domain of the PMT includes both square-integrable functions and functions where the integral of \Eq{metTRANS} does not converge, such as $e^q$. The cost of this expanded domain is the loss of global unitarity, albeit for functions whose $L^2$ norms are undefined.

We proceed to analyze the NIMT. Applied once, the NIMT of $\psi_\kappa(q)$ is
\begin{equation}
	\NIMT{\Mat{S}} \left[\psi_\kappa(q) \right] = \frac{1+\frac{iB}{2A} \, \kappa^2}{\sqrt{A}} \, e^{\frac{i C}{2A}Q^2} \, \psi_\kappa\left(\frac{Q}{A} \right) \, .
\end{equation}

\noindent Reintroducing $w$, the NIMT is stable where
\begin{equation}
	\left|1+\frac{i}{2}w^2 \right| \le 1 \, ,
	\label{NIMTstabCond}
\end{equation}

\noindent which is shown alongside the stability region of the PMT in \Fig{1StepStab}. This region makes up only a small subset of the stability region of the PMT. 

Notably, the NIMT is no longer $L^2$-stable; as such, square-integrable functions will be magnified. There are three ways to minimize the magnification: (i) reduce the step size to $B/A \lesssim 1/2\kappa_i$, with $\kappa_i$ the largest Fourier mode number; (ii) apply a low-pass filter to remove fast-growing Fourier modes; or (iii) increase the truncation order. However, we show shortly that increasing the truncation order of the NIMT increases its vulnerability to numerical noise, so only (i) and (ii) are recommended.

Let us now assess how subsequent iterations of the NIMT affect its stability. We first observe that each iteration of the NIMT adds the overall phase $Q^2C/2A$ that contributes to the derivatives of subsequent NIMT iterations. This sequence will very quickly become unwieldy as the iteration number increases. To achieve an analytical estimate of the iterated NIMT stability, we shall therefore neglect the contributions of the phase to all derivatives. This is consistent with the near-identity limit, where $C/A$ is vanishingly small.

In this approximation, the norm of the $K$-iterated NIMT is
\begin{align}
	&\left\| \NIMT{\Mat{S}_{K-1}^{-1} \Mat{S}_K}\left\{\ldots \NIMT{\Mat{S}_{1}}\left[\psi_\kappa(q) \nullFrac \right] \right\}\right\|\nonumber\\
	& = \left|\prod_{n=1}^K \frac{1+\frac{iB_n}{2A_n} \frac{\kappa^2}{\prod_{j=1}^{n-1} A^2_j}}{\sqrt{A_n}} \right| \left\| \psi_\kappa\left(\frac{Q}{\prod_{n=1}^{K} A_n} \right)\right\| \, ,
	\label{iterNIMT}
\end{align}

\noindent where for $n = 1$, we define $\prod_{j=1}^{n-1} A^2_j = 1$. When the iteration is uniform, \ie $A_n = A$ and $B_n = B$,
\Eq{iterNIMT} simplifies to
\begin{align}
	&\left\| \NIMT{\Mat{S}_{K-1}^{-1} \Mat{S}_K}\left\{\ldots \NIMT{\Mat{S}_{1}}\left[\psi_\kappa(q) \nullFrac \right] \right\}\right\| \nonumber\\
	&= \left| A^{-\frac{K}{2}}\prod_{n=0}^{K-1} \left(1 + \frac{i w^2}{2} A^{-2n} \right) \right| \, \left\| \psi_\kappa\left(\frac{Q}{A^K} \right)\right\| \, ,
\end{align}

\noindent where we have reintroduced $w \doteq \kappa \sqrt{B/A}$. Hence, the $K$-iterated NIMT is stable where
\begin{equation}
	\left| \left(-\frac{iw^2}{2};A^{-2} \right)_K \right| \le 1 \, ,
	\label{IterStab}
\end{equation}

\noindent where $(a;q)_K \doteq \prod_{n=0}^{K-1} \left(1 - aq^n \right)$ is the $q$-Pochhammer symbol~\cite{Olver10}, \ie the $q$-analog of the rising factorial.

Figure~\ref{IterNIMTstab} shows the stability region at four different iteration numbers: $K = 2$, $K = 5$, $K = 10$, and $K = 20$. As \Eq{IterStab} indicates, the stability of the iterated NIMT now explicitly depends on $A$, so each subplot of \Fig{IterNIMTstab} includes stability diagrams for $A = 0.9$, $A = 1$, $A = 1.1$, and $A = 2$. These values were chosen to emphasize the near-identity behavior of the iterated NIMT, when $A \approx 1$. There are two notable observations. First, the stability region for $A=1$ is independent of $K$. For other values of $A$, the stability region changes significantly with $K$, decreasing for $A < 1$ and increasing for $A > 1$. Second, the sensitivity of the iterated NIMT increases with $K$, as seen by considering the rate at which the $A = 1.1$ and $A = 0.9$ contours separate.

\begin{figure}[t!]
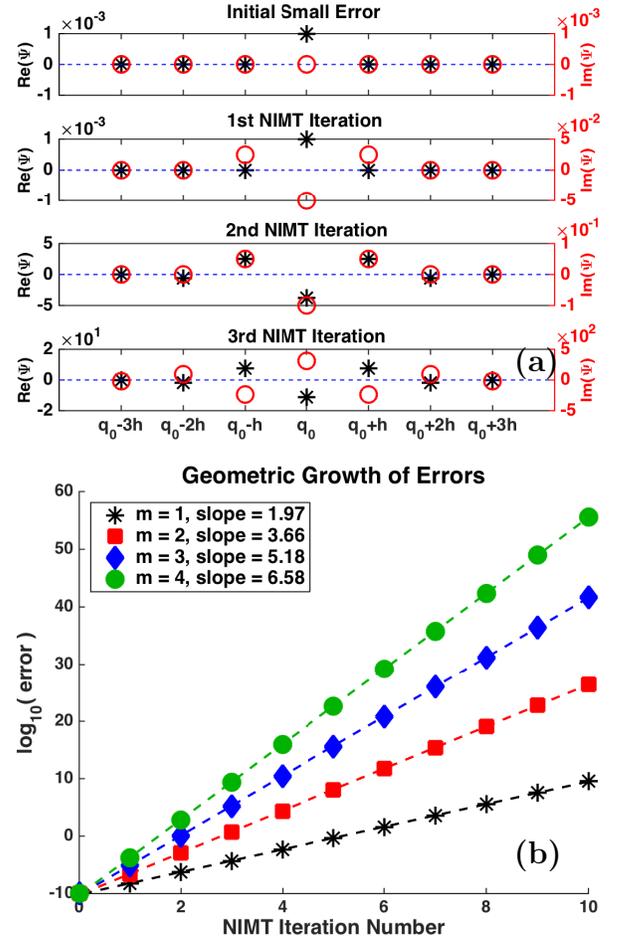

	\begin{overpic}[width=0.9\linewidth,trim={13mm 7mm 4mm 6mm},clip]{NIMT_NoiseInstability.eps}
		\put(85,12){\textbf{\large(a)}}
	\end{overpic}
	
	\vspace{3mm}
	\begin{overpic}[width=0.9\linewidth,trim={7mm 4mm 15mm 3mm},clip]{NIMT_NoiseGrowthRate.eps}
		\put(85,12){\textbf{\large(b)}}
	\end{overpic}
	\caption{\textbf{(a)} The d-instability increases an initial error of $10^{-3}$ by multiple orders of magnitude after only three iterations. The error also propagates away from the initial location, with width proportional to the iteration number. \textbf{(b)} Error from the d-instability as the NIMT truncation number $m$ is varied. The error is defined as $\| \boldsymbol{\Psi}\|$. Notable parameters are $A = D = 1$, $B = 1/2$, $C = 0$, $h = 0.1$, and initial error $10^{-10}$. Using \Eq{growthRATE} (\Eq{incGAMMA}) to estimate the growth rate and then taking the base-10 logarithm, the slopes for $m = 1$, $m = 2$, $m = 3$, and $m = 4$ are estimated respectively as $2.000$ ($2.004$), $3.699$ ($3.708$), $5.222$ ($5.235$), and $6.620$ ($6.637$). These values agree well with those obtained by a best fit line (dashed line).}
	\label{noiseINSTAB}
\end{figure}

Consequently, a step size $B/A$ that is initially stable, but with $A < 1$, will become quickly and increasingly unstable as the NIMT is iterated. This introduces an interesting tradeoff consideration when computing a finite transformation: is it better to use a coarse discretization with a large step size but few iterations, or a fine discretization with a small step size but many iterations? The answer depends largely on implementation specifics; we find in the following subsection that a fine discretization is preferable for our chosen example, but this is not necessarily indicative of a general principle.


Although we shall not dwell much on implementation details, we must make one cautionary remark regarding the finite-difference scheme used to discretize the NIMT. Because discrete differentiation is poorly conditioned, any noise in the original function $\psi(q)$ will be magnified when its derivatives are computed. Since derivatives are computed with each iteration of the NIMT, this noise will grow geometrically. We call this instability the \textit{d-instability} (with `d' standing for discretization). As shown in \Fig{noiseINSTAB}, it is particularly disastrous for iterated NIMTs with large truncation order. 

A basic description of the d-instability is afforded by the transformation of a constant function. Suppose one attempts to transform a function that is identically zero everywhere except at a single grid point, where the function is erroneously non-zero by some unspecified noise source. When the grid spacing $h$ is uniform, the growth rate of the d-instability, $\gamma$, can be estimated analytically. Let $\Delta_{k}$ be a $k$-th order finite-difference matrix such that $h^{-k} \Delta_k \Vect{f}$ equals the $k$-th discrete derivative of $\Vect{f}$. In this specific test problem, any non-zero norm is due to noise; hence, the error of the $K$-th iterated, $m$-th order NIMT is bounded with the triangle inequality as
\begin{equation}
	\|\boldsymbol{\Psi} \| \le \left[\frac{1}{\sqrt{|A|}} \sum_{n=0}^m \frac{\|\Delta_{2n} \|}{h^{2n}} \frac{|B/2A|^n}{n!}\right]^K \|\boldsymbol{\psi}\| \, ,
	\label{noiseINEQ}
\end{equation}

\noindent where $\boldsymbol{\psi}$ and $\boldsymbol{\Psi}$ are the discretized versions of $\psi(q)$ and $\Psi(Q)$ respectively, and $\|\Delta_{2n} \|$ is the subordinate matrix norm of $\Delta_{2n}$. There is a freedom to choose the norm with which \Eq{noiseINEQ} is evaluated; we choose the $\infty$-norm, denoted $\|\ldots \|_\infty$, as it yields the readily-evaluated matrix row norm as its subordinate~\cite{Trefethen97}. Considering only the leading order in $1/h$, $\gamma$ is estimated as
\begin{equation}
	\gamma \approx \frac{1}{\sqrt{|A|}} \, \frac{\|\Delta_{2m} \|_\infty}{h^{2m}} \, \frac{|B/2A|^m}{m!} \, .
	\label{growthRATE}
\end{equation}

Equation \eq{growthRATE} has been purposefully separated: for increasing $m$, the factor $\|\Delta_{2m} \|/h^{2m}$ increases while the factor $|B/2A|^m/m!$ decreases. In fact, as defined, $\gamma \to 0$ as $m \to \infty$ for any reasonable class of $\Delta_{2m}$; this does not mean the d-instability disappears for high truncation orders, but rather that the d-instability is not dominated by the leading order in $1/h$ when $m$ is large. Instead, a subset of intermediate-order terms dominate, which are not included in \Eq{growthRATE}. For central finite-difference schemes with homogeneous boundary conditions, $\|\Delta_{2n} \|_\infty = 2^{2n}$~\cite{Abramowitz70}, and the growth rate is uniformly estimated to be
\begin{equation}
    \gamma \approx \frac{1}{\sqrt{|A|}} \, \exp\left(\left| \frac{2B}{Ah^2} \right| \right) \, \frac{\Gamma\left(m+1, \left| \frac{2B}{Ah^2} \right| \right)}{\Gamma\left(m+1 \right)} \, ,
    \label{incGAMMA}
\end{equation}

\noindent where $\Gamma\left(s,x \right)$ is the incomplete gamma function~\cite{Olver10}. Notably, $\Gamma\left(s,x \right) \to \Gamma\left(s\right)$ as $s \to \infty$. Equation \eq{growthRATE} is sufficient for the error estimation of low truncation order schemes; for large $m$, however, \Eq{incGAMMA} should be used instead.

Thus far, our discussion of the d-instability has been contingent on a maliciously designed initial condition. Such a specific state will not likely arise in practical applications; nevertheless, local d-instabilities can certainly arise. For example, consider the NIMT of a function $\psi(q)$ that asymptotes to $0$ at the domain edge. Near the domain edge, $\psi(q)$ is nearly constant, but a source of error, interpolation or otherwise, will inevitably cause at least one data point to deviate. The local d-instability will then grow rapidly, and will propagate inward from the domain edge until the transformed function is entirely dominated by noise. Since the d-instability growth rate scales with truncation order, using low $m$ schemes will minimize its deleterious effects. Marginally smoothing the input data before taking derivatives will also delay its onset. 


\subsection{Numerical example: Time evolution of the QHO}

To demonstrate the iterated NIMT, let us consider once again the time evolution of the $1$-D QHO, introduced in \Sec{SecMET}. There, the symplectic matrix $\Mat{S}$ can be expressed as
\begin{equation}
	\Mat{S}_t = 
	\begin{pmatrix}
		\cos{t} & \sin{t}\\
		-\sin{t} & \cos{t}
	\end{pmatrix} \, ,
	\label{SEvo}
\end{equation}

\noindent where we have added the index $t$ to emphasize the dependence on time. This matrix can be represented as $\Mat{S}_t = (\Mat{S}_{\Delta t})^K$, where $\Mat{S}_{\Delta t}$ evolves the system by $\Delta t \ll 1$ and $K = t/\Delta t$. From \Eq{CanonTRANS}, the scalar functions $A_{\Delta t}$, $B_{\Delta t}$, $C_{\Delta t}$, and $D_{\Delta t}$ to be used in the iterated NIMT are
\begin{equation}
	A_{\Delta t} = D_{\Delta t} = \cos(\Delta t) \, , \quad B_{\Delta t} = -C_{\Delta t} = \sin(\Delta t) \, .
\end{equation}

For visualization, it is useful to introduce the \textit{Wigner function} that corresponds to $\psi(q)$, defined as~\cite{Wigner32}:
\begin{equation}
	W_\psi(q,p) = \frac{1}{2\pi}\int \limits_{-\infty}^\infty \dd y~\psi\left(q-\frac{y}{2}\right)\psi^*\left(q+\frac{y}{2}\right)e^{ipy} \, .
	\label{wigFUNC}
\end{equation}

\noindent As shown in \Refs{deGosson06,Littlejohn86a,Lohmann93}, the Wigner function of the metaplectic image $\Psi$ is simply the Wigner function of the original $\psi$ correspondingly rotated. This is also readily understood from the physical meaning of $W_\psi$. Specifically, if $\psi$ is a wave field, then $W_\psi$ can be interpreted as the phase-space quasiprobability distribution function of the wave quanta. The prefix `quasi' marks the fact that $W_\psi$ is not positive-definite unless averaged over a phase space volume of size $\Delta q \, \Delta p \gtrsim 2\pi$~\cite{Cartwright76, OConnell81}; nonetheless, $W_\psi$ is always real by definition, even for complex $\psi$.

\begin{figure*}[t!]
    \centering
    \begin{overpic}[width=0.32\linewidth,trim={-4mm 7mm 4mm 8mm},clip]{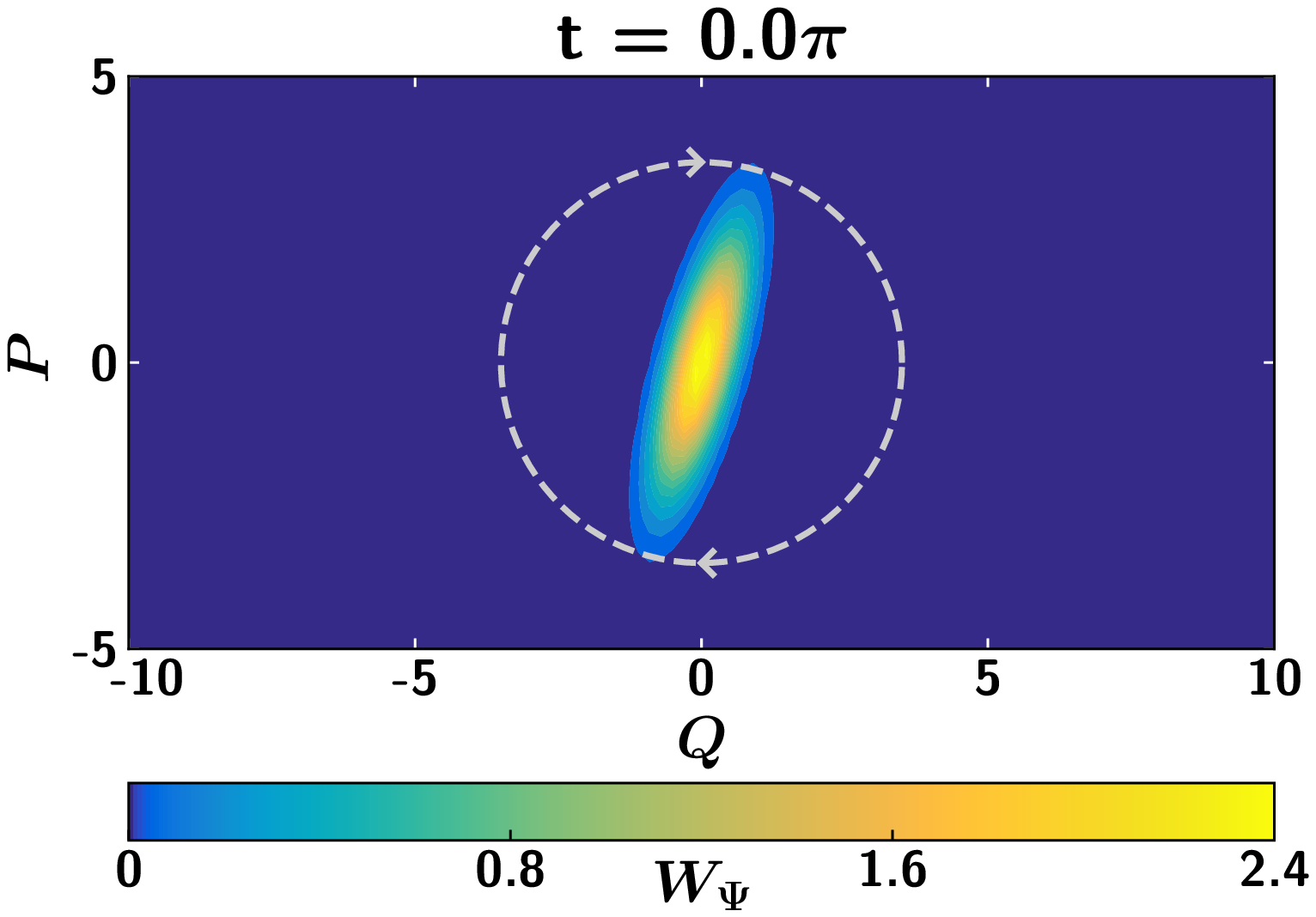}
		\put(85,25){\textbf{\color{white} \large(a)}}
	\end{overpic}
	\begin{overpic}[width=0.32\linewidth,trim={-4mm 7mm 4mm 8mm},clip]{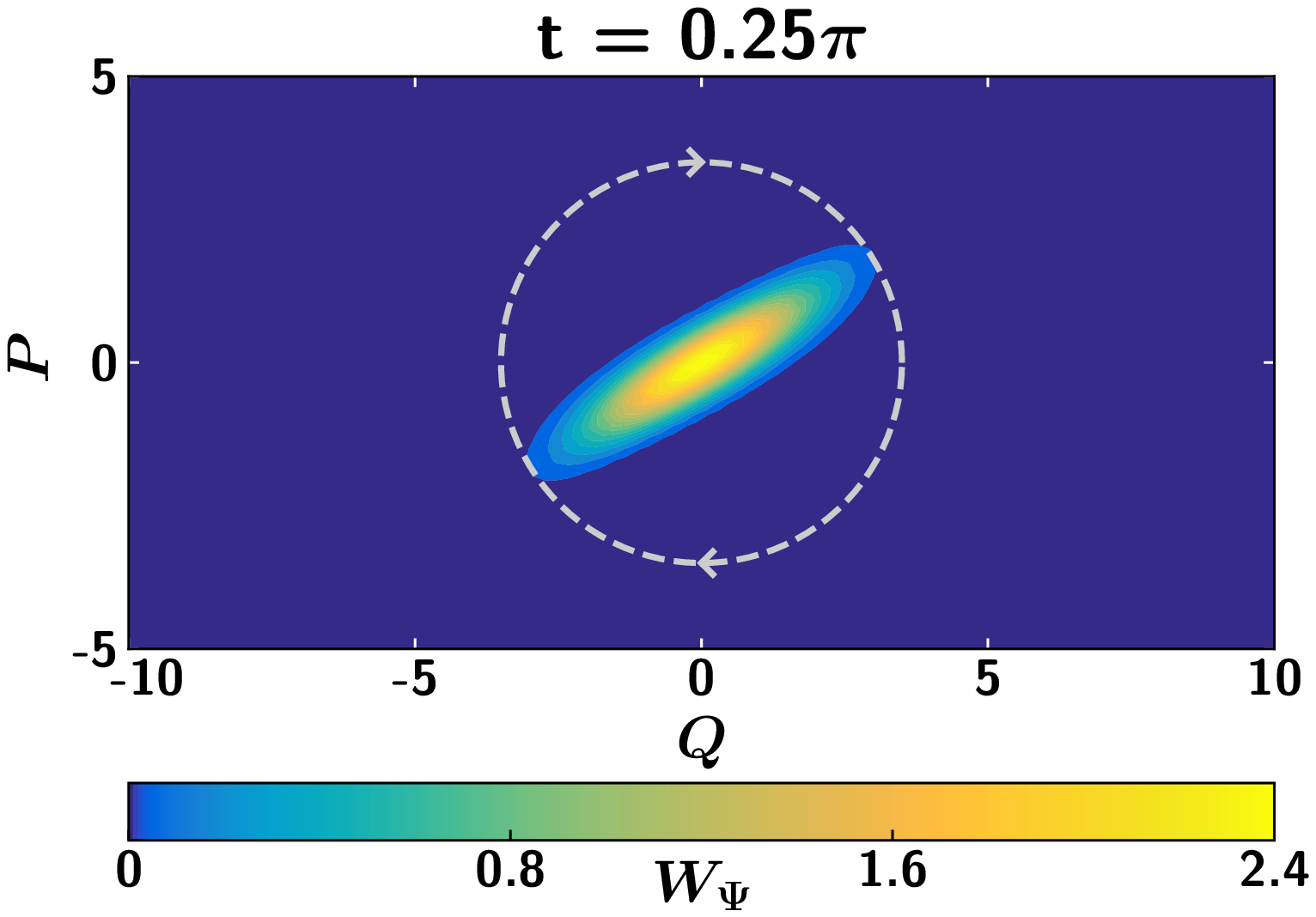}
		\put(85,25){\textbf{\color{white} \large(b)}}
	\end{overpic}
	\begin{overpic}[width=0.32\linewidth,trim={-4mm 7mm 4mm 8mm},clip]{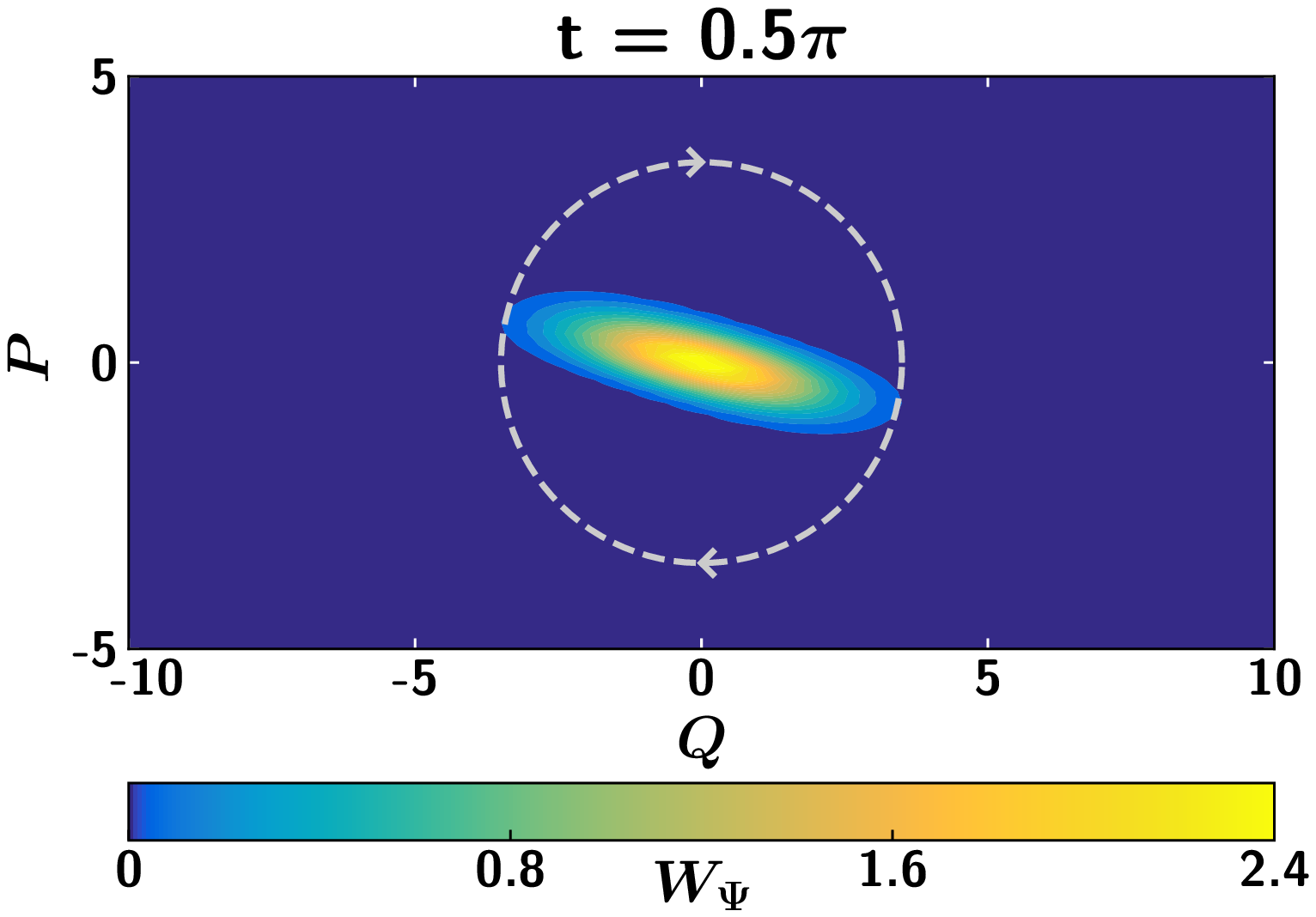}
		\put(85,25){\textbf{\color{white} \large(c)}}
	\end{overpic}
	
	\begin{overpic}[width=0.32\linewidth,trim={-4mm 7mm 4mm 12mm},clip]{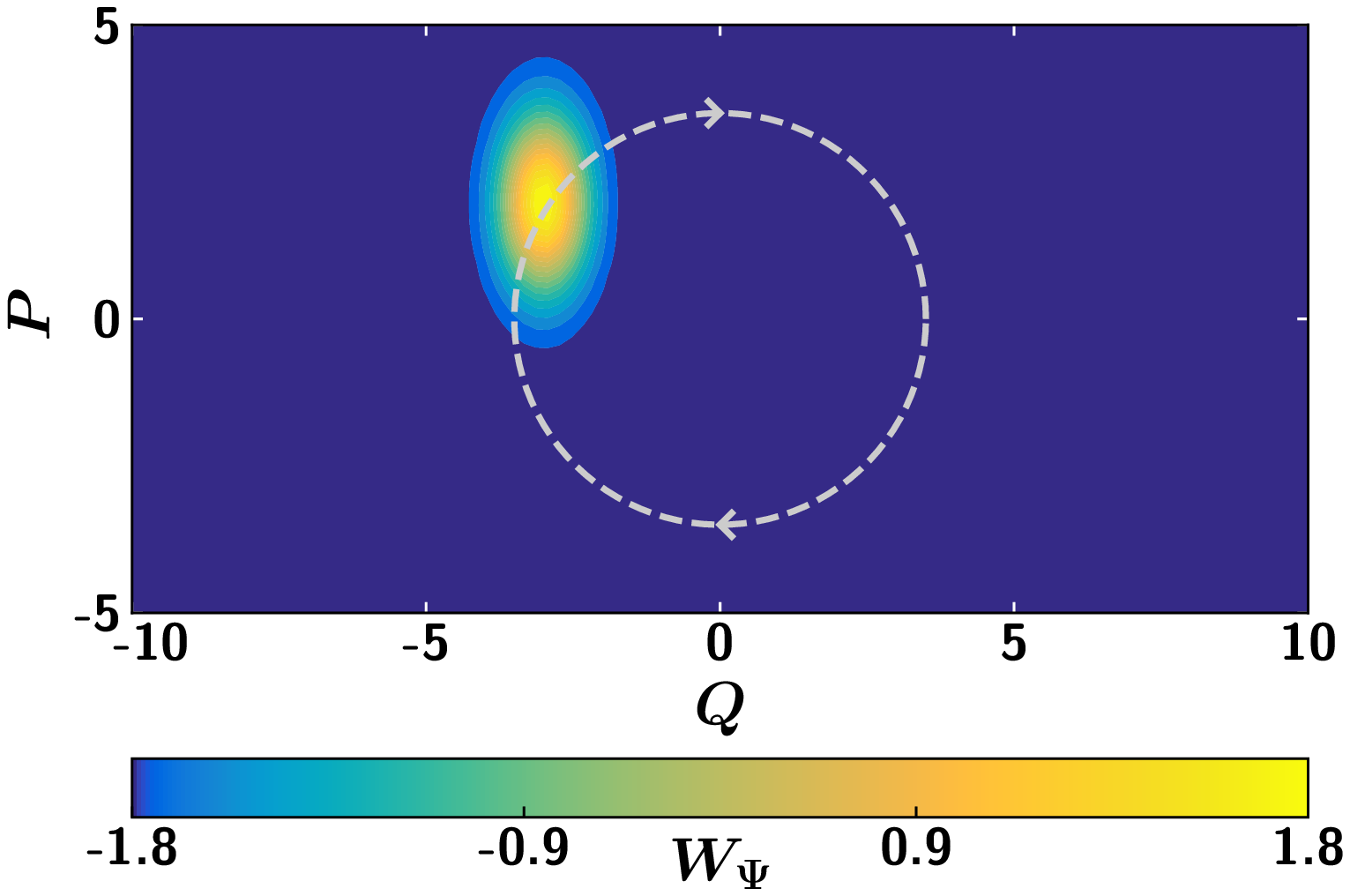}
		\put(85,25){\textbf{\color{white} \large(d)}}
	\end{overpic}
	\begin{overpic}[width=0.32\linewidth,trim={-4mm 7mm 4mm 12mm},clip]{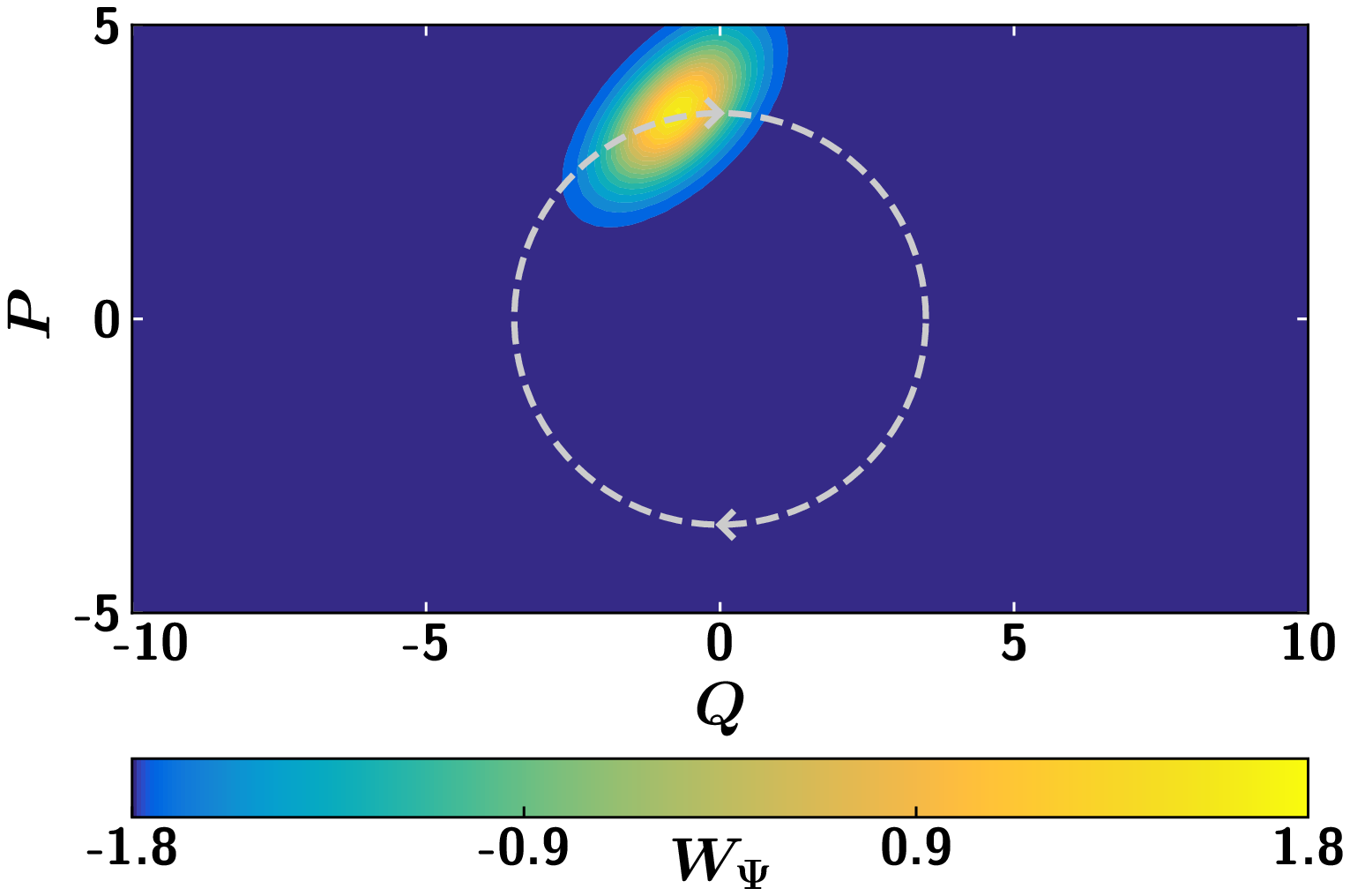}
		\put(85,25){\textbf{\color{white} \large(e)}}
	\end{overpic}
	\begin{overpic}[width=0.32\linewidth,trim={-4mm 7mm 4mm 12mm},clip]{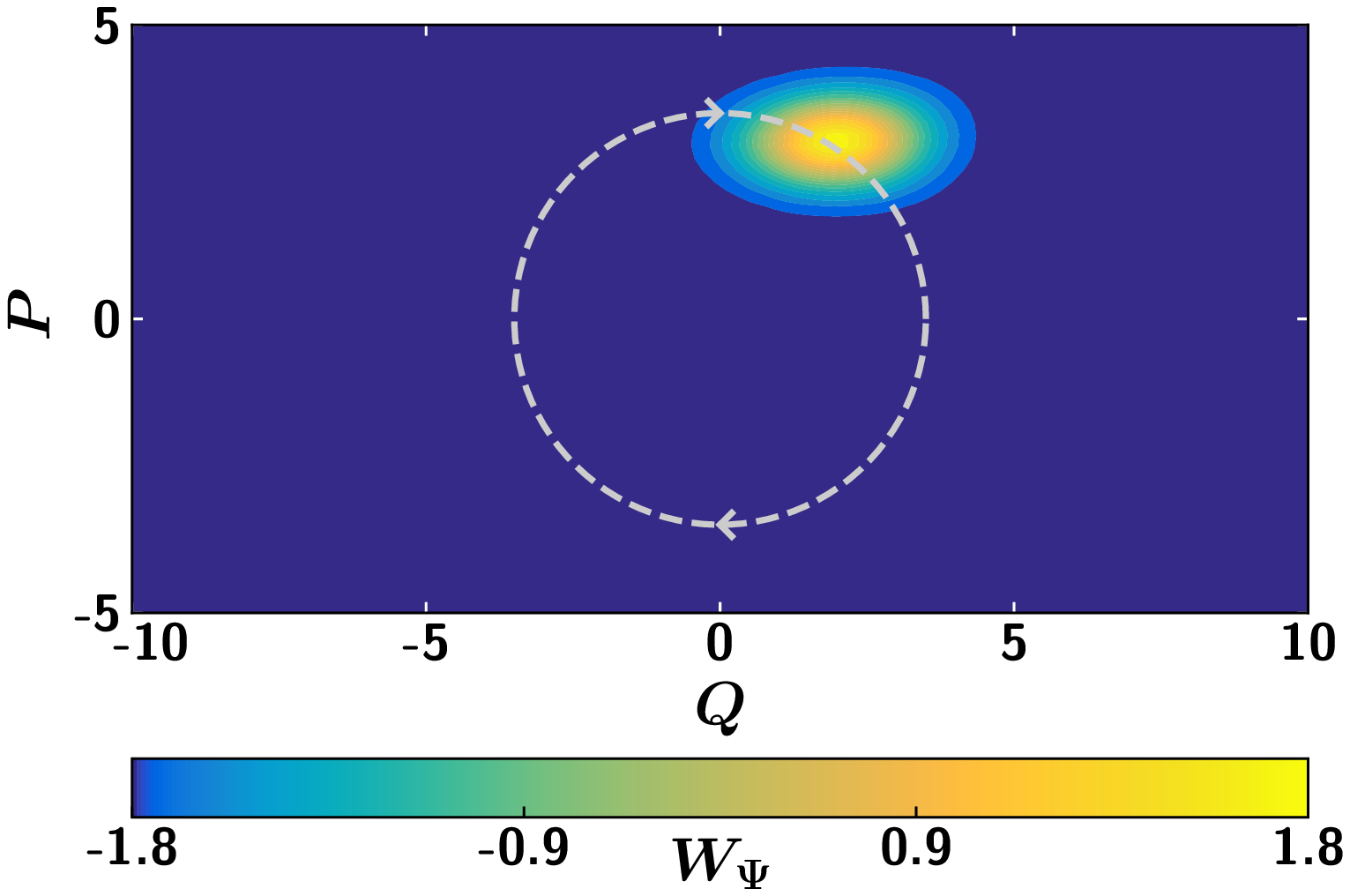}
		\put(85,25){\textbf{\color{white} \large(f)}}
	\end{overpic}
	
	\begin{overpic}[width=0.32\linewidth,trim={-4mm 7mm 4mm 12mm},clip]{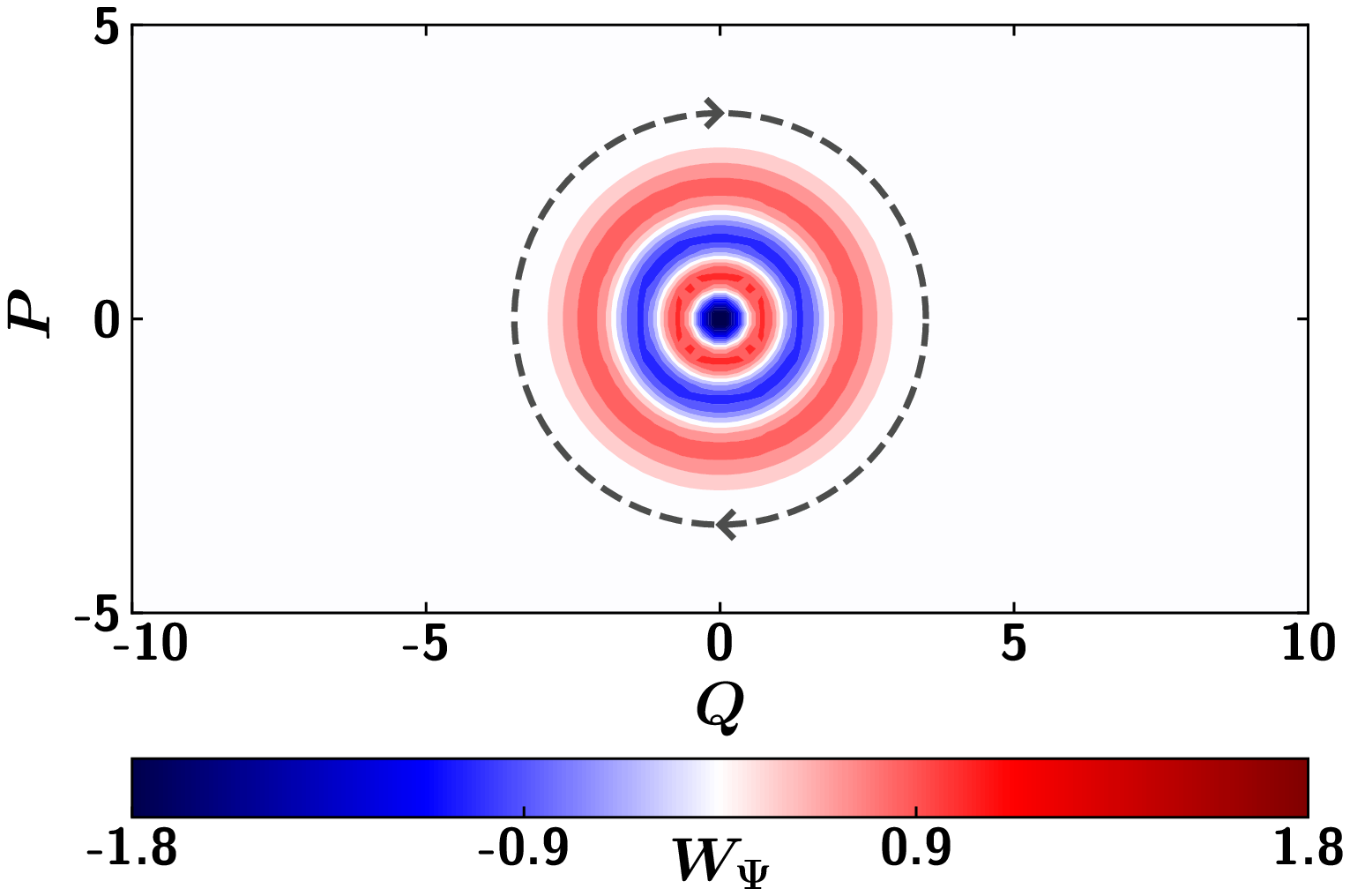}
		\put(85,25){\textbf{\color{black} \large(g)}}
	\end{overpic}
	\begin{overpic}[width=0.32\linewidth,trim={-4mm 7mm 4mm 12mm},clip]{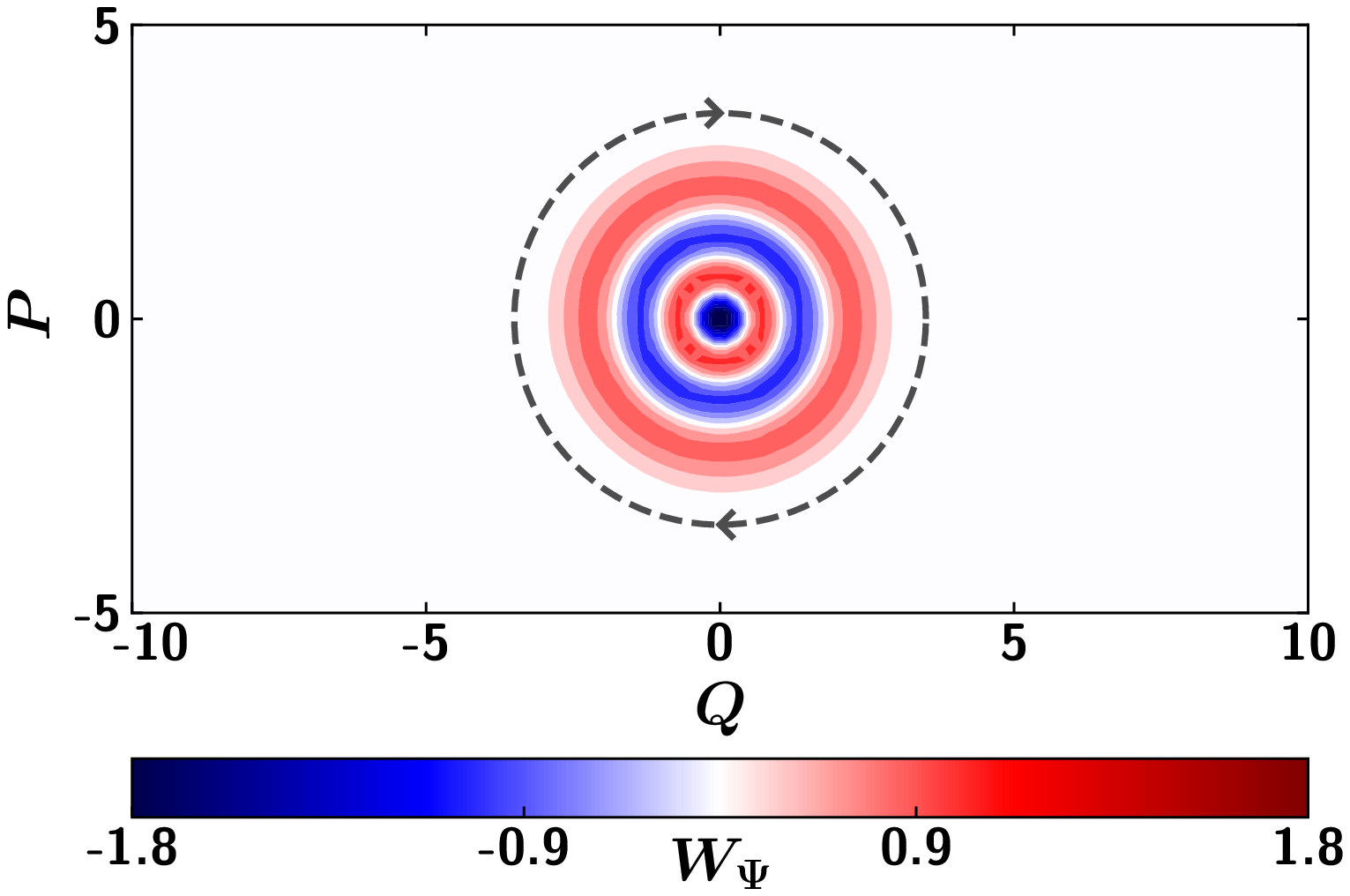}
		\put(85,25){\textbf{\color{black} \large(h)}}
	\end{overpic}
	\begin{overpic}[width=0.32\linewidth,trim={-4mm 7mm 4mm 12mm},clip]{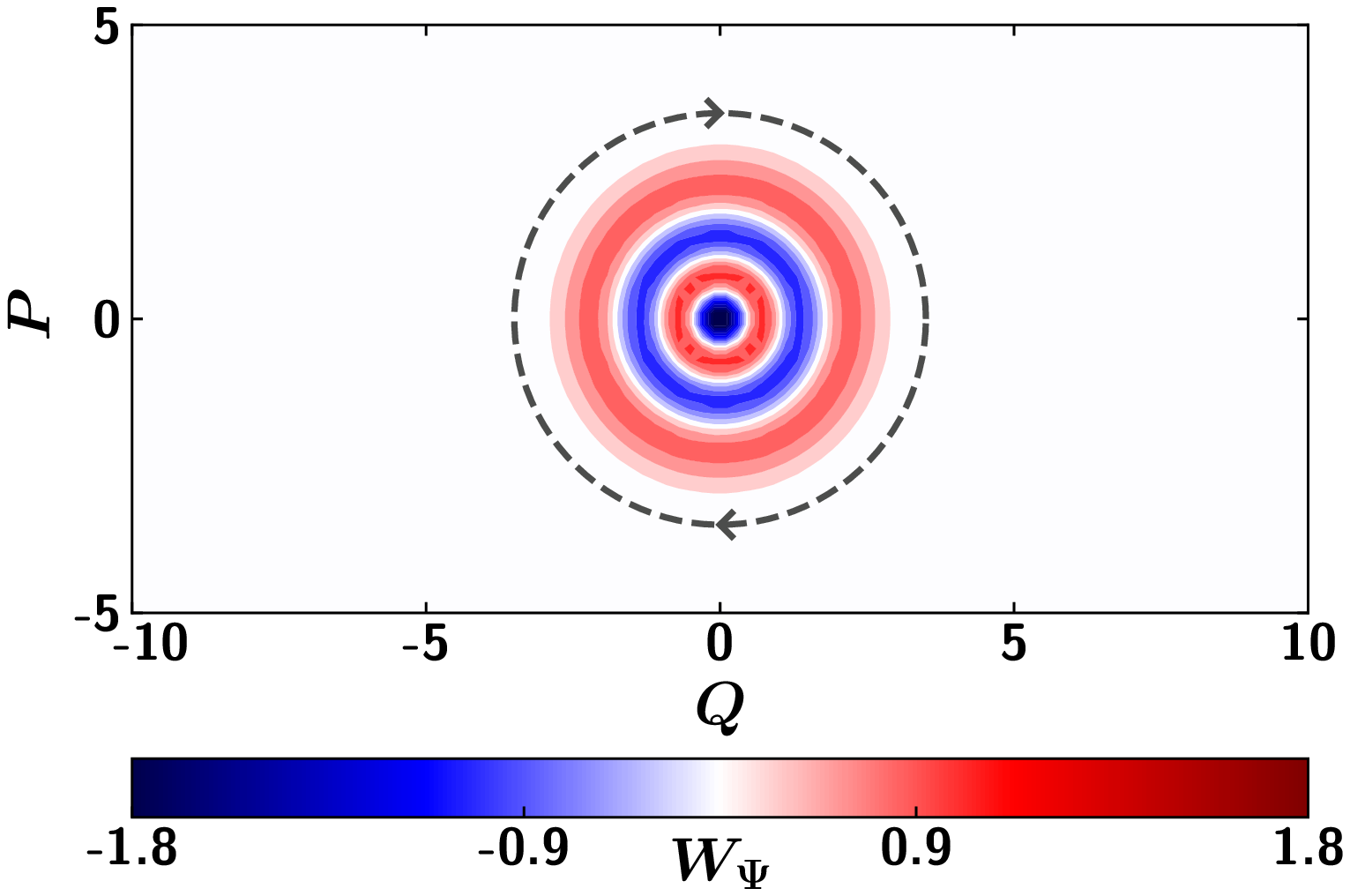}
		\put(85,25){\textbf{\color{black} \large(i)}}
	\end{overpic}
	\caption{\textbf{(a)-(c)} Time evolution of the Wigner function $W_\psi$ for the QHO with initial wavefunction $\psi(q) = e^{(i-1)q^2}$. The wavefunction is evolved using the iterated NIMT with $\Mat{S}_t$ given by \Eq{SEvo} and a step size of $\pi/2000$. $W_\psi$ is subsequently calculated via \Eq{wigFUNC}. Shown are snapshots at (a) $t = 0$, (b) $t = \pi/4$, and (c) $t = \pi/2$. \textbf{(d)-(f)} Same as (a)-(c), but with $\psi(q) = e^{-(q+3)^2+2iq}$. \textbf{(g)-(i)} Same as (a)-(c), but with $\psi(q) = (48\sqrt{\pi})^{-1/2}e^{-q^2/2}H_3(q)$. As expected, the Wigner functions are rotated by the MT along classical trajectories, shown in grey. In the third example, $W_\psi$ is rotationally-symmetric, so it is preserved by the MT.}
	\label{FT}
\end{figure*}

For our example, we consider the time evolution of three initial states: (i) a chirped Gaussian profile, $\psi(q) = e^{(i-1)q^2}$, which is relevant for bit-flip operations in chirp-modulated communication~\cite{Kaminsky05}, (ii) a squeezed coherent state, $\psi(q) = e^{-(q+3)^2+2iq}$, which is relevant for high-sensitivity detectors~\cite{Xiao87}, and (iii) the QHO eigenstate corresponding to $n = 3$, namely, $\psi(q) = (48\sqrt{\pi})^{-1/2}e^{-q^2/2}H_3(q)$, where $H_n(q)$ is the $n$-th degree Hermite polynomial. For these choices, the exact metaplectic image of $\psi(q)$ can be found explicitly from \Eq{metTRANS}, which facilitates benchmarking of our algorithm. They are given respectively by
\begin{subequations}
	\begin{align}
		\label{GAUSS}
		&\hspace{-5mm}\Psi_t(Q) = \pm \frac{\exp\left[i\frac{(2+2i)\sin(2t) + \cos(2t) - 1}{(8+8i)\sin^2(t)+2\sin(2t)} Q^2 \right] }{\sqrt{\cos(t) + (2+2i)\sin(t)}} \, , \\
		\label{SQCOH}
		&\hspace{-5mm}\Psi_t(Q) = \pm \frac{\exp\left\{ \frac{i}{2}\cot\left(t \right)Q^2 - i\frac{\left[\csc\left(t \right) Q - 2 - 6i \right]^2}{2\cot\left(t \right) + 4i} - 9\right\}}{ \sqrt{\cos\left( t \right) +2i \sin\left( t \right)} }\, ,\\
		\label{HERM3}
		&\hspace{-5mm}\Psi_t(Q) = (48\sqrt{\pi})^{-1/2}\exp\left(-\frac{Q^2 + 7it}{2}\right) H_3(Q) \, .
	\end{align}
\end{subequations}

\noindent The overall sign is chosen based on the winding number of $\Mat{S}_t$ as discussed in \Fig{riemann}: an odd winding number corresponds to the $-$ sign, while an even winding number corresponds to the $+$ sign. Each of these three example functions is evolved in time using the iterated NIMT with a uniform step size of $\Delta t = \pi/2000$, and the resulting Wigner functions are shown in \Fig{FT}. Here, $\Psi_t(Q)$ is discretized on an equally-spaced grid ranging from $[-10,10]$, and in the final example, the second-order NIMT was used in place of the first-order NIMT. 

As the NIMT is sequentially applied, \Fig{FT} shows that the resultant Wigner functions indeed rotate in phase space as expected. (In the third example, $W_\psi$ is rotationally-symmetric, so it is preserved by the MT.) This shows that the iterated NIMT can indeed perform finite transformations with high accuracy. For computing the Fourier transform, which corresponds to $t=\pi/2$, the iterated NIMT is robust to changes in the step size; discretizing the trajectory into $10^2$, $10^3$, and $10^4$ steps all yielded well-behaved solutions. The same is not true for changes in grid resolution, nor in changes of truncation order. Indeed, $\Psi_t(Q)$ quickly succumbed to amplified noise when (i) a Chebyshev-spaced grid was used in place of the equally-spaced grid, and (ii) the truncation order was increased beyond third-order.

\begin{figure*}[t]
    \centering
	\begin{overpic}[width=0.32\linewidth,trim={3mm 5mm 8mm 3mm},clip]{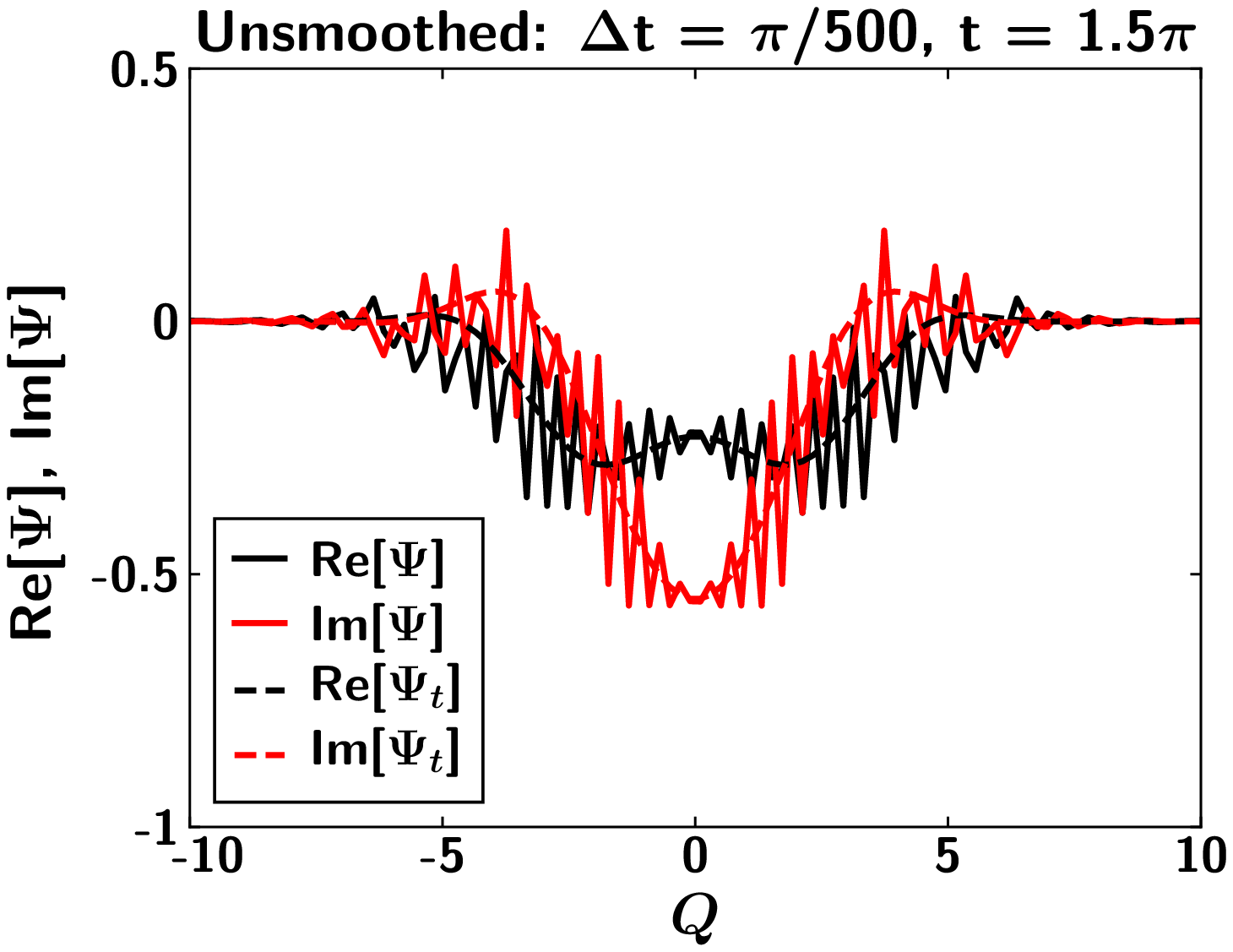}
		\put(85,14){\textbf{\large(a)}}
	\end{overpic}
	\begin{overpic}[width=0.32\linewidth,trim={3mm 5mm 8mm 3mm},clip]{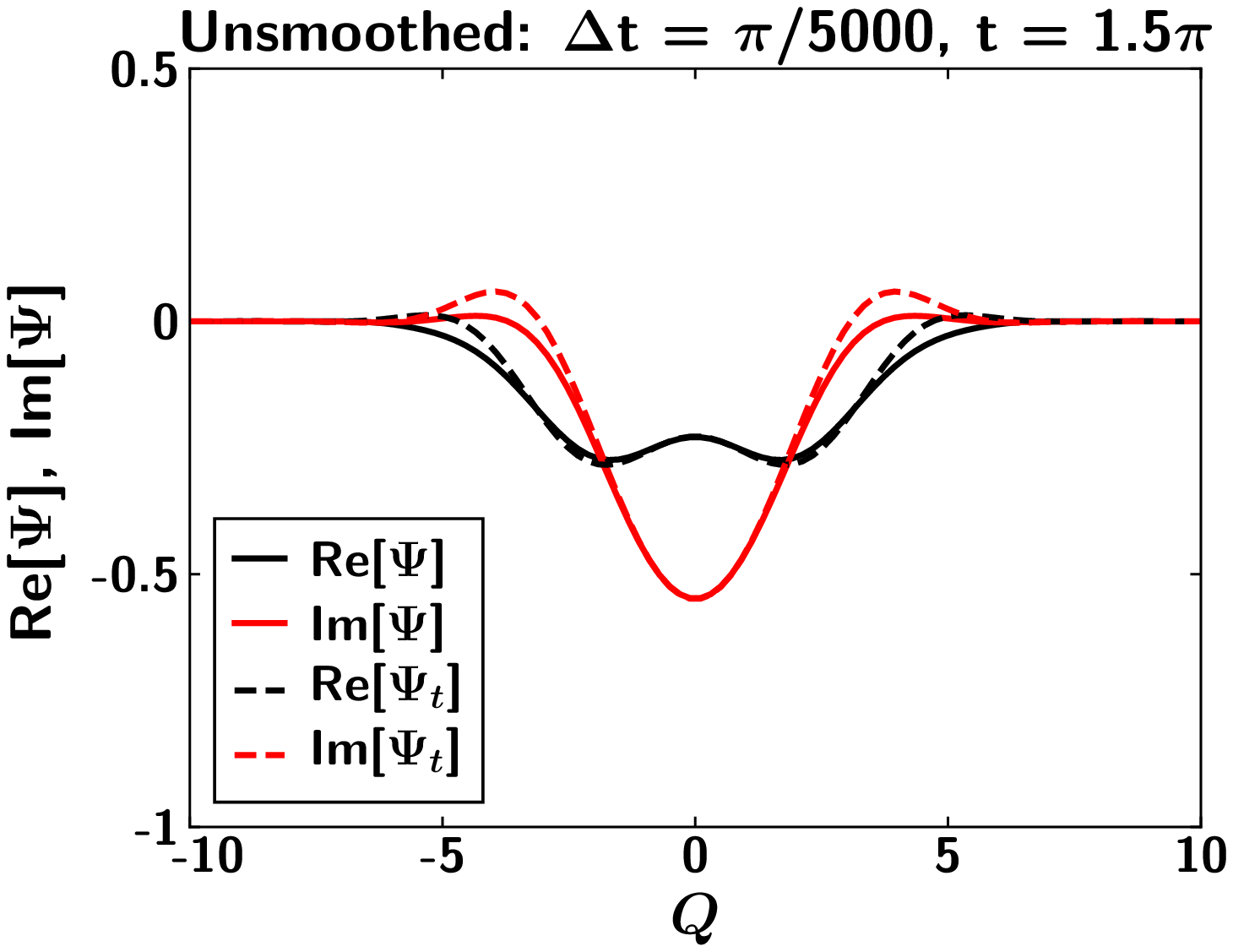}
		\put(85,14){\textbf{\large(b)}}
	\end{overpic}
	\begin{overpic}[width=0.32\linewidth,trim={3mm 5mm 8mm 3mm},clip]{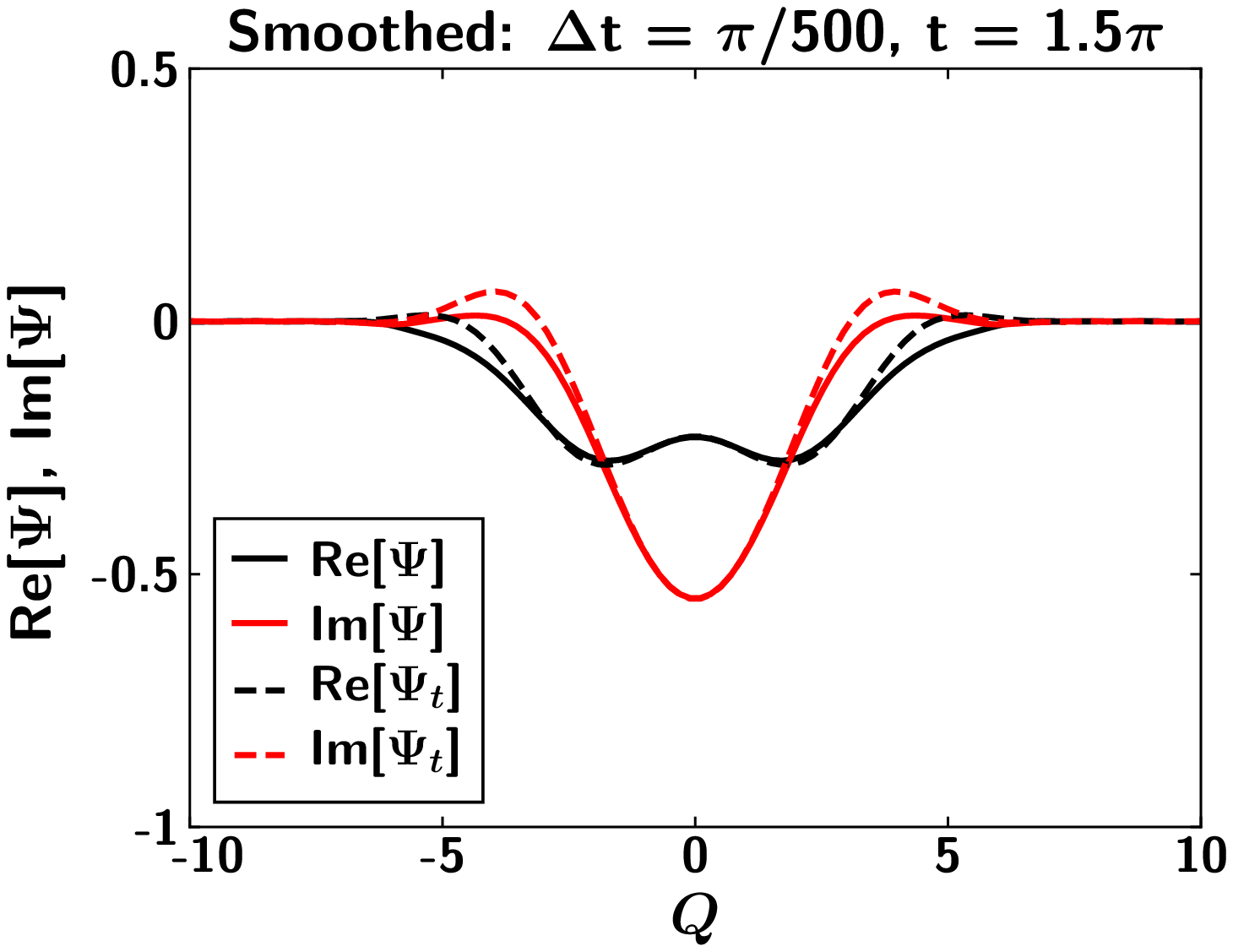}
		\put(85,14){\textbf{\large(c)}}
	\end{overpic}
	\caption{Time evolution of $\psi(q) = e^{(i-1)q^2}$ using the iterated NIMT for different choices of $\Delta t$, with and without smoothing. The final time is $t = 3\pi/2$ for all three figures. \textbf{(a)} Result of applying the iterated NIMT at a relatively large step size of $\Delta t = \pi/500$ without smoothing. In this case, high-frequency noise is amplified and quickly dominates the signal. \textbf{(b)} Same as (a) but with a smaller step size of $\Delta t = \pi/5000$. \textbf{(c)} Same as (a) but with low-pass filtering. As is clearly seen, low-pass filtering suppresses noise amplification.}
	\label{disruption}
\end{figure*}

\begin{figure*}[t!]
    \centering
    \begin{overpic}[width=0.24\linewidth,trim={2mm 24mm 1mm 17mm},clip]{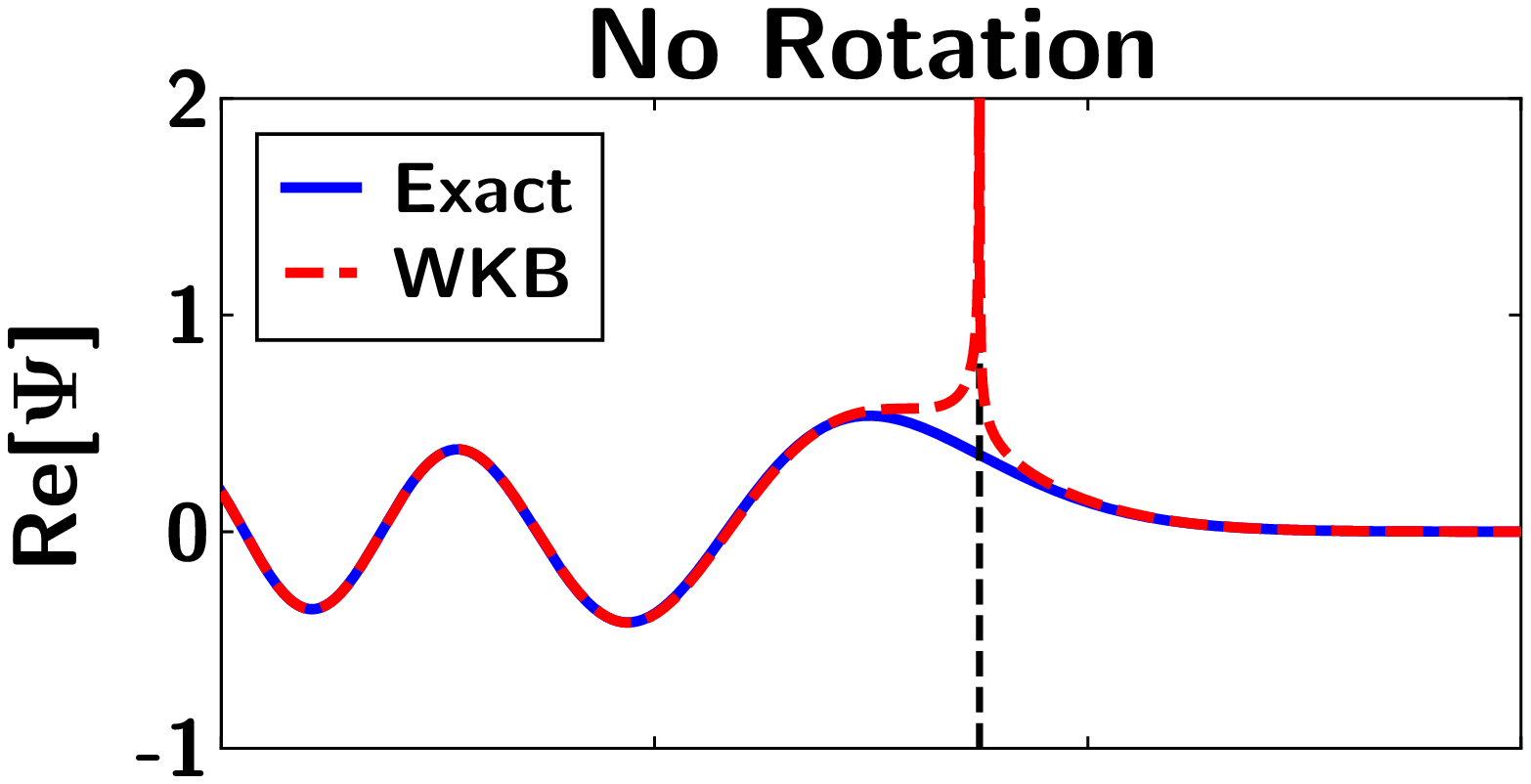}
		\put(85,5){\textbf{\color{black} \large(a)}}
	\end{overpic}
	\begin{overpic}[width=0.24\linewidth,trim={2mm 24mm 1mm 17mm},clip]{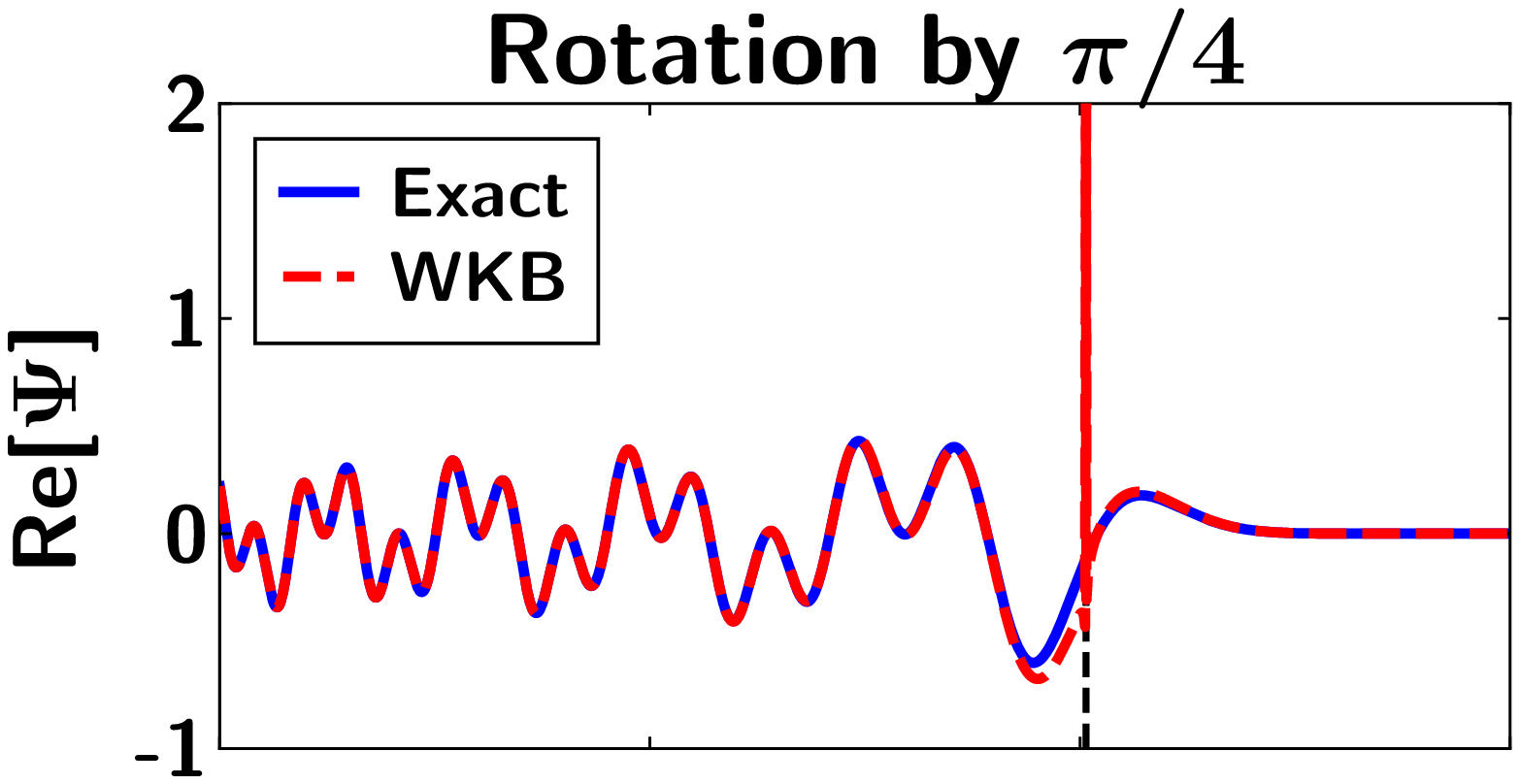}
		\put(85,5){\textbf{\color{black} \large(b)}}
	\end{overpic}
	\begin{overpic}[width=0.24\linewidth,trim={2mm 24mm 1mm 17mm},clip]{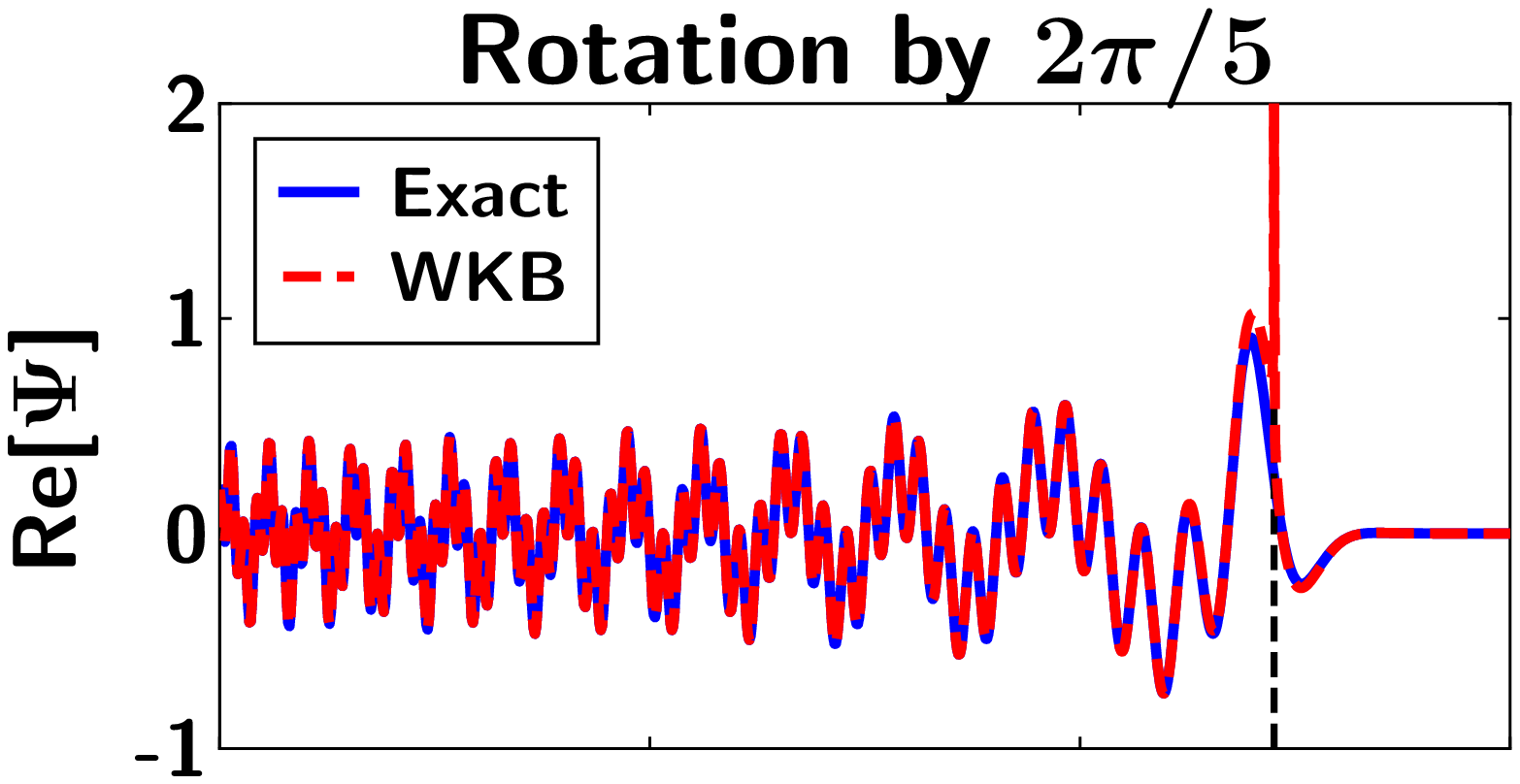}
		\put(85,5){\textbf{\color{black} \large(c)}}
	\end{overpic}
	\begin{overpic}[width=0.24\linewidth,trim={2mm 24mm 1mm 17mm},clip]{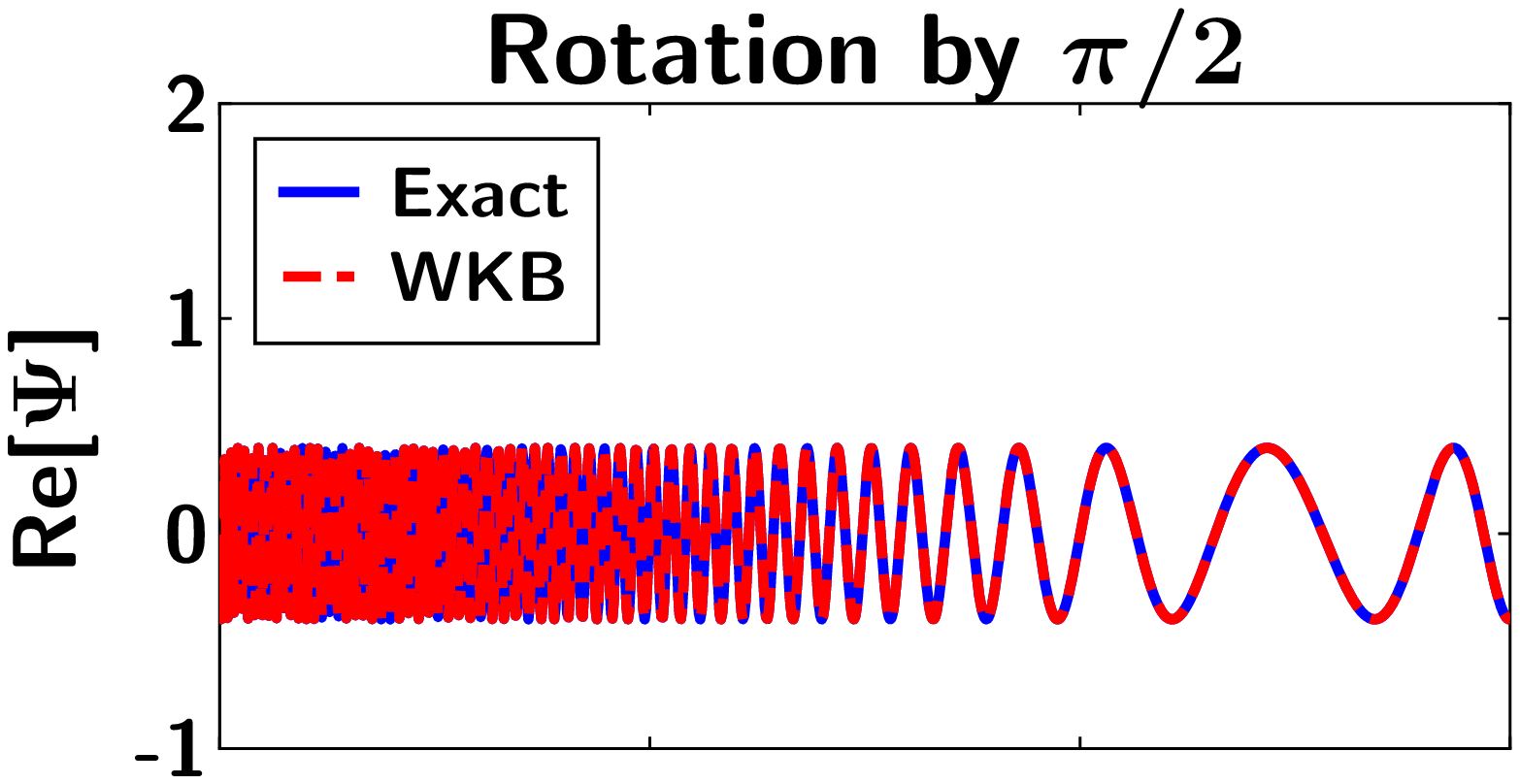}
		\put(85,5){\textbf{\color{black} \large(d)}}
	\end{overpic}
	
	\begin{overpic}[width=0.24\linewidth,trim={2mm 11mm 3mm 21mm},clip]{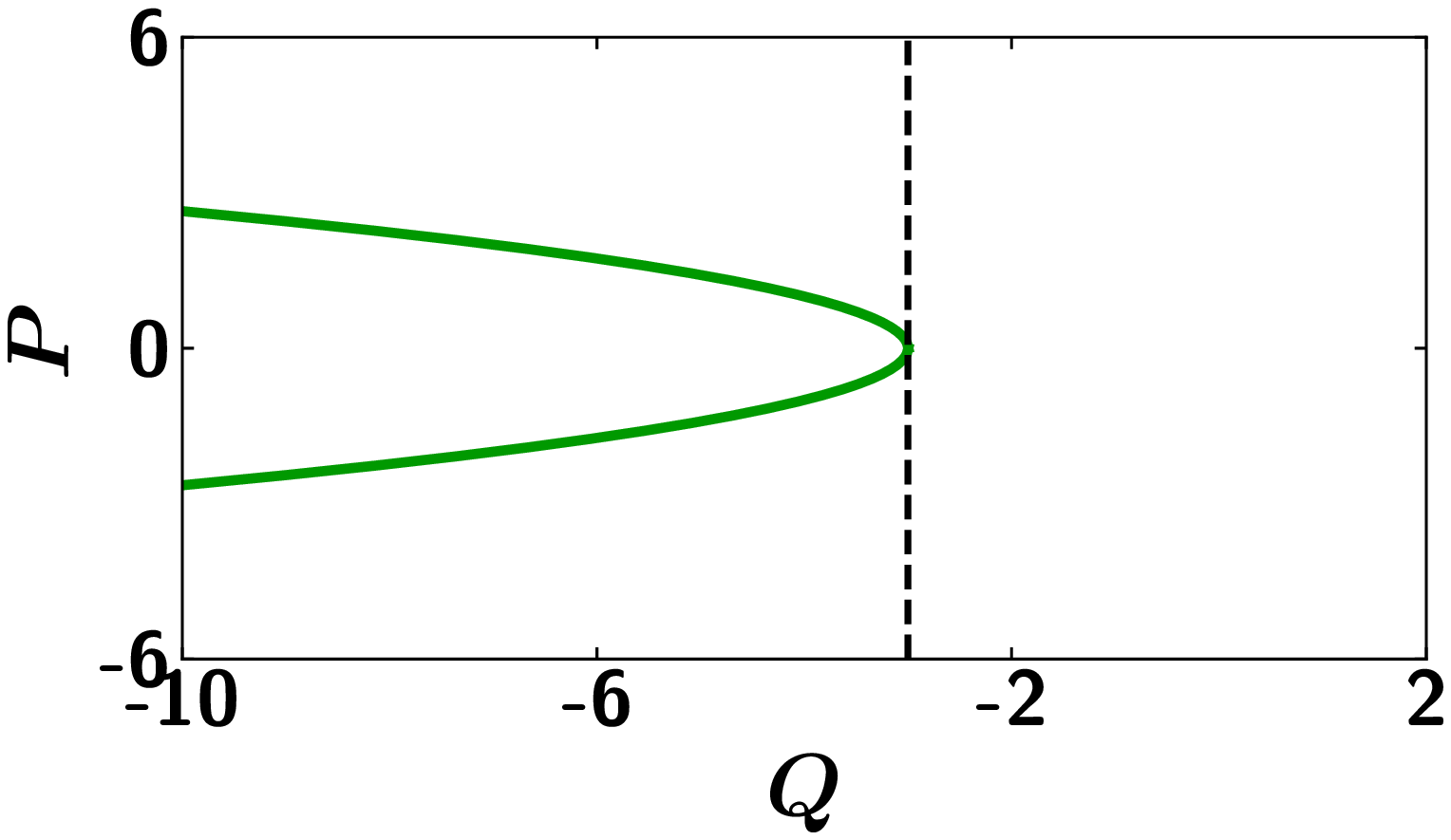}
		\put(85,15){\textbf{\color{black} \large(e)}}
	\end{overpic}
	\begin{overpic}[width=0.24\linewidth,trim={2mm 11mm 3mm 21mm},clip]{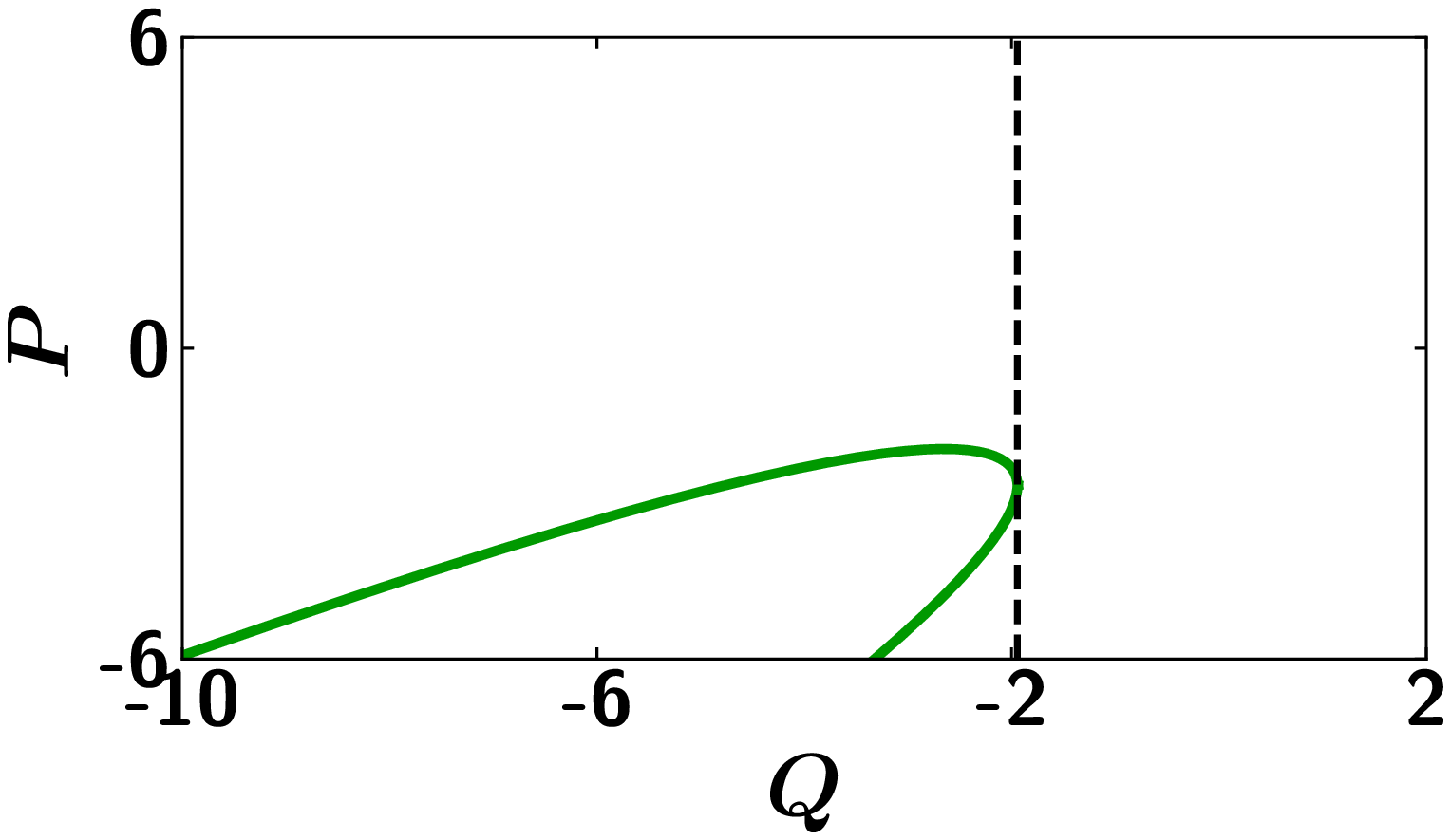}
		\put(85,15){\textbf{\color{black} \large(f)}}
	\end{overpic}
	\begin{overpic}[width=0.24\linewidth,trim={2mm 11mm 3mm 21mm},clip]{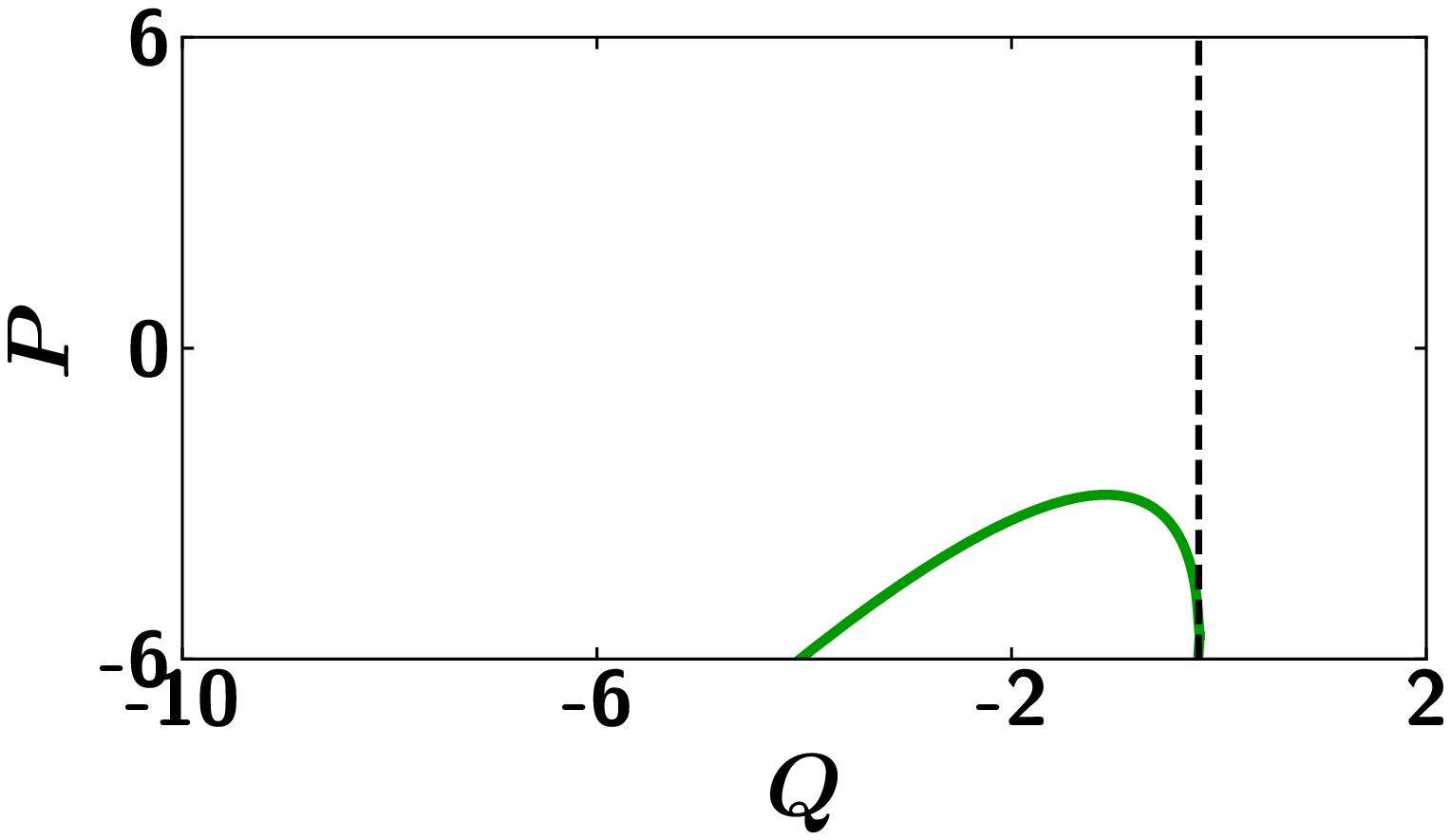}
		\put(85,15){\textbf{\color{black} \large(g)}}
	\end{overpic}
	\begin{overpic}[width=0.24\linewidth,trim={2mm 11mm 3mm 21mm},clip]{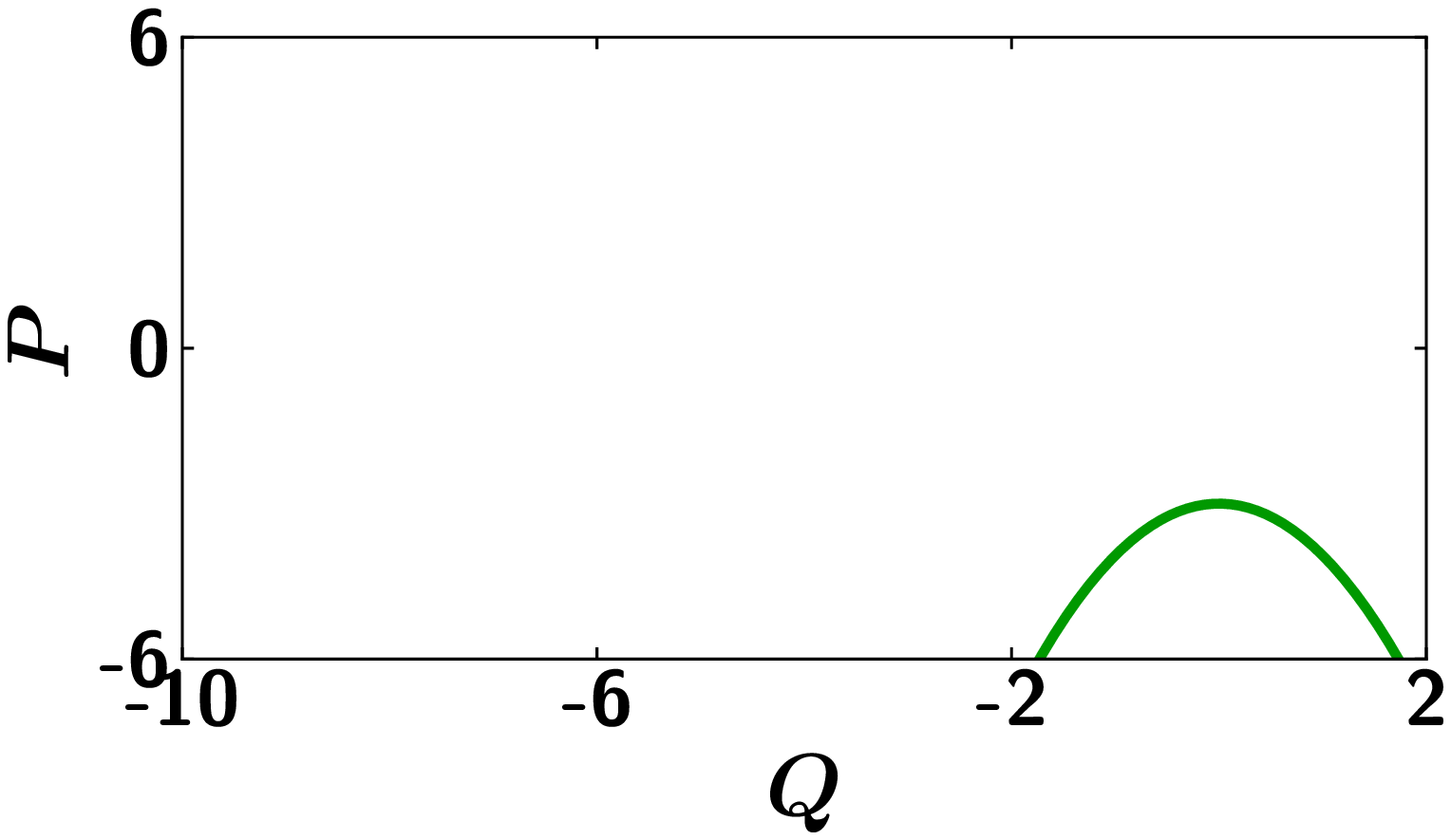}
		\put(85,15){\textbf{\color{black} \large(h)}}
	\end{overpic}
	\caption{\textbf{(a)-(d)} Comparison of the exact and WKB solution of the metaplectically-rotated Airy's equation \eq{metAiry} for $a = -3$ and different rotation angles $t$: (a) $t = 0$, (b) $t = -\pi/4$, (c) $t = -2\pi/5$, and (d) $t = -\pi/2$. \textbf{(e)-(h)} Corresponding WKB dispersion surfaces for \Eq{metAiry}. The caustic at $Q = \tan(t)\sin(t)/4 + a\cos(t)$ coincides with the location on the dispersion surface where $\dd P/ \dd Q \to \infty$.}
	\label{rotAiry}
\end{figure*}

Recall from \Fig{IterNIMTstab} that the iterated NIMT is typically a magnifying transformation whose magnification factor depends in a complicated manner on both the path discretization and the input function. For our chosen examples, the magnification is reduced by refining the discretization of the path $\Mat{S}_t$ (\Fig{disruption}). When a step size of $\pi/500$ is used, $\Psi_t(Q)$ quickly disrupts and becomes dominated by noise. However, refining the discretization by a factor of $10$ avoids the numerical instability and leads to a well-behaved solution. 

We reiterate that the magnification of the NIMT is not reduced for \textit{every} input function by refining the discretization; a rigorous profiling should be performed to determine how the magnification scales with path discretization when using the iterated NIMT in a new application. Alternatively, since the magnification scales with Fourier mode number, occasionally smoothing the signal between NIMT iterations will suppress high-frequency growth. This approach is shown in the final column of \Fig{disruption}, where a third-degree Savitzky--Golay filter~\cite{Savitzky64} with a window size of $5$ is applied every $50$ iterations.


\section{Phase-space rotation for cutoff removal}
\label{SecRot}

In addition to the time integration of quadratic Hamiltonian systems, the NIMT is also naturally suited for modeling caustics that arise in geometrical optics near cutoffs. As motivated in the introduction, such caustics can be resolved by rotating the phase space using metaplectic operators. Although this idea of using phase-space rotations to avoid caustics is not entirely new~\cite{Tracy14,Littlejohn85}, it is not yet a common practice, and therefore merits brief discussion.

Consider a wave field incident on an isolated cutoff in a $1$-D inhomogeneous medium. As is well-known, the corresponding wave field $\psi$ near the cutoff is approximately described by Airy's equation~\cite{Tracy14}
\begin{equation}
	\pd^2_q \psi(q) - (q - a) \psi(q) = 0 \, ,
	\label{airy}
\end{equation}

\noindent with the cutoff located at $a$. Applying the Wentzel--Kramers--Brillouin (WKB) approximation to \Eq{airy} yields the dispersion surface $p(q) = \sqrt{a-q}$ on which the wave `quanta' is asymptotically confined, as well as the divergent wave envelope $\phi(q) \sim (a-q)^{-1/4}$. Thus, the caustic at $q = a$ manifests as a singularity in the WKB envelope, as illustrated in the first column of \Fig{rotAiry}.

\begin{figure*}[t!]
    \centering
    \begin{overpic}[width=0.24\linewidth,trim={2mm 24mm 1mm 17mm},clip]{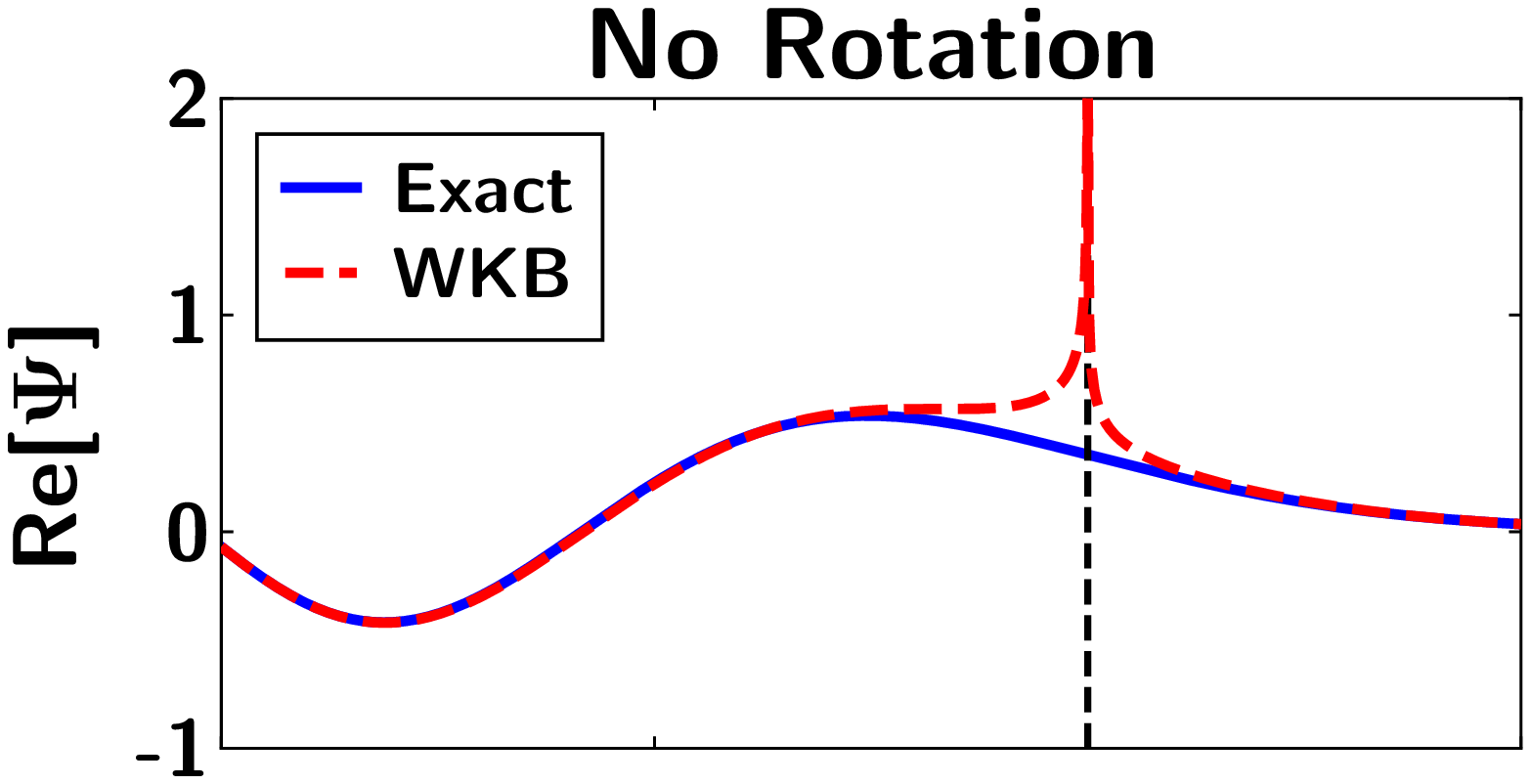}
		\put(85,5){\textbf{\color{black} \large(a)}}
	\end{overpic}
	\begin{overpic}[width=0.24\linewidth,trim={2mm 24mm 1mm 17mm},clip]{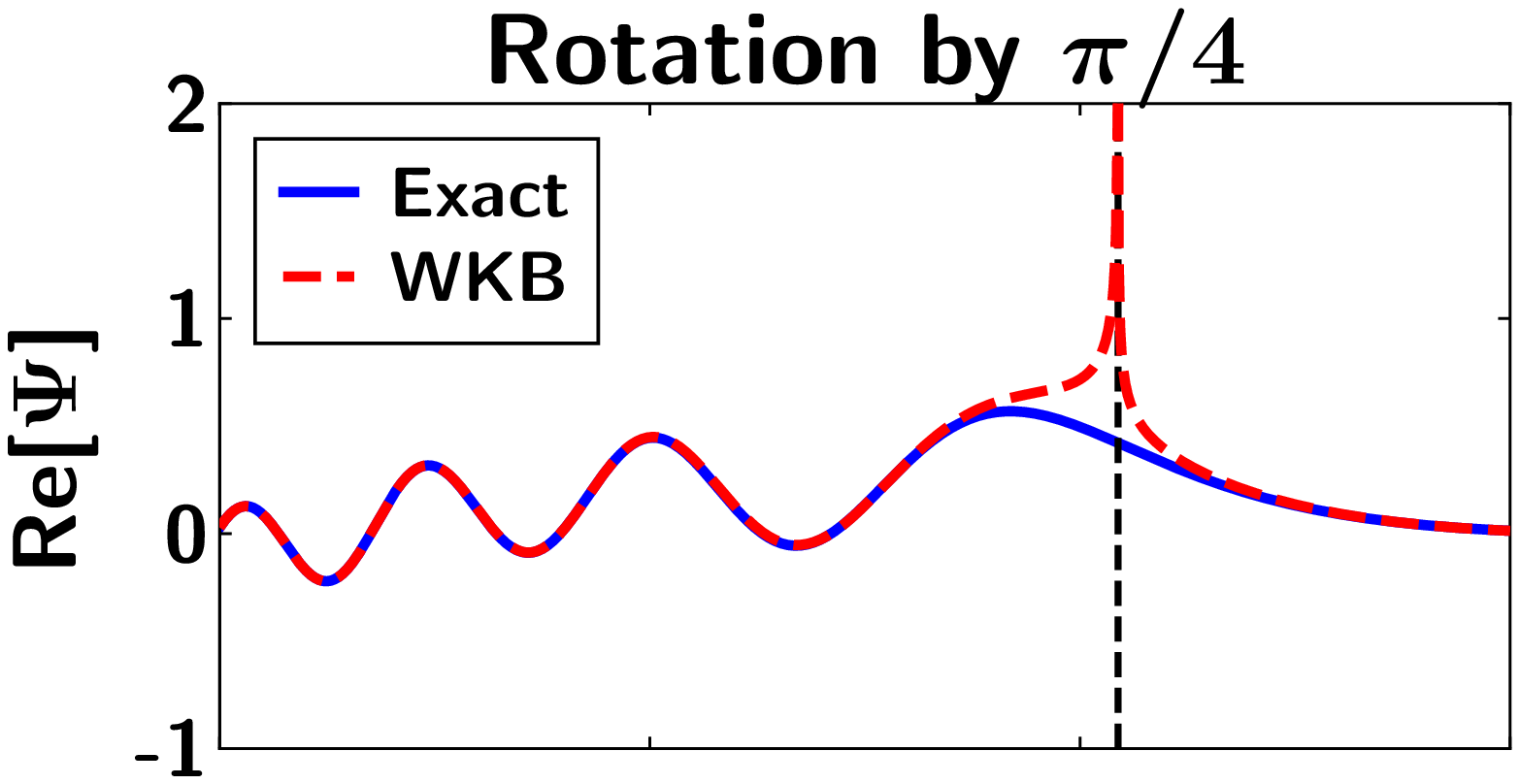}
		\put(85,5){\textbf{\color{black} \large(b)}}
	\end{overpic}
	\begin{overpic}[width=0.24\linewidth,trim={2mm 24mm 1mm 17mm},clip]{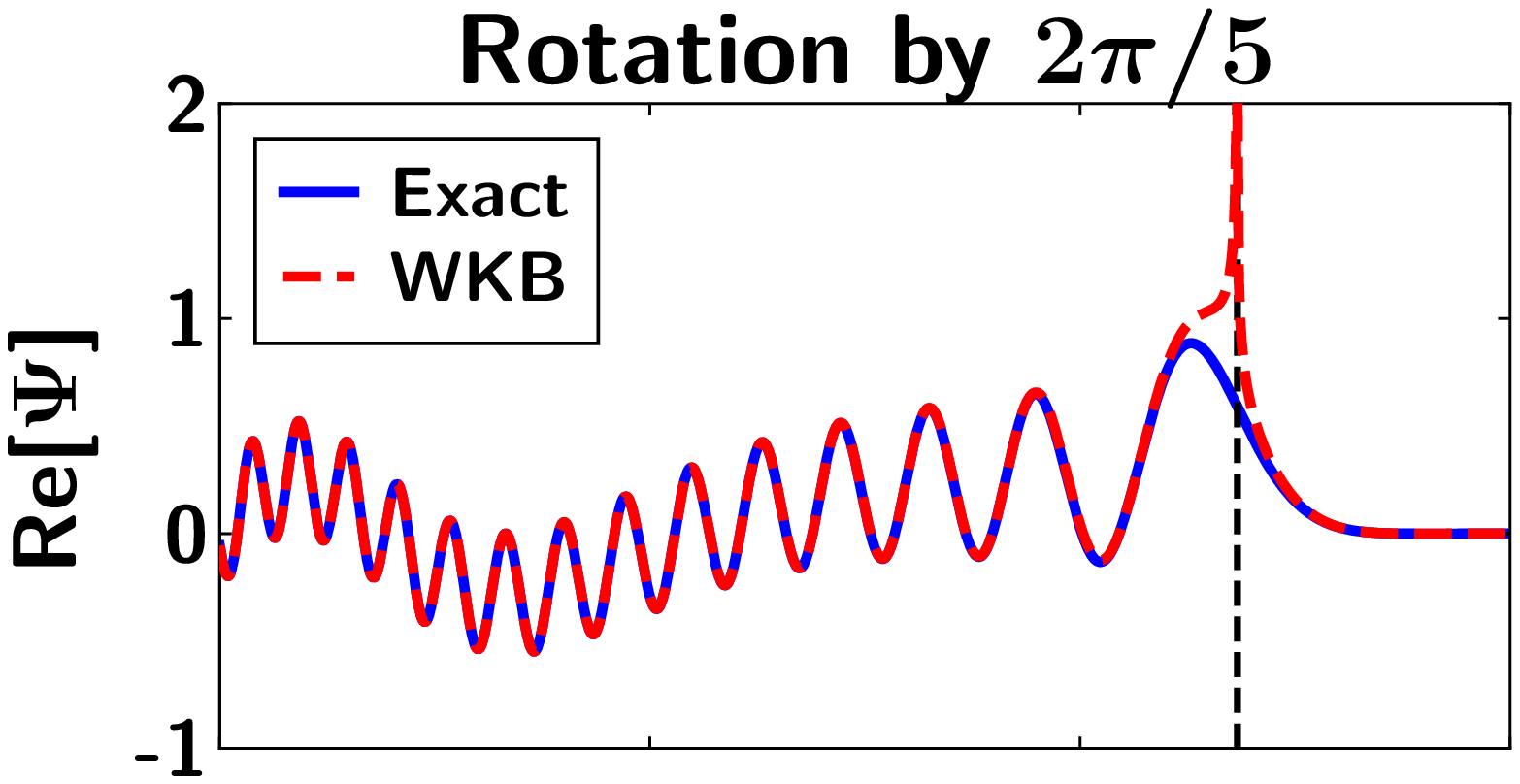}
		\put(85,5){\textbf{\color{black} \large(c)}}
	\end{overpic}
	\begin{overpic}[width=0.24\linewidth,trim={2mm 24mm 1mm 17mm},clip]{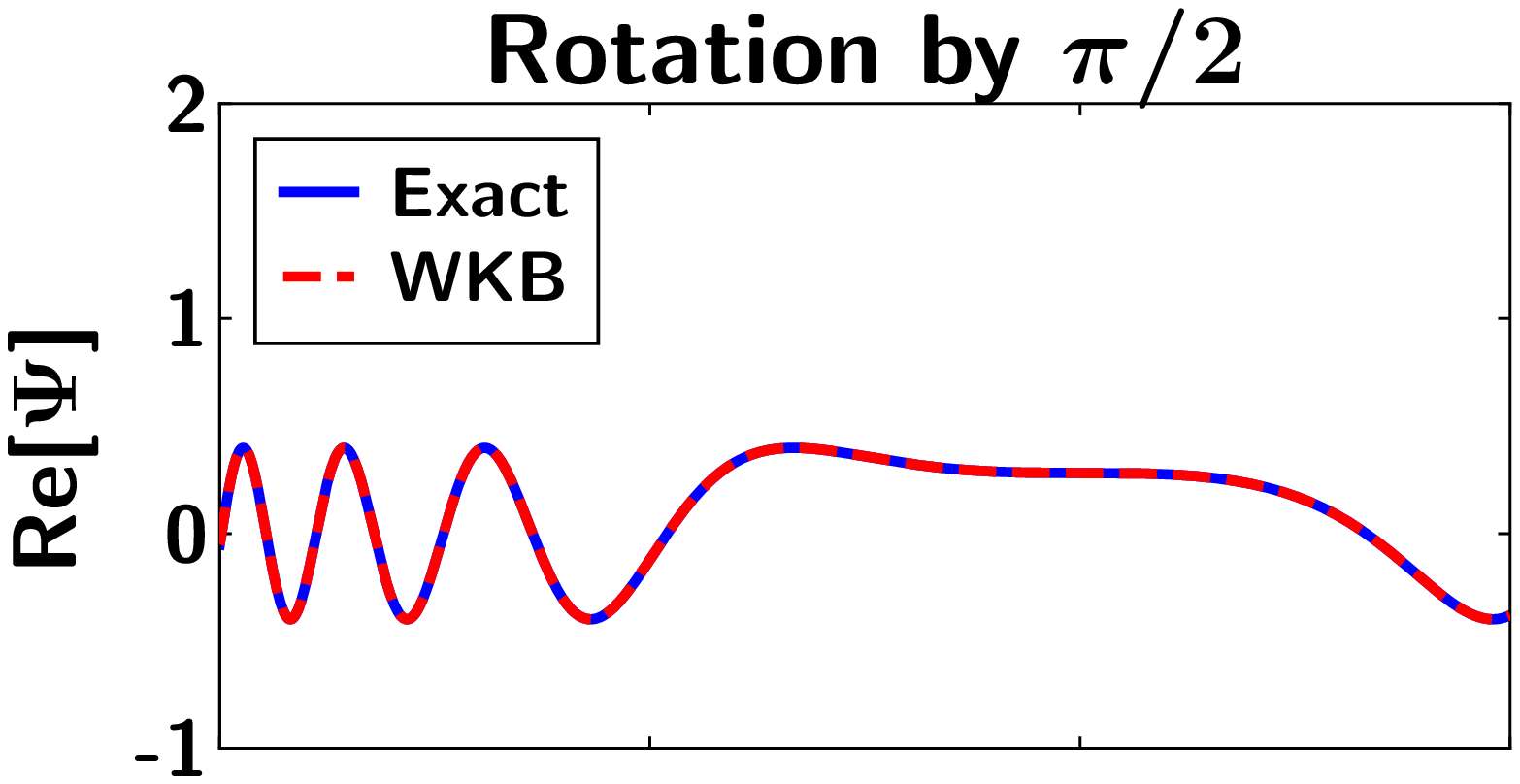}
		\put(85,5){\textbf{\color{black} \large(d)}}
	\end{overpic}
	
	\begin{overpic}[width=0.24\linewidth,trim={2mm 11mm 3mm 21mm},clip]{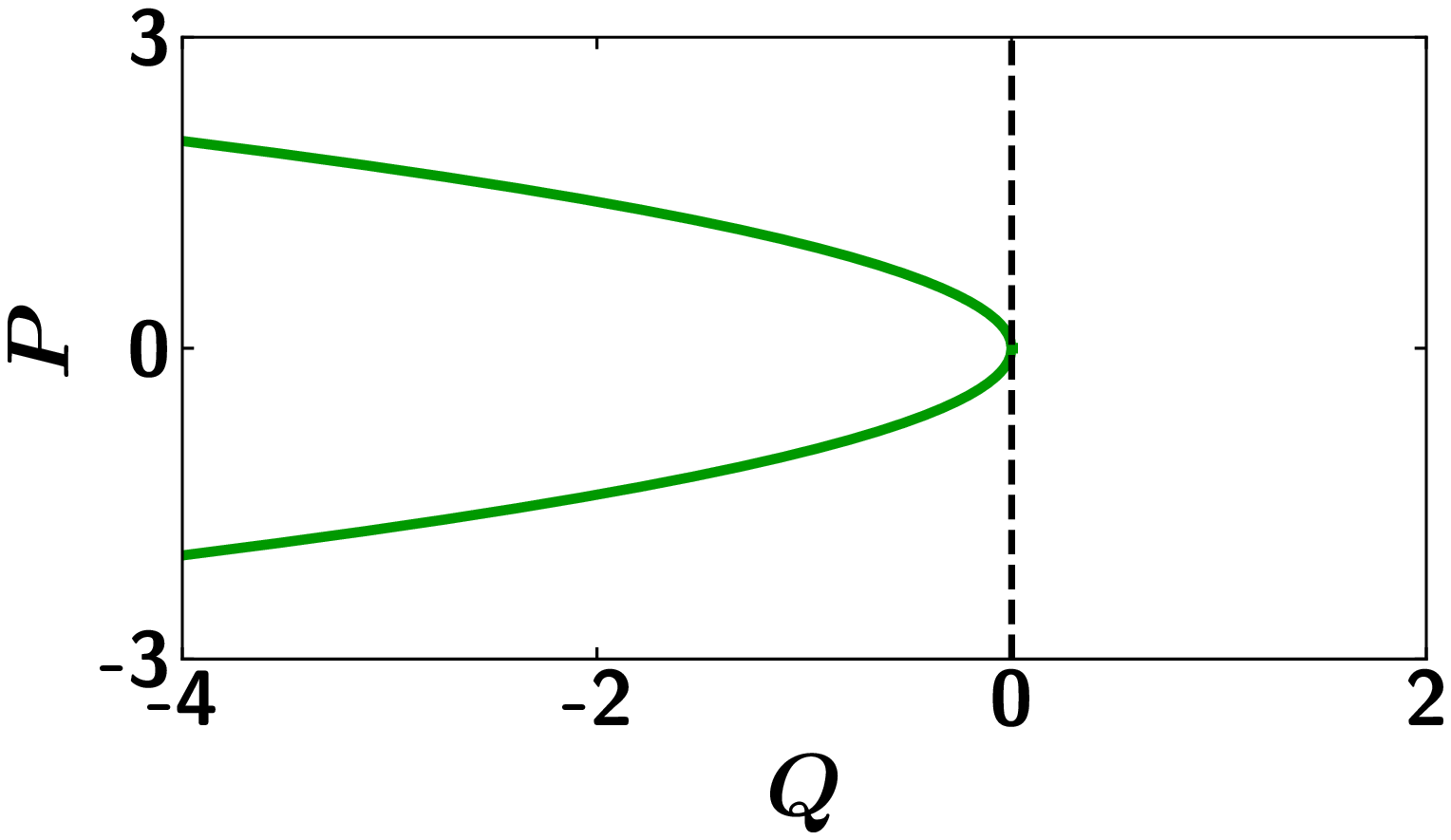}
		\put(85,15){\textbf{\color{black} \large(e)}}
	\end{overpic}
	\begin{overpic}[width=0.24\linewidth,trim={2mm 11mm 3mm 21mm},clip]{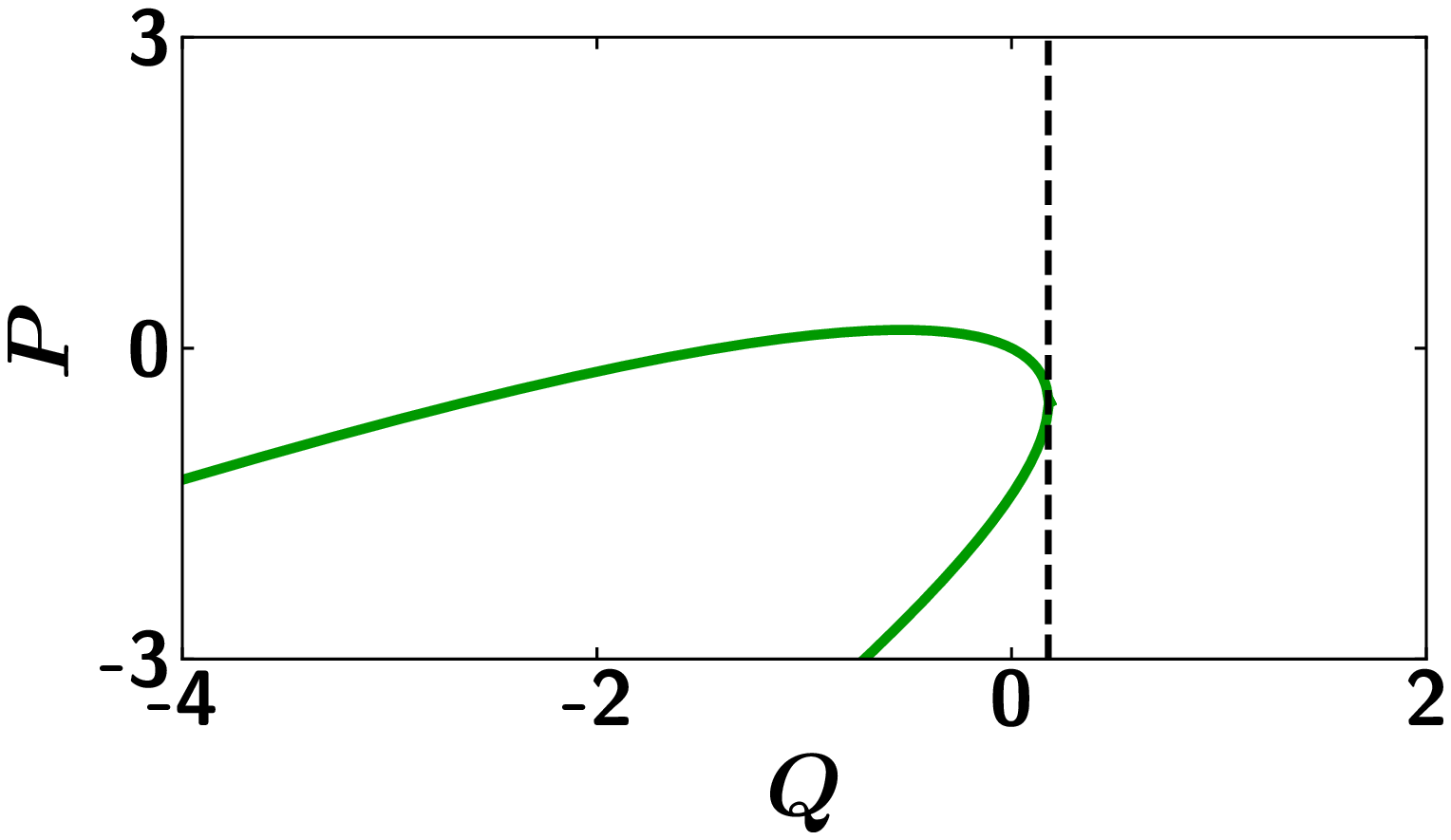}
		\put(85,15){\textbf{\color{black} \large(f)}}
	\end{overpic}
	\begin{overpic}[width=0.24\linewidth,trim={2mm 11mm 3mm 21mm},clip]{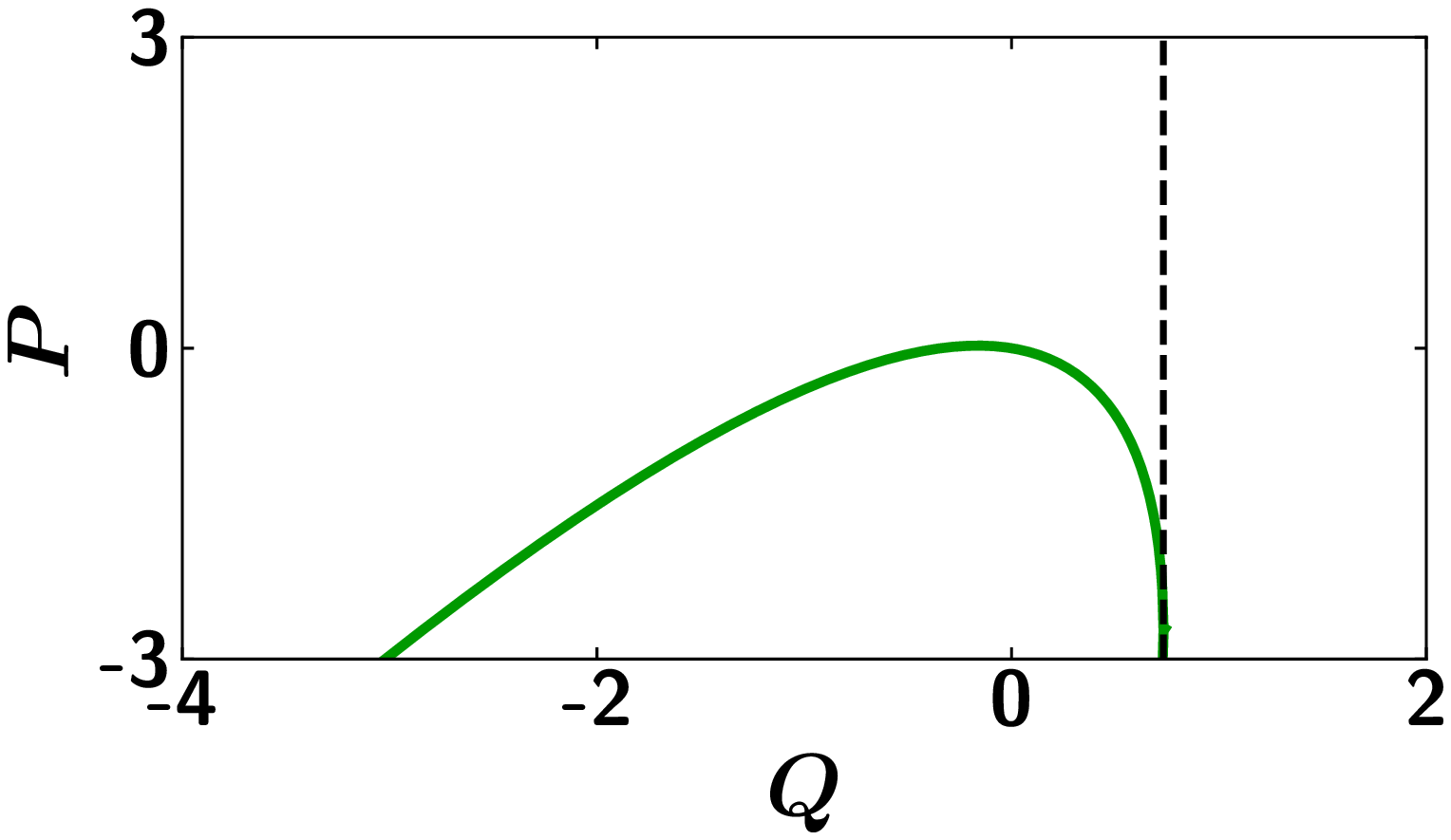}
		\put(85,15){\textbf{\color{black} \large(g)}}
	\end{overpic}
	\begin{overpic}[width=0.24\linewidth,trim={2mm 11mm 3mm 21mm},clip]{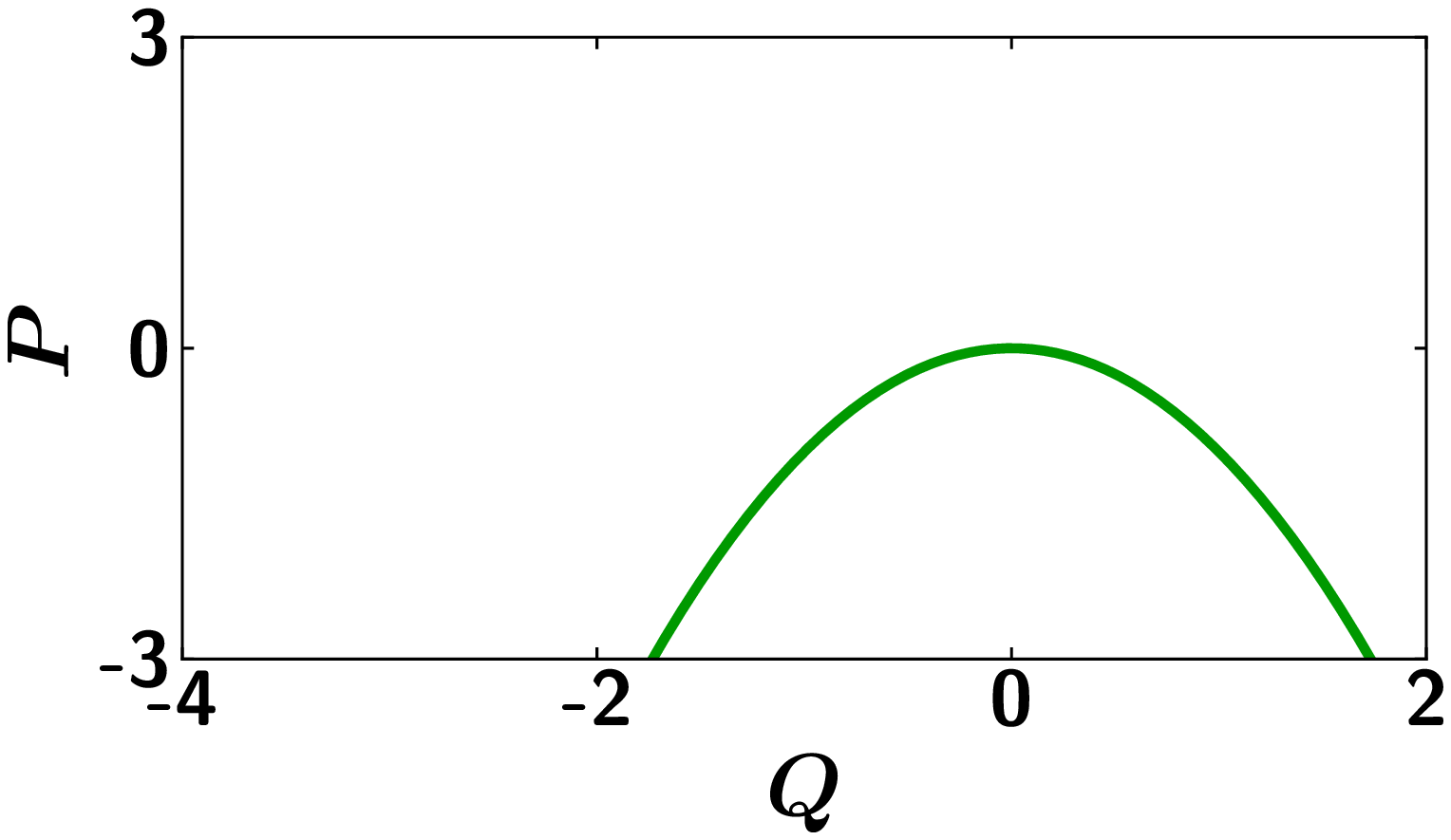}
		\put(85,15){\textbf{\color{black} \large(h)}}
	\end{overpic}
	\caption{Same as \Fig{rotAiry}, but with $a = 0$.}
	\label{rotAiry0}
\end{figure*}

Let us now rotate the phase space using the MT corresponding to \Eq{SEvo} as
\begin{subequations}
    \begin{align}
	    q &= \cos(t) Q + i \sin(t) \frac{\dd}{\dd Q} \, , \\
	    \frac{\dd}{\dd q} &= i \sin(t) Q + \cos(t) \frac{\dd}{\dd Q} \, ,
    \end{align}
\end{subequations}

\begin{widetext}
    \noindent with $t$ now specifying the (negative) rotation angle rather than time. Then, \Eq{airy} becomes
    \begin{equation}
	    \cos^2(t) \pd^2_Q \Psi(Q) + i\left[\sin(2t) Q - \sin(t)\right] \pd_Q \Psi(Q) - \left[ \sin^2(t) Q^2 + \cos(t) Q - \frac{i}{2}\sin(2t) - a\right] \Psi(Q) = 0 \, ,
	    \label{metAiry}
    \end{equation}

    \noindent where $\Psi(Q)$ is the metaplectic image of $\psi(q)$. Applying the WKB approximation to \Eq{metAiry} yields
    \begin{subequations}
        \label{metAiryWKB}
	    \begin{align}
		    \Psi_\text{WKB}(Q) &=  \frac{\alpha_+ e^{i\Theta_+(Q)} + \alpha_- e^{i\Theta_-(Q)}}{\left[ \sin^2(t) - 4Q \cos(t) + 4a\cos^2(t) \right]^{1/4}} \, , \\
		    \Theta_\pm(Q) &= \frac{\tan(t) Q - \sin(t) Q^2}{2 \cos(t)} \pm \frac{\left[ \sin^2(t) - 4Q \cos(t) + 4a\cos^2(t) \right]^{3/2}}{12 \cos^3(t)} \, ,
	    \end{align}
    \end{subequations}

    \noindent with $\alpha_\pm$ constants determined by boundary conditions, which should be matched on either side of the caustic separately due to Stokes phenomenon~\cite{Heading62}. 

    In \Fig{rotAiry}, the WKB result is compared to the exact result, which can be computed via \Eq{metTRANS} as
    \begin{align}
	    \Psi_t(Q) = \pm \frac{1}{\sqrt{\cos(t)}} \, \mathrm{Ai}\left[\frac{Q}{\cos(t)} - \frac{\tan^2(t)}{4} - a\right] \exp\left[i\frac{\tan(t) Q - \sin(t) Q^2}{2 \cos(t)} - i\frac{\tan^3(t)}{12} - i\frac{a\tan(t)}{2}\right] \, , 
    \end{align}
\end{widetext}

\noindent where $\mathrm{Ai}$ is the Airy function~\cite{Olver10}. As the phase space is rotated, the caustic moves steadily towards increasing $Q$. At a rotation angle of $\pi/2$, \Eq{metAiry} becomes
\begin{equation}
    -i\pd_Q \Psi(Q) + (Q^2 - a)\Psi(Q) = 0 \, .
    \label{airyFT}
\end{equation}

\noindent In this case, the caustic disappears entirely and the WKB approximation, obtained from the ($+$) solution to \Eqs{metAiryWKB} as
\begin{equation}
    \Psi_\text{WKB}(Q) = \alpha e^{iaQ - iQ^3/3} \, ,
    \label{airyFTwkb}
\end{equation}

\noindent becomes exact. Importantly, the WKB approximation to \Eq{airyFT} holds at any $a$, even though for $a \ge 0$ there are values of $Q$ at which the wavenumber $P$ approaches zero. (For example, see \Fig{rotAiry0} for the case $a = 0$.) This is to be expected because: (i) $a$ can be removed from \Eq{airy} by a simple variable transformation, and hence has no fundamental meaning, and (ii) it is $\dd P/ \dd Q$ that determines the validity of geometrical optics, not the value of $P$ \textit{per~se}.

\begin{figure}
    \includegraphics[width = \linewidth]{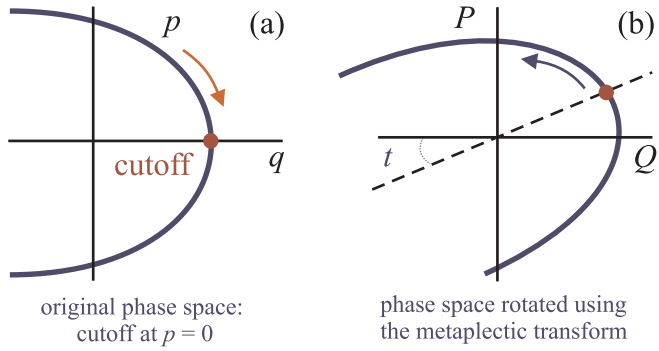}
    \caption{A schematic of the dispersion curves of a one-dimensional wave incident on a cutoff: (a) in the original phase-space coordinates, (b) in phase-space coordinates rotated by some nonzero angle $t$ (blue arrow). In the original coordinates, the geometrical-optics envelope travels along the dispersion curve (red arrow) and becomes singular when it reaches the cutoff, where $\dd p/\dd q \to \infty$ (red circle). In the rotated coordinates, $\dd P/\dd Q$ is finite at the location corresponding to the former cutoff (red circle), and the envelope remains analytic at this location, i.e., the caustic is avoided.}
    \label{causticRot}
\end{figure}

For other equations, a single MT is not sufficient to reinstate geometrical optics for the entire field. However, multiple MTs applied sequentially can. Specifically, a phase space can be continually rotated using the NIMT such that $\dd P/\dd Q$ always remains finite (\Fig{causticRot}). In that frame, the WKB approximation will hold indefinitely and there will be no caustics. 

For example, consider the wave equation
\begin{equation}
    \pd_q^2 \psi + (a^2 - q^2) \psi = 0 \, ,
    \label{hermEQ}
\end{equation}

\noindent with $a$ constant. The WKB approximation is applicable only far enough from the cutoffs located at $q = \pm a$, and there is no single MT for which the image of \Eq{hermEQ} will be free of cutoffs. However, note that \Eq{hermEQ} is the time-independent limit of the QHO \eq{QHO}, whose solutions are eigenfunctions of the phase-space rotation operator (\Sec{SecMET}). Therefore, in the appropriately-rotating frame which maintains the wavenumber constant (say, $P = a$), the WKB approximation holds perfectly. More general wave equations can be handled in a similar manner, but require introducing additional machinery. Hence, the corresponding discussion is postponed until future publications.

\section{Conclusion}
\label{SecCONCL}

In this work, we derive a pseudo-differential representation of the MT in arbitrary dimensions. This is a general result which can be useful for both analytical and numerical applications. An important example is the simulation of a wavepacket evolving in a quadratic potential, whose propagator is a metaplectic operator. Evolving the system by $\Delta t$ would invoke an MT that is near-identity, which is not a common consideration in MT-algorithm design. In contrast, the pseudo-differential representation that we propose here readily displays the simplicity of the MT in the near-identity limit, suggestive of a new algorithm.

Specifically, in the near-identity limit the pseudo-differential series can be accurately truncated; the correspondingly finite stencil width then enables local, pointwise transformations. This is useful when transforming `incomplete' functions, \eg signals measured over finite intervals; it also leads naturally to a linear time algorithm called the NIMT. When applied once, the NIMT performs a fast, near-identity transformation; when iterated, the NIMT can perform an arbitrary MT by synthesizing a series of near-identity transformations. With a computational efficiency of $O(K N^3 N_p)$, the NIMT is potentially faster than existing MT algorithms, which often scale as $O(N_p \log N_p)$ from their similarity with the fast Fourier transform. Moreover, unlike these other algorithms, the NIMT is the same algorithm regardless the number of dimensions and the structure of~$\Mat{S}$. Hence, the NIMT is flexible in its application, and should thereby complement the existing collection of MT algorithms.

We assess the stability of the iterated NIMT and identify two dominant instabilities: the loss of unitarity via truncation error (magnification), and the poor conditioning of discrete derivatives (d-instability). One might expect the NIMT magnification to be suppressed by reducing the transformation `step size', \ie its deviation from identity, or by increasing the number of terms retained; however, this is not true. Reducing the step size increases the number of iterations needed to perform a finite transformation, and it is not clear whether this tradeoff is beneficial in the general case. Increasing the truncation order indeed decreases the NIMT magnification, but also increases its susceptibility to the d-instability. The most robust avenue to NIMT stability therefore appears to be the combined use of a low-order truncation with occasional smoothing, which we demonstrate in a numerical example.


\section*{Acknowledgements}

This work was supported by the U.S. DOE through Contract No. DE-AC02-09CH11466.


\appendix


\section{Derivation of Eqs. (13)}
\label{AppABCDderiv}

Here, we present the derivation of \Eqs{SymplecABCD} from \Eq{SymplecS}, which is known~\cite{Tracy93} but included here for completeness.
Consider
\begin{equation}
    \Mat{J} \doteq \begin{pmatrix}
            \OMat{N} & \IMat{N}\\
            -\IMat{N} & \OMat{N}
        \end{pmatrix} \, .
        \label{JDEF}
\end{equation}

\noindent Since $\Mat{J}\Mat{J} = -\IMat{2N}$, \Eq{SymplecS} implies that
\begin{equation}
    \Mat{S}^{-1} = - \Mat{J} \Mat{S}^\intercal \Mat{J} = \begin{pmatrix}
            \Mat{D}^\intercal & -\Mat{B}^\intercal\\
            -\Mat{C}^\intercal & \Mat{A}^\intercal
        \end{pmatrix}
\end{equation}

\noindent and also that $\Mat{S}^{-1}$ is symplectic, \ie
\begin{equation}
    \Mat{S}^{-1}\Mat{J} \left(\Mat{S}^{-1}\right)^\intercal = \Mat{J} \, .
    \label{SInvJeq}
\end{equation}

Using
\begin{equation}
    \Mat{S} \Mat{J} \Mat{S}^\intercal = \begin{pmatrix}
            \Mat{A}\Mat{B}^\intercal - \Mat{B}\Mat{A}^\intercal & \Mat{A}\Mat{D}^\intercal - \Mat{B}\Mat{C}^\intercal\\
            \Mat{C}\Mat{B}^\intercal - \Mat{D}\Mat{A}^\intercal & \Mat{C}\Mat{D}^\intercal - \Mat{D}\Mat{C}^\intercal
        \end{pmatrix}
\end{equation}

\noindent together with \Eq{SymplecS} leads to \Eqs{symplec1}, \eq{symplec3}, and \eq{symplec6}. Likewise, since
\begin{equation}
    \Mat{S}^{-1}\Mat{J} \left(\Mat{S}^{-1}\right)^\intercal = \begin{pmatrix}
            \Mat{B}^\intercal\Mat{D} - \Mat{D}^\intercal\Mat{B} & \Mat{D}^\intercal\Mat{A} - \Mat{B}^\intercal\Mat{C}\\
            \Mat{C}^\intercal\Mat{B} - \Mat{A}^\intercal\Mat{D} & \Mat{A}^\intercal\Mat{C} - \Mat{C}^\intercal\Mat{A}
        \end{pmatrix} \, ,
\end{equation}

\noindent \Eq{SInvJeq} readily yields \Eqs{symplec2}, \eq{symplec4}, and \eq{symplec5}. 


\section{Deriving the metaplectic transform from its pseudo-differential representation}
\label{NIMT2MT}

Here, we show the pseudo-differential representation \eq{NIMTsymbND} leads to the original integral representation \eq{metTRANS} regardless the size of $\|\Lambda\|$. This proves that the PMT is in fact exact, even though it was originally derived in \Sec{SecPMT} using an expansion in $\|\Lambda\|$.

As a starting point, let us rewrite \Eq{NIMTsymbND} as
\begin{align}
	\Psi(\Vect{Q}) = &\pm \frac{e^{i\Vect{Q}^\intercal \Mat{G} \Vect{Q}}}{\sqrt{\det{\Mat{A}}}}\nonumber\\
	&\times \int \dd \Vect{q}'~ \delta \left(\Vect{q}' - \Mat{A}^{-1}\Vect{Q} \right) e^{\frac{i}{4} \Lambda \dubDot \pd^2_{\Vect{q}'\Vect{q}'}} \, \psi(\Vect{q}') \, ,
	\label{DeltaFunc}
\end{align}

\noindent where we have replaced $\nabla$ with $\pd_{\Vect{q}'}$ to avoid ambiguities. We introduce the Fourier representation of $\psi(\Vect{q})$ as
\begin{subequations}
	\begin{align}
		\psi(\Vect{q}) &= \frac{1}{(2\pi)^N} \int \dd \Vect{p} ~\fourier{\psi}(\Vect{p}) \, e^{i\Vect{q}^\intercal \Vect{p}} \, ,\\
		\fourier{\psi}(\Vect{p}) &= \int \dd \Vect{q} ~\psi(\Vect{q}) \, e^{-i\Vect{q}^\intercal \Vect{p}} \, ,
	\end{align}
\end{subequations}

\noindent which, when substituted into \Eq{DeltaFunc}, yields
\begin{align}
	\Psi(\Vect{Q}) =& \pm \frac{e^{i \Vect{Q}^\intercal \Mat{G} \Vect{Q}}}{(2\pi)^N\sqrt{\det{\Mat{A}}}} \int \dd \Vect{q}'~ \delta \left(\Vect{q}' - \Mat{A}^{-1}\Vect{Q} \right) \nonumber\\
	&\times \int \dd \Vect{q} ~\psi(\Vect{q}) \int \dd \Vect{p} ~ e^{-\frac{i}{4} \Vect{p}^\intercal \Lambda \Vect{p} + i(\Vect{q}'-\Vect{q})^\intercal \Vect{p}} \, .
\end{align}

\noindent The Gaussian integral can be performed explicitly,
\begin{align}
	&\int \dd \Vect{p} ~ e^{-\frac{i}{4} \Vect{p}^\intercal \Lambda \Vect{p} + i(\Vect{q}'-\Vect{q})^\intercal \Vect{p}}\nonumber\\
	&= \frac{(-2\pi i)^{N/2}}{\sqrt{\det{\Mat{A}^{-1} \Mat{B}}}} \, e^{\frac{i}{2} \left(\Delta \Vect{q}\right)^\intercal \Mat{B}^{-1} \Mat{A} \left(\Delta \Vect{q}\right) } \, ,
\end{align}

\noindent with the branch cut chosen to restrict all complex phases to the interval $(-\pi,\pi]$, and with $\Delta \Vect{q} \doteq \Vect{q}' - \Vect{q}$. Then, performing the trivial integration over $\dd \Vect{q}'$ yields \Eq{metTRANS}.

Note that neither the smoothness nor even the differentiability of $\psi$ is invoked in the above argument; the existence of the Fourier image of $\psi$ is enough.


\section{Asymptotic parameterization of near-identity symplectic matrices}
\label{AppMat}

The $N$-D NIMT involves computing the quantities $\det{\Mat{A}}$, $\Mat{A}^{-1}$, $\Mat{A}^{-1}\Mat{B}$, and $\Mat{C} \Mat{A}^{-1}$. However, when $\Mat{S}$ is near-identity, one can derive approximate asymptotic formulas for these quantities which help calculate them more efficiently. In particular, calculating the lowest-order terms does not require any explicit matrix multiplications.

Generally speaking, the near-identity behavior of a group is governed by its Lie algebra. For the group of $2N\times 2N$ real symplectic matrices, denoted $\Sp$, the Lie algebra is the space of all $2N\times2N$ real Hamiltonian matrices~\cite{Dragt05}. Note that a matrix $\Mat{H}$ is Hamiltonian if and only if $\Mat{J} \Mat{H}$ is symmetric, with $\Mat{J}$ defined in \Eq{JDEF}.

By the connectivity of $\Sp$ and the polar decomposition, any symplectic matrix $\Mat{S}$ can be parameterized as~\cite{Littlejohn86a, Dragt82, Hall15}
\begin{equation}
    \Mat{S} = e^{\epsilon \Mat{H}_s}e^{\epsilon \Mat{H}_a} \, ,
    \label{polarS}
\end{equation}

\noindent where $\Mat{H}_s$ and $\Mat{H}_a$ are symmetric and antisymmetric Hamiltonian matrices, respectively. The formal parameter $\epsilon$ has been introduced to aid with ordering the forthcoming expansions when $\Mat{H}_s$ and $\Mat{H}_a$ are small. Note that if $\Mat{H}$ is Hamiltonian, then $\Mat{H}^\intercal$ also is; hence, $\Mat{H}_s$ and $\Mat{H}_a$ can be uniquely represented as
\begin{equation}
    \Mat{H}_s = \frac{\Mat{H} + \Mat{H}^\intercal}{2} \, , \quad \Mat{H}_a = \frac{\Mat{H} - \Mat{H}^\intercal}{2}
\label{HDecomp}
\end{equation}

\noindent for some Hamiltonian matrix $\Mat{H}$~\cite{Dragt82}. In this sense, $\Mat{S}$ is parameterized by a single Hamiltonian matrix $\Mat{H}$.

Let us consider the case when $\Mat{S}$ is near-identity, meaning $\Mat{H}$ is close to $\OMat{2N}$. Expanding \Eq{polarS} in $\epsilon$ yields
\begin{equation}
    \Mat{S} = \IMat{2N} + \epsilon \Mat{H} + \frac{\epsilon^2}{4}\left(2\Mat{H} \Mat{H} - \Mat{H} \Mat{H}^\intercal + \Mat{H}^\intercal \Mat{H} \right) + O(\epsilon^3) \, .
    \label{SMATexpand}
\end{equation}

\noindent Since any Hamiltonian matrix can be decomposed as
\begin{equation}
    \Mat{H} = \begin{pmatrix}
        \Mat{V}^\intercal & \Mat{W}\\
        -\Mat{U} & -\Mat{V}
    \end{pmatrix} = \Mat{J}
    \begin{pmatrix}
        \Mat{U} & \Mat{V}\\
        \Mat{V}^\intercal & \Mat{W}
    \end{pmatrix} \, ,
    \label{Hblock}
\end{equation}

\noindent with $\Mat{U}$ and $\Mat{W}$ being symmetric matrices, we obtain the following expansions from \Eq{SMATexpand}:
\begin{subequations}
\begin{align}
	&\Mat{A} \approx \IMat{N} + \epsilon\Mat{V}^\intercal + \epsilon^2 \frac{\Mat{V}_s \Mat{V}^\intercal - \Mat{V}^\intercal \Mat{V}_a+ \TmMat \Mat{U} - \Mat{W} \TpMat}{2} ,\\
	&\Mat{B} \approx \epsilon\Mat{W} + \epsilon^2 \frac{\Mat{V}_s\Mat{W} - \Mat{W} \, \Mat{V}_a + \TmMat \Mat{V} + \Mat{V}^\intercal \TpMat}{2} ,\\
	&\Mat{C} \approx -\epsilon\Mat{U} + \epsilon^2 \frac{\Mat{V}_s \Mat{U} + \Mat{U} \, \Mat{V}_a - \TmMat \Mat{V}^\intercal + \Mat{V} \, \TpMat}{2} ,\\
	&\Mat{D} \approx \IMat{N}-\epsilon\Mat{V} + \epsilon^2 \frac{\Mat{V}_s \Mat{V} + \Mat{V} \, \Mat{V}_a - \TmMat \Mat{W} - \Mat{U} \TpMat}{2} ,
\end{align}
\end{subequations}

\noindent where
\begin{subequations}
	\begin{align}
		\Mat{V}_s &\doteq \frac{1}{2}\left(\Mat{V} + \Mat{V}^\intercal \right) \, , \quad \Mat{V}_a \doteq \frac{1}{2}\left(\Mat{V} - \Mat{V}^\intercal \right) \, ,\\
		\TpMat &\doteq \frac{1}{2}\left(\Mat{U} + \Mat{W} \right) \, , \quad \TmMat \doteq \frac{1}{2}\left(\Mat{U} - \Mat{W} \right) \, .
	\end{align}
\end{subequations}

One can show that
\begin{equation}
	\Mat{A}^{-1} \approx \IMat{N} - \epsilon \Mat{V}^\intercal + \epsilon^2 \frac{\Mat{V}^\intercal \Mat{V}_s - \Mat{V}_a\Mat{V}^\intercal - \TmMat \Mat{U} + \Mat{W} \TpMat}{2}
	\label{AInvExpans}
\end{equation}

\noindent satisfies both $\Mat{A}^{-1} \Mat{A} = \IMat{N}$ and $\Mat{A} \Mat{A}^{-1} = \IMat{N}$ to $O(\epsilon^3)$. By direct multiplication one also obtains
\begin{subequations}
	\begin{align}
		\Mat{A}^{-1}\Mat{B} &\approx \epsilon\Mat{W} + \epsilon^2\left(\Mat{V}_a\Mat{W} + \Mat{V}^\intercal\TmMat \right)_s \, ,\\
		\Mat{C} \Mat{A}^{-1} &\approx -\epsilon\Mat{U} + \epsilon^2\left(\Mat{V}_s \Mat{U} + \Mat{V}\TpMat \right)_s \, ,
	\end{align}
\end{subequations}

\noindent where the subscript $_s$ denotes the symmetric part. Notably, the expansions of both $\Mat{A}^{-1} \Mat{B}$ and $\Mat{C} \Mat{A}^{-1}$ are symmetric at each order of $\epsilon$, as required by \Eqs{symplec3} and \eq{symplec5}. Finally, let us approximate $\det{\Mat{A}}$ as
\begin{subequations}
    \begin{gather}
        \label{CharPoly}
	    \det{\Mat{A}} \approx \det\left(\IMat{N} + \epsilon \Mat{M} \right) \, ,\\
	    \Mat{M} \doteq \Mat{V}^\intercal + \epsilon \frac{\Mat{V}_s \Mat{V}^\intercal - \Mat{V}^\intercal \Mat{V}_a+ \TmMat \Mat{U} - \Mat{W} \TpMat}{2} \, .
    \end{gather}
\end{subequations}

\noindent Up to the factor $\epsilon^N$, the right-hand side of \Eq{CharPoly} is simply the characteristic polynomial of $-\Mat{M}$. Using, for example, Faddeev--LeVerrier's method leads to
\begin{align}
	\det{\Mat{A}} \approx &\, 1 + \epsilon \, \Tr \left( \Mat{V} \right)\nonumber\\
	&+ \epsilon^2 \frac{ \left[ \Tr \left( \Mat{V} \right) \right]^2 + \Tr \left( \TmMat \Mat{U} - \Mat{W} \TpMat \right)}{2} \, .
\end{align}


\section{Reducing the PMT to an envelope equation for eikonal functions}
\label{AppEik}

Often, the function $\psi(\Vect{q})$ can be characterized by a rapidly-varying phase $\theta(\Vect{q})$, and a complex envelope $\phi(\Vect{q})$ which varies much slower than $\theta(\Vect{q})$. If such a partition is defined, then we call $\psi(\Vect{q})$ an \textit{eikonal} function. Eikonal solutions to physical systems are frequently sought as a means to develop approximate, reduced models; an example is the WKB approximation for quantum particles~\cite{Heading62}. In reduced models, phase and envelope dynamics are typically governed by separate equations, which often makes it convenient to consider the phase and envelope as separate entities~\cite{Tracy14}. Let us therefore explore how the PMT partitions eikonal functions.

Let $\psi(\Vect{q}) = \phi(\Vect{q})e^{i\theta(\Vect{q})}$, and let $\Vect{k}(\Vect{q}) \doteq \nabla \theta(\Vect{q})$ with component functions $\{k_j (\Vect{q})\}$. Then, by induction
\begin{equation}
    \pd^n_{q_j} \, \psi(\Vect{q}) = e^{i\theta(\Vect{q})}\left[i k_j(\Vect{q}) + \pd_{q_j} \right]^n \phi(\Vect{q}) \, .
\end{equation}

\noindent An analogous result is obtained in the case of mixed partial derivatives, which implies that $\nabla$ and $\nablaNEW \doteq i\Vect{k}(\Vect{q}) + \nabla$ have the same commutation relations among their vector components; hence, the phase function effects a formal mapping from a differential operator acting on the full function $\psi(\Vect{q})$ to the differential operator acting solely on the envelope $\phi(\Vect{q})$. For example, see the definition of the envelope dispersion operator in \Ref{Dodin19a}.

For an eikonal function, the PMT is
\begin{equation}
	\Psi(\Vect{Q}) = \pm \frac{e^{i \Vect{Q}^\intercal \Mat{G} \Vect{Q} + i\theta(\Vect{q})}}{\sqrt{\det{\Mat{A}}}}\left. \exp\left( \frac{i}{4} \Lambda \dubDot \nablaNEW \nablaNEW \right) \phi(\Vect{q}) \right|_{\Vect{q} = \Mat{A}^{-1}\Vect{Q}} \, .
	\label{NIMTEik}
\end{equation}

\noindent At least for near-identity transformations, $\Psi$ can also be cast in the eikonal form. Let $\Psi(\Vect{Q}) = \Phi(\Vect{Q})e^{i\Theta(\Vect{Q})}$, then
\begin{align}
	\Phi(\Vect{Q}) =& \pm \frac{e^{i\Vect{Q}^\intercal \Mat{G} \Vect{Q} + i\theta(\Vect{q}) - i \Theta(\Vect{Q})}}{\sqrt{\det{\Mat{A}}}} \nonumber\\
	&\left. \times \exp\left(\frac{i}{4} \Lambda \dubDot \nablaNEW \nablaNEW \right) \phi(\Vect{q}) \right|_{\Vect{q} = \Mat{A}^{-1}\Vect{Q}} \, .
	\label{alphaTRANS}
\end{align}

\noindent Since $\Phi(\Vect{Q})$ is generally complex, the definition of $\Theta(\Vect{Q})$ is not unique, so choosing it is a matter of convenience (as long as $\Theta$ remains fast compared to $\Phi$). Here, we choose to define $\Theta(\Vect{Q})$ such that it is (i) real, (ii) independent of $\phi(\Vect{q})$, and (iii) simplifies the resultant expression for $\Phi(\Vect{Q})$ as much as possible. Then, the first-order truncation of \Eq{alphaTRANS} yields the eikonal partition
\begin{subequations}
    \label{NIET}
	\begin{align}
        \label{phaseTRANS}
		\Theta(\Vect{Q}) &\approx \theta\left(\Vect{q} \right) + \frac{1}{2} \Vect{Q}^\intercal \Mat{C} \Mat{A}^{-1} \Vect{Q}\nonumber\\
		&\hspace{3mm} \left.- \frac{1}{2} \left[ \Vect{k}\left(\Vect{q} \right) \right]^\intercal \Mat{A}^{-1}\Mat{B} \left[ \Vect{k}\left(\Vect{q} \right) \right]\right|_{\Vect{q} = \Mat{A}^{-1}\Vect{Q}} \, ,\\
        \Phi(\Vect{Q}) &\approx \left.\frac{\phi\left(\Vect{q} \right) + \frac{i}{2}\Tr \left\{ \Mat{A}^{-1}\Mat{B} \nullFrac \, \Mat{F}(\Vect{q}) \right\} }{\sqrt{\det{\Mat{A}}}} \right|_{\Vect{q} = \Mat{A}^{-1}\Vect{Q}} \, ,\\
        \Mat{F}(\Vect{q}) &\doteq \nabla\nabla \phi\left(\Vect{q} \right) + i\Vect{k}\left(\Vect{q} \right) \otimes \nabla \phi\left(\Vect{q} \right) \nonumber\\
        &\hspace{3mm}+ i\nabla \phi\left(\Vect{q} \right) \otimes \Vect{k}\left(\Vect{q} \right) + i\phi\left(\Vect{q} \right) \nabla \Vect{k}\left(\Vect{q} \right) \, ,
    \end{align}
\end{subequations}
    
\noindent where $\otimes$ is the tensor product. If one prefers, additional approximations can be placed on \Eqs{NIET} that are consistent with the eikonal ordering ansatz, such as neglecting $\nabla\nabla \phi$ in favor of the terms involving $\Vect{k}$.

Let us also calculate the local wavevector in the new coordinates, $\Vect{K} \doteq \pd_{\Vect{Q}} \Theta$. From \Eq{phaseTRANS} one obtains
\begin{equation}
	\Vect{K}(\Vect{Q}) = \Mat{C} \Mat{A}^{-1} \Vect{Q} + (\Mat{A}^{-1})^\intercal \Vect{R}(\Mat{A}^{-1}\Vect{Q}) \, ,
\label{KTrans}
\end{equation}

\noindent where
\begin{equation}
    \Vect{R}(\Vect{q}) \doteq \Vect{k}\left(\Vect{q} \right) - \left[\nabla \Vect{k} \left(\Vect{q} \right) \right] \Mat{A}^{-1}\Mat{B} \, \Vect{k} \left(\Vect{q} \right) \, .
\end{equation}

\noindent When $\Vect{Q}$ is obtained as $\Vect{Q} = \Mat{A}\Vect{q} + \Mat{B} \Vect{k}(\Vect{q})$, \Eq{KTrans} becomes
\begin{align}
    \Vect{K}(\Vect{q}) &= \Mat{C} \Vect{q} + \Mat{D} \Vect{k}(\Vect{q})\nonumber\\
    &\hspace{3mm} - (\Mat{A}^{-1})^\intercal \left\{\Vect{k}(\Vect{q}) -\Vect{R}\left[\Vect{q} + \Mat{A}^{-1}\Mat{B}\Vect{k}(\Vect{q}) \right] \nullFrac \right\} \, .
	\label{KqTrans}
\end{align}

\noindent Assuming that $\epsilon \doteq \|\Mat{A}^{-1}\Mat{B} \|$ is small, then $\Vect{R}\left[\Vect{q} + \Mat{A}^{-1}\Mat{B}\Vect{k}(\Vect{q})\right] \approx \Vect{k}(\Vect{q}) + O(\epsilon^2)$. Substituting this into \Eq{KqTrans} yields
\begin{align}
    \Vect{K}(\Vect{q}) = \Mat{C} \Vect{q} + \Mat{D} \Vect{k}(\Vect{q}) + O(\epsilon^2) = \Vect{P} + O(\epsilon^2) \, ,
\end{align}
    
\noindent where $\Vect{P}$ is defined in \Eq{CanonTRANS}. This shows that the transform \eq{NIET} maps ($\Vect{q}$,$\Vect{k}(\Vect{q})$) to ($\Vect{Q}$,$\Vect{K}(\Vect{Q})$) with $O(\epsilon^2)$ accuracy, which is consistent with the accuracy of \Eqs{NIET}. In this sense, this transform is natural and can be useful for modeling the propagation of eikonal waves, as we shall discuss in a separate paper.

\bibliography{../../../../../../Documents/Biblio.bib}

\begin{thebibliography}{39}%
\makeatletter
\providecommand \@ifxundefined [1]{%
 \@ifx{#1\undefined}
}%
\providecommand \@ifnum [1]{%
 \ifnum #1\expandafter \@firstoftwo
 \else \expandafter \@secondoftwo
 \fi
}%
\providecommand \@ifx [1]{%
 \ifx #1\expandafter \@firstoftwo
 \else \expandafter \@secondoftwo
 \fi
}%
\providecommand \natexlab [1]{#1}%
\providecommand \enquote  [1]{``#1''}%
\providecommand \bibnamefont  [1]{#1}%
\providecommand \bibfnamefont [1]{#1}%
\providecommand \citenamefont [1]{#1}%
\providecommand \href@noop [0]{\@secondoftwo}%
\providecommand \href [0]{\begingroup \@sanitize@url \@href}%
\providecommand \@href[1]{\@@startlink{#1}\@@href}%
\providecommand \@@href[1]{\endgroup#1\@@endlink}%
\providecommand \@sanitize@url [0]{\catcode `\\12\catcode `\$12\catcode
  `\&12\catcode `\#12\catcode `\^12\catcode `\_12\catcode `\%12\relax}%
\providecommand \@@startlink[1]{}%
\providecommand \@@endlink[0]{}%
\providecommand \url  [0]{\begingroup\@sanitize@url \@url }%
\providecommand \@url [1]{\endgroup\@href {#1}{\urlprefix }}%
\providecommand \urlprefix  [0]{URL }%
\providecommand \Eprint [0]{\href }%
\providecommand \doibase [0]{http://dx.doi.org/}%
\providecommand \selectlanguage [0]{\@gobble}%
\providecommand \bibinfo  [0]{\@secondoftwo}%
\providecommand \bibfield  [0]{\@secondoftwo}%
\providecommand \translation [1]{[#1]}%
\providecommand \BibitemOpen [0]{}%
\providecommand \bibitemStop [0]{}%
\providecommand \bibitemNoStop [0]{.\EOS\space}%
\providecommand \EOS [0]{\spacefactor3000\relax}%
\providecommand \BibitemShut  [1]{\csname bibitem#1\endcsname}%
\let\auto@bib@innerbib\@empty
\bibitem [{\citenamefont {Littlejohn}(1986)}]{Littlejohn86a}%
  \BibitemOpen
  \bibfield  {author} {\bibinfo {author} {\bibfnamefont {R.~G.}\ \bibnamefont
  {Littlejohn}},\ }\href {\doibase 10.1016/0370-1573(86)90103-1} {\bibfield
  {journal} {\bibinfo  {journal} {Phys. Rep.}\ }\textbf {\bibinfo {volume}
  {138}},\ \bibinfo {pages} {193} (\bibinfo {year} {1986})}\BibitemShut
  {NoStop}%
\bibitem [{\citenamefont {{de Gosson}}(2006)}]{deGosson06}%
  \BibitemOpen
  \bibfield  {author} {\bibinfo {author} {\bibfnamefont {M.}~\bibnamefont {{de
  Gosson}}},\ }\href {\doibase 10.1007/3-7643-7575-2} {\emph {\bibinfo {title}
  {Symplectic Geometry and Quantum Mechanics}}}\ (\bibinfo  {publisher} {Basel:
  Birkh{\"a}user},\ \bibinfo {year} {2006})\BibitemShut {NoStop}%
\bibitem [{\citenamefont {Tracy}\ and\ \citenamefont
  {Kaufman}(1993)}]{Tracy93}%
  \BibitemOpen
  \bibfield  {author} {\bibinfo {author} {\bibfnamefont {E.~R.}\ \bibnamefont
  {Tracy}}\ and\ \bibinfo {author} {\bibfnamefont {A.~N.}\ \bibnamefont
  {Kaufman}},\ }\href {\doibase 10.1103/PhysRevE.48.2196} {\bibfield  {journal}
  {\bibinfo  {journal} {Phys. Rev. E}\ }\textbf {\bibinfo {volume} {48}},\
  \bibinfo {pages} {2196} (\bibinfo {year} {1993})}\BibitemShut {NoStop}%
\bibitem [{\citenamefont {Tracy}\ \emph {et~al.}(2007)\citenamefont {Tracy},
  \citenamefont {Kaufman},\ and\ \citenamefont {Jaun}}]{Tracy07}%
  \BibitemOpen
  \bibfield  {author} {\bibinfo {author} {\bibfnamefont {E.~R.}\ \bibnamefont
  {Tracy}}, \bibinfo {author} {\bibfnamefont {A.~N.}\ \bibnamefont {Kaufman}},
  \ and\ \bibinfo {author} {\bibfnamefont {A.}~\bibnamefont {Jaun}},\ }\href
  {\doibase 10.1063/1.2748051} {\bibfield  {journal} {\bibinfo  {journal}
  {Phys. Plasmas}\ }\textbf {\bibinfo {volume} {14}},\ \bibinfo {pages}
  {082102} (\bibinfo {year} {2007})}\BibitemShut {NoStop}%
\bibitem [{\citenamefont {Gopinathan}\ \emph {et~al.}(2008)\citenamefont
  {Gopinathan}, \citenamefont {Situ}, \citenamefont {Naughton},\ and\
  \citenamefont {Sheridan}}]{Gopinathan08}%
  \BibitemOpen
  \bibfield  {author} {\bibinfo {author} {\bibfnamefont {U.}~\bibnamefont
  {Gopinathan}}, \bibinfo {author} {\bibfnamefont {G.}~\bibnamefont {Situ}},
  \bibinfo {author} {\bibfnamefont {T.~J.}\ \bibnamefont {Naughton}}, \ and\
  \bibinfo {author} {\bibfnamefont {J.~T.}\ \bibnamefont {Sheridan}},\ }\href
  {\doibase 10.1364/JOSAA.25.000108} {\bibfield  {journal} {\bibinfo  {journal}
  {J. Opt. Soc. Am. A}\ }\textbf {\bibinfo {volume} {25}},\ \bibinfo {pages}
  {108} (\bibinfo {year} {2008})}\BibitemShut {NoStop}%
\bibitem [{\citenamefont {Camara}\ \emph {et~al.}(2011)\citenamefont {Camara},
  \citenamefont {Alieva}, \citenamefont {Rodrigo},\ and\ \citenamefont
  {Calvo}}]{Camara11}%
  \BibitemOpen
  \bibfield  {author} {\bibinfo {author} {\bibfnamefont {A.}~\bibnamefont
  {Camara}}, \bibinfo {author} {\bibfnamefont {T.}~\bibnamefont {Alieva}},
  \bibinfo {author} {\bibfnamefont {J.~A.}\ \bibnamefont {Rodrigo}}, \ and\
  \bibinfo {author} {\bibfnamefont {M.~L.}\ \bibnamefont {Calvo}},\ }\href
  {\doibase 10.1364/OL.36.002441} {\bibfield  {journal} {\bibinfo  {journal}
  {Opt. Lett.}\ }\textbf {\bibinfo {volume} {36}},\ \bibinfo {pages} {2441}
  (\bibinfo {year} {2011})}\BibitemShut {NoStop}%
\bibitem [{\citenamefont {Bazarov}(2012)}]{Bazarov12}%
  \BibitemOpen
  \bibfield  {author} {\bibinfo {author} {\bibfnamefont {I.~V.}\ \bibnamefont
  {Bazarov}},\ }\href {\doibase 10.1103/PhysRevSTAB.15.050703} {\bibfield
  {journal} {\bibinfo  {journal} {Phys. Rev. ST Accel. Beams}\ }\textbf
  {\bibinfo {volume} {15}},\ \bibinfo {pages} {050703} (\bibinfo {year}
  {2012})}\BibitemShut {NoStop}%
\bibitem [{\citenamefont {Child}(2014)}]{Child14}%
  \BibitemOpen
  \bibfield  {author} {\bibinfo {author} {\bibfnamefont {M.~S.}\ \bibnamefont
  {Child}},\ }\href@noop {} {\emph {\bibinfo {title} {Semiclassical Mechanics
  with Molecular Applications}}},\ \bibinfo {edition} {2nd}\ ed.\ (\bibinfo
  {publisher} {Oxford: Oxford University Press},\ \bibinfo {year}
  {2014})\BibitemShut {NoStop}%
\bibitem [{\citenamefont {Ozaktas}\ \emph {et~al.}(1996)\citenamefont
  {Ozaktas}, \citenamefont {Arikan}, \citenamefont {Kutay},\ and\ \citenamefont
  {Bozdagi}}]{Ozaktas96}%
  \BibitemOpen
  \bibfield  {author} {\bibinfo {author} {\bibfnamefont {H.~M.}\ \bibnamefont
  {Ozaktas}}, \bibinfo {author} {\bibfnamefont {O.}~\bibnamefont {Arikan}},
  \bibinfo {author} {\bibfnamefont {M.~A.}\ \bibnamefont {Kutay}}, \ and\
  \bibinfo {author} {\bibfnamefont {G.}~\bibnamefont {Bozdagi}},\ }\href
  {\doibase 10.1109/78.536672} {\bibfield  {journal} {\bibinfo  {journal} {IEEE
  Trans. Signal Processing}\ }\textbf {\bibinfo {volume} {44}},\ \bibinfo
  {pages} {2141} (\bibinfo {year} {1996})}\BibitemShut {NoStop}%
\bibitem [{\citenamefont {Hennelly}\ and\ \citenamefont
  {Sheridan}(2005)}]{Hennelly05b}%
  \BibitemOpen
  \bibfield  {author} {\bibinfo {author} {\bibfnamefont {B.~M.}\ \bibnamefont
  {Hennelly}}\ and\ \bibinfo {author} {\bibfnamefont {J.~T.}\ \bibnamefont
  {Sheridan}},\ }\href {\doibase 10.1364/JOSAA.22.000928} {\bibfield  {journal}
  {\bibinfo  {journal} {J. Opt. Soc. Am. A}\ }\textbf {\bibinfo {volume}
  {22}},\ \bibinfo {pages} {928} (\bibinfo {year} {2005})}\BibitemShut
  {NoStop}%
\bibitem [{\citenamefont {Healy}\ and\ \citenamefont
  {Sheridan}(2010)}]{Healy10}%
  \BibitemOpen
  \bibfield  {author} {\bibinfo {author} {\bibfnamefont {J.~J.}\ \bibnamefont
  {Healy}}\ and\ \bibinfo {author} {\bibfnamefont {J.~T.}\ \bibnamefont
  {Sheridan}},\ }\href {\doibase 10.1364/JOSAA.27.000021} {\bibfield  {journal}
  {\bibinfo  {journal} {J. Opt. Soc. Am. A}\ }\textbf {\bibinfo {volume}
  {27}},\ \bibinfo {pages} {21} (\bibinfo {year} {2010})}\BibitemShut {NoStop}%
\bibitem [{\citenamefont {Koc}\ \emph {et~al.}(2010)\citenamefont {Koc},
  \citenamefont {Ozaktas},\ and\ \citenamefont {Hesselink}}]{Koc10a}%
  \BibitemOpen
  \bibfield  {author} {\bibinfo {author} {\bibfnamefont {A.}~\bibnamefont
  {Koc}}, \bibinfo {author} {\bibfnamefont {H.~M.}\ \bibnamefont {Ozaktas}}, \
  and\ \bibinfo {author} {\bibfnamefont {L.}~\bibnamefont {Hesselink}},\ }\href
  {\doibase 10.1364/JOSAA.27.001288} {\bibfield  {journal} {\bibinfo  {journal}
  {J. Opt. Soc. Am. A}\ }\textbf {\bibinfo {volume} {27}},\ \bibinfo {pages}
  {1288} (\bibinfo {year} {2010})}\BibitemShut {NoStop}%
\bibitem [{\citenamefont {Ding}\ \emph {et~al.}(2012)\citenamefont {Ding},
  \citenamefont {Pei},\ and\ \citenamefont {Liu}}]{Ding12}%
  \BibitemOpen
  \bibfield  {author} {\bibinfo {author} {\bibfnamefont {J.-J.}\ \bibnamefont
  {Ding}}, \bibinfo {author} {\bibfnamefont {S.-C.}\ \bibnamefont {Pei}}, \
  and\ \bibinfo {author} {\bibfnamefont {C.-L.}\ \bibnamefont {Liu}},\ }\href
  {\doibase 10.1364/JOSAA.29.001615} {\bibfield  {journal} {\bibinfo  {journal}
  {J. Opt. Soc. Am. A}\ }\textbf {\bibinfo {volume} {29}},\ \bibinfo {pages}
  {1615} (\bibinfo {year} {2012})}\BibitemShut {NoStop}%
\bibitem [{\citenamefont {Pei}\ and\ \citenamefont {Huang}(2016)}]{Pei16}%
  \BibitemOpen
  \bibfield  {author} {\bibinfo {author} {\bibfnamefont {S.-C.}\ \bibnamefont
  {Pei}}\ and\ \bibinfo {author} {\bibfnamefont {S.-G.}\ \bibnamefont
  {Huang}},\ }\href {\doibase 10.1364/JOSAA.33.000214} {\bibfield  {journal}
  {\bibinfo  {journal} {J. Opt. Soc. Am. A}\ }\textbf {\bibinfo {volume}
  {33}},\ \bibinfo {pages} {214} (\bibinfo {year} {2016})}\BibitemShut
  {NoStop}%
\bibitem [{\citenamefont {Sun}\ and\ \citenamefont {Li}(2018)}]{Sun18a}%
  \BibitemOpen
  \bibfield  {author} {\bibinfo {author} {\bibfnamefont {Y.-N.}\ \bibnamefont
  {Sun}}\ and\ \bibinfo {author} {\bibfnamefont {B.-Z.}\ \bibnamefont {Li}},\
  }\href {\doibase 10.1364/JOSAA.35.001346} {\bibfield  {journal} {\bibinfo
  {journal} {J. Opt. Soc. Am. A}\ }\textbf {\bibinfo {volume} {35}},\ \bibinfo
  {pages} {1346} (\bibinfo {year} {2018})}\BibitemShut {NoStop}%
\bibitem [{\citenamefont {Healy}(2018)}]{Healy18}%
  \BibitemOpen
  \bibfield  {author} {\bibinfo {author} {\bibfnamefont {J.~J.}\ \bibnamefont
  {Healy}},\ }\href {\doibase 10.1088/2040-8986/aa9e20} {\bibfield  {journal}
  {\bibinfo  {journal} {J. Opt.}\ }\textbf {\bibinfo {volume} {20}},\ \bibinfo
  {pages} {014008} (\bibinfo {year} {2018})}\BibitemShut {NoStop}%
\bibitem [{\citenamefont {Tracy}\ \emph {et~al.}(2014)\citenamefont {Tracy},
  \citenamefont {Brizard}, \citenamefont {Richardson},\ and\ \citenamefont
  {Kaufman}}]{Tracy14}%
  \BibitemOpen
  \bibfield  {author} {\bibinfo {author} {\bibfnamefont {E.~R.}\ \bibnamefont
  {Tracy}}, \bibinfo {author} {\bibfnamefont {A.~J.}\ \bibnamefont {Brizard}},
  \bibinfo {author} {\bibfnamefont {A.~S.}\ \bibnamefont {Richardson}}, \ and\
  \bibinfo {author} {\bibfnamefont {A.~N.}\ \bibnamefont {Kaufman}},\ }\href
  {\doibase 10.1017/CBO9780511667565} {\emph {\bibinfo {title} {Ray Tracing and
  Beyond: Phase Space Methods in Plasma Wave Theory}}}\ (\bibinfo  {publisher}
  {Cambridge: Cambridge University Press},\ \bibinfo {year} {2014})\BibitemShut
  {NoStop}%
\bibitem [{\citenamefont {Littlejohn}(1985)}]{Littlejohn85}%
  \BibitemOpen
  \bibfield  {author} {\bibinfo {author} {\bibfnamefont {R.~G.}\ \bibnamefont
  {Littlejohn}},\ }\href {\doibase 10.1103/PhysRevLett.54.1742} {\bibfield
  {journal} {\bibinfo  {journal} {Phys. Rev. Lett.}\ }\textbf {\bibinfo
  {volume} {54}},\ \bibinfo {pages} {1742} (\bibinfo {year}
  {1985})}\BibitemShut {NoStop}%
\bibitem [{\citenamefont {Stoler}(1981)}]{Stoler81}%
  \BibitemOpen
  \bibfield  {author} {\bibinfo {author} {\bibfnamefont {D.}~\bibnamefont
  {Stoler}},\ }\href {\doibase 10.1364/JOSA.71.000334} {\bibfield  {journal}
  {\bibinfo  {journal} {J. Opt. Soc. Am.}\ }\textbf {\bibinfo {volume} {71}},\
  \bibinfo {pages} {334} (\bibinfo {year} {1981})}\BibitemShut {NoStop}%
\bibitem [{\citenamefont {Shankar}(1994)}]{Shankar94}%
  \BibitemOpen
  \bibfield  {author} {\bibinfo {author} {\bibfnamefont {R.}~\bibnamefont
  {Shankar}},\ }\href@noop {} {\emph {\bibinfo {title} {Principles of Quantum
  Mechanics}}},\ \bibinfo {edition} {2nd}\ ed.\ (\bibinfo  {publisher} {New
  York: Plenum},\ \bibinfo {year} {1994})\BibitemShut {NoStop}%
\bibitem [{\citenamefont {Goldstein}\ \emph {et~al.}(2002)\citenamefont
  {Goldstein}, \citenamefont {Poole},\ and\ \citenamefont
  {Safko}}]{Goldstein02}%
  \BibitemOpen
  \bibfield  {author} {\bibinfo {author} {\bibfnamefont {H.}~\bibnamefont
  {Goldstein}}, \bibinfo {author} {\bibfnamefont {C.~P.}\ \bibnamefont
  {Poole}}, \ and\ \bibinfo {author} {\bibfnamefont {J.~L.}\ \bibnamefont
  {Safko}},\ }\href@noop {} {\emph {\bibinfo {title} {Classical Mechanics}}},\
  \bibinfo {edition} {3rd}\ ed.\ (\bibinfo  {publisher} {New York:
  Addison-Wesley},\ \bibinfo {year} {2002})\BibitemShut {NoStop}%
\bibitem [{\citenamefont {Luneburg}(1964)}]{Luneburg64}%
  \BibitemOpen
  \bibfield  {author} {\bibinfo {author} {\bibfnamefont {R.~K.}\ \bibnamefont
  {Luneburg}},\ }\href@noop {} {\emph {\bibinfo {title} {Mathematical Theory of
  Optics}}}\ (\bibinfo  {publisher} {Berkeley: U. California Press},\ \bibinfo
  {year} {1964})\BibitemShut {NoStop}%
\bibitem [{\citenamefont {Moshinsky}\ and\ \citenamefont
  {Quesne}(1971)}]{Moshinsky71}%
  \BibitemOpen
  \bibfield  {author} {\bibinfo {author} {\bibfnamefont {M.}~\bibnamefont
  {Moshinsky}}\ and\ \bibinfo {author} {\bibfnamefont {C.}~\bibnamefont
  {Quesne}},\ }\href {\doibase 10.1063/1.1665805} {\bibfield  {journal}
  {\bibinfo  {journal} {J. Math. Phys.}\ }\textbf {\bibinfo {volume} {12}},\
  \bibinfo {pages} {1772} (\bibinfo {year} {1971})}\BibitemShut {NoStop}%
\bibitem [{\citenamefont {Collins}(1970)}]{Collins70}%
  \BibitemOpen
  \bibfield  {author} {\bibinfo {author} {\bibfnamefont {S.~A.}\ \bibnamefont
  {Collins}},\ }\href {\doibase 10.1364/JOSA.60.001168} {\bibfield  {journal}
  {\bibinfo  {journal} {J. Opt. Soc. Am.}\ }\textbf {\bibinfo {volume} {60}},\
  \bibinfo {pages} {1168} (\bibinfo {year} {1970})}\BibitemShut {NoStop}%
\bibitem [{\citenamefont {Olver}\ \emph {et~al.}(2010)\citenamefont {Olver},
  \citenamefont {Lozier}, \citenamefont {Boisvert},\ and\ \citenamefont
  {Clark}}]{Olver10}%
  \BibitemOpen
  \bibfield  {author} {\bibinfo {author} {\bibfnamefont {F.~W.~J.}\
  \bibnamefont {Olver}}, \bibinfo {author} {\bibfnamefont {D.~W.}\ \bibnamefont
  {Lozier}}, \bibinfo {author} {\bibfnamefont {R.~F.}\ \bibnamefont
  {Boisvert}}, \ and\ \bibinfo {author} {\bibfnamefont {C.~W.}\ \bibnamefont
  {Clark}},\ }\href@noop {} {\emph {\bibinfo {title} {NIST Handbook of
  Mathematical Functions}}}\ (\bibinfo  {publisher} {Cambridge: Cambridge
  University Press},\ \bibinfo {year} {2010})\BibitemShut {NoStop}%
\bibitem [{\citenamefont {Trefethen}\ and\ \citenamefont {{Bau,
  III}}(1997)}]{Trefethen97}%
  \BibitemOpen
  \bibfield  {author} {\bibinfo {author} {\bibfnamefont {L.~N.}\ \bibnamefont
  {Trefethen}}\ and\ \bibinfo {author} {\bibfnamefont {D.}~\bibnamefont {{Bau,
  III}}},\ }\href@noop {} {\emph {\bibinfo {title} {Numerical Linear
  Algebra}}}\ (\bibinfo  {publisher} {Philadelphia: SIAM},\ \bibinfo {year}
  {1997})\BibitemShut {NoStop}%
\bibitem [{\citenamefont {Abramowitz}\ and\ \citenamefont
  {Stegun}(1970)}]{Abramowitz70}%
  \BibitemOpen
  \bibfield  {author} {\bibinfo {author} {\bibfnamefont {M.}~\bibnamefont
  {Abramowitz}}\ and\ \bibinfo {author} {\bibfnamefont {I.~A.}\ \bibnamefont
  {Stegun}},\ }\href@noop {} {\emph {\bibinfo {title} {Handbook of Mathematical
  Functions}}}\ (\bibinfo  {publisher} {New York: Dover},\ \bibinfo {year}
  {1970})\BibitemShut {NoStop}%
\bibitem [{\citenamefont {Wigner}(1932)}]{Wigner32}%
  \BibitemOpen
  \bibfield  {author} {\bibinfo {author} {\bibfnamefont {E.}~\bibnamefont
  {Wigner}},\ }\href {\doibase 10.1103/PhysRev.40.749} {\bibfield  {journal}
  {\bibinfo  {journal} {Phys. Rev.}\ }\textbf {\bibinfo {volume} {40}},\
  \bibinfo {pages} {749} (\bibinfo {year} {1932})}\BibitemShut {NoStop}%
\bibitem [{\citenamefont {Lohmann}(1993)}]{Lohmann93}%
  \BibitemOpen
  \bibfield  {author} {\bibinfo {author} {\bibfnamefont {A.~W.}\ \bibnamefont
  {Lohmann}},\ }\href {\doibase 10.1364/JOSAA.10.002181} {\bibfield  {journal}
  {\bibinfo  {journal} {J. Opt. Soc. Am. A}\ }\textbf {\bibinfo {volume}
  {10}},\ \bibinfo {pages} {2181} (\bibinfo {year} {1993})}\BibitemShut
  {NoStop}%
\bibitem [{\citenamefont {Cartwright}(1976)}]{Cartwright76}%
  \BibitemOpen
  \bibfield  {author} {\bibinfo {author} {\bibfnamefont {N.~D.}\ \bibnamefont
  {Cartwright}},\ }\href {\doibase 10.1016/0378-4371(76)90145-X} {\bibfield
  {journal} {\bibinfo  {journal} {Physica A}\ }\textbf {\bibinfo {volume}
  {83}},\ \bibinfo {pages} {210} (\bibinfo {year} {1976})}\BibitemShut
  {NoStop}%
\bibitem [{\citenamefont {{O'Connell}}\ and\ \citenamefont
  {Wigner}(1981)}]{OConnell81}%
  \BibitemOpen
  \bibfield  {author} {\bibinfo {author} {\bibfnamefont {R.~F.}\ \bibnamefont
  {{O'Connell}}}\ and\ \bibinfo {author} {\bibfnamefont {E.~P.}\ \bibnamefont
  {Wigner}},\ }\href {\doibase 10.1016/0375-9601(81)90881-1} {\bibfield
  {journal} {\bibinfo  {journal} {Phys. Lett. A}\ }\textbf {\bibinfo {volume}
  {85}},\ \bibinfo {pages} {121} (\bibinfo {year} {1981})}\BibitemShut
  {NoStop}%
\bibitem [{\citenamefont {Kaminsky}\ and\ \citenamefont
  {Simanjuntak}(2005)}]{Kaminsky05}%
  \BibitemOpen
  \bibfield  {author} {\bibinfo {author} {\bibfnamefont {E.~J.}\ \bibnamefont
  {Kaminsky}}\ and\ \bibinfo {author} {\bibfnamefont {L.}~\bibnamefont
  {Simanjuntak}},\ }\href {\doibase 10.1117/12.605426} {\bibfield  {journal}
  {\bibinfo  {journal} {Proc. SPIE}\ }\textbf {\bibinfo {volume} {5778}},\
  \bibinfo {pages} {894} (\bibinfo {year} {2005})}\BibitemShut {NoStop}%
\bibitem [{\citenamefont {Xiao}\ \emph {et~al.}(1987)\citenamefont {Xiao},
  \citenamefont {Wu},\ and\ \citenamefont {Kimble}}]{Xiao87}%
  \BibitemOpen
  \bibfield  {author} {\bibinfo {author} {\bibfnamefont {M.}~\bibnamefont
  {Xiao}}, \bibinfo {author} {\bibfnamefont {L.-A.}\ \bibnamefont {Wu}}, \ and\
  \bibinfo {author} {\bibfnamefont {H.~J.}\ \bibnamefont {Kimble}},\ }\href
  {\doibase 10.1103/PhysRevLett.59.278} {\bibfield  {journal} {\bibinfo
  {journal} {Phys. Rev. Lett.}\ }\textbf {\bibinfo {volume} {59}},\ \bibinfo
  {pages} {278} (\bibinfo {year} {1987})}\BibitemShut {NoStop}%
\bibitem [{\citenamefont {Savitzky}\ and\ \citenamefont
  {Golay}(1964)}]{Savitzky64}%
  \BibitemOpen
  \bibfield  {author} {\bibinfo {author} {\bibfnamefont {A.}~\bibnamefont
  {Savitzky}}\ and\ \bibinfo {author} {\bibfnamefont {M.~J.~E.}\ \bibnamefont
  {Golay}},\ }\href {\doibase 10.1021/ac60214a047} {\bibfield  {journal}
  {\bibinfo  {journal} {Anal. Chem.}\ }\textbf {\bibinfo {volume} {36}},\
  \bibinfo {pages} {1627} (\bibinfo {year} {1964})}\BibitemShut {NoStop}%
\bibitem [{\citenamefont {Heading}(1962)}]{Heading62}%
  \BibitemOpen
  \bibfield  {author} {\bibinfo {author} {\bibfnamefont {J.}~\bibnamefont
  {Heading}},\ }\href {http://store.doverpublications.com/0486497429.html}
  {\emph {\bibinfo {title} {An Introduction to Phase-Integral Methods}}}\
  (\bibinfo  {publisher} {London: Methuen},\ \bibinfo {year}
  {1962})\BibitemShut {NoStop}%
\bibitem [{\citenamefont {Dragt}(2005)}]{Dragt05}%
  \BibitemOpen
  \bibfield  {author} {\bibinfo {author} {\bibfnamefont {A.~J.}\ \bibnamefont
  {Dragt}},\ }\href {\doibase 10.1196/annals.1350.025} {\bibfield  {journal}
  {\bibinfo  {journal} {Ann. N. Y. Acad. Sci.}\ }\textbf {\bibinfo {volume}
  {1045}},\ \bibinfo {pages} {291} (\bibinfo {year} {2005})}\BibitemShut
  {NoStop}%
\bibitem [{\citenamefont {Dragt}(1982)}]{Dragt82}%
  \BibitemOpen
  \bibfield  {author} {\bibinfo {author} {\bibfnamefont {A.~J.}\ \bibnamefont
  {Dragt}},\ }\href {\doibase 10.1063/1.33615} {\bibfield  {journal} {\bibinfo
  {journal} {AIP Conf. Proc.}\ }\textbf {\bibinfo {volume} {87}},\ \bibinfo
  {pages} {147} (\bibinfo {year} {1982})}\BibitemShut {NoStop}%
\bibitem [{\citenamefont {Hall}(2015)}]{Hall15}%
  \BibitemOpen
  \bibfield  {author} {\bibinfo {author} {\bibfnamefont {B.~C.}\ \bibnamefont
  {Hall}},\ }\href {\doibase 10.1007/978-3-319-13467-3} {\emph {\bibinfo
  {title} {Lie Groups, Lie Algebras, and Representations: An Elementary
  Introduction}}}\ (\bibinfo  {publisher} {Berlin: Springer},\ \bibinfo {year}
  {2015})\BibitemShut {NoStop}%
\bibitem [{\citenamefont {Dodin}\ \emph {et~al.}(2019)\citenamefont {Dodin},
  \citenamefont {Ruiz}, \citenamefont {Yanagihara}, \citenamefont {Zhou},\ and\
  \citenamefont {Kubo}}]{Dodin19a}%
  \BibitemOpen
  \bibfield  {author} {\bibinfo {author} {\bibfnamefont {I.~Y.}\ \bibnamefont
  {Dodin}}, \bibinfo {author} {\bibfnamefont {D.~E.}\ \bibnamefont {Ruiz}},
  \bibinfo {author} {\bibfnamefont {K.}~\bibnamefont {Yanagihara}}, \bibinfo
  {author} {\bibfnamefont {Y.}~\bibnamefont {Zhou}}, \ and\ \bibinfo {author}
  {\bibfnamefont {S.}~\bibnamefont {Kubo}},\ }\href {\doibase
  10.1063/1.5095076} {\bibfield  {journal} {\bibinfo  {journal} {Phys.
  Plasmas}\ }\textbf {\bibinfo {volume} {26}},\ \bibinfo {pages} {072110}
  (\bibinfo {year} {2019})}\BibitemShut {NoStop}%
\end{thebibliography}%
\bibliographystyle{apsrev4-1}
\end{document}